\documentclass{article}
\usepackage{amssymb,amsmath}
\usepackage[dvipdfmx]{graphicx}
\usepackage{bbm}
\usepackage{bm}
\usepackage{caption}
\usepackage{color}
\usepackage{comment}
\usepackage{comment}
\usepackage{enumerate}
\usepackage{ytableau}
\usepackage{here}
\usepackage{subfig}
\DeclareMathOperator*{\Tr}{{\rm Tr}}

\numberwithin{equation}{section}

\begin{document}

\thispagestyle{empty}
\begin{flushright}

\end{flushright}
\vskip1.5cm
\begin{center}
{\Large \bf Mirror symmetry of 3d $\mathcal{N}=4$ gauge theories \\
\vskip0.35cm
and supersymmetric indices \\
}

\vskip2cm
Tadashi Okazaki\footnote{tokazaki@perimeterinstitute.ca}

\bigskip
{\it
Perimeter Institute for Theoretical Physics,\\
Waterloo, Ontario, Canada N2L 2Y5
}

\end{center}

\vskip1cm
\begin{abstract}
We compute supersymmetric indices 
to test mirror symmetry of three-dimensional $\mathcal{N}=4$ gauge theories 
and dualities of half-BPS enriched boundary conditions and interfaces in four-dimensional $\mathcal{N}=4$ Super Yang-Mills theory. 
We find the matching of indices as strong evidences for various dualities of the 3d interfaces conjectured by Gaiotto and Witten 
under the action of S-duality in Type IIB string theory. 
\end{abstract}


\newpage
\tableofcontents

\section{Introduction and conclusions}
\label{sec_intro}

Three-dimensional $\mathcal{N}=4$ supersymmetric gauge theories have 
a moduli space of supersymmetric vacua consisting of a Higgs branch $\mathcal{M}_{H}$ and a Coulomb branch $\mathcal{M}_{C}$ 
which are hyperk\"{a}hler manifolds. 
The theories have an intriguing duality, known as mirror symmetry, 
which relates theories with completely different UV descriptions 
where the Higgs branch and the Coulomb branch are exchanged 
and FI parameters and mass parameters are exchanged \cite{Intriligator:1996ex}. 
This mapping is very non-trivial. 
The Higgs branch is a hyperk\"{a}hler quotient realized as a zero locus of the $D$-term constraints divided by the gauge group. 
On the Higgs branch the gauge symmetry is broken completely and the Higgs branch is not affected by the quantum correction. 
On the Coulomb branch the gauge group is generically broken to its maximal torus. 
Unlike the Higgs branch, the Coulomb branch receives perturbative and non-perturbative quantum corrections. 

The quantum Coulomb branch is studied in terms of Hilbert series in \cite{Cremonesi:2013lqa}. 
The Hilbert series is a generating function that counts chiral operators on the branches $\mathcal{M}_{H/C}$ of vacua, 
graded by their dimensions and quantum numbers under global symmetry. 
It encodes the quantum numbers of generators and relations of the chiral ring $\mathbb{C}[\mathcal{M}_{H/C}]$ of the corresponding branches $\mathcal{M}_{H/C}$ of vacua. 
Subsequently, the quantum Coulomb branch has been described in \cite{Bullimore:2015lsa} in terms of the Abelianization map 
that translates the vev of a monopole operator into a linear combination of Abelian monopole operator vevs in the low-energy effective theory. 
Furthermore, the quantum Coulomb branch has been mathematically defined in \cite{Nakajima:2015txa,Braverman:2016wma,Braverman:2016pwk}. 

Three-dimensional $\mathcal{N}=4$ gauge theories can be realized in Type IIB string theory using the brane configuration \cite{Hanany:1996ie}. 
Mirror symmetry can be viewed as arising from S-duality of Type IIB string theory. 
Type IIB brane construction is extended to further study of 
half-BPS boundary conditions and interfaces for four-dimensional $\mathcal{N}=4$ Super Yang-Mills (SYM) theory \cite{Gaiotto:2008sa, Gaiotto:2008sd, Gaiotto:2008ak}, 
quarter-BPS corner configuration for four-dimensional $\mathcal{N}=4$ SYM theory \cite{Gaiotto:2017euk, Gaiotto:2019jvo}, 
half-BPS boundary conditions for three-dimensional $\mathcal{N}=4$ gauge theory \cite{Chung:2016pgt} 
and two-dimensional $\mathcal{N}=(0,4)$ supersymmetric gauge theories \cite{Hanany:2018hlz}. 
S-duality turns out to give a physical underpinning to Geometric Langlands Program 
\cite{Kapustin:2006pk, Gaiotto:2017euk, Creutzig:2017uxh, Frenkel:2018dej} 
and Symplectic Duality \cite{MR3594664, Braden:2014iea}. 

In this paper we compute supersymmetric full- and half-indices 
to test mirror symmetry of 3d $\mathcal{N}=4$ gauge theories 
and to extend the analysis in \cite{Gaiotto:2019jvo} for the dualities of half-BPS boundary conditions and interfaces in 4d $\mathcal{N}=4$ SYM theory. 
\footnote{See \cite{Okazaki:2019bok} for further dualities of $\mathcal{N}=(0,4)$ half-BPS boundaries for 3d $\mathcal{N}=4$ gauge theories 
and quarter-BPS corners for 4d $\mathcal{N}=4$ gauge theories. }
These dualities were originally conjectured by Gaiotto and Witten \cite{Gaiotto:2008ak}. 
There have been plenty of works on the subject of 3d $\mathcal{N}=4$ full superconformal indices 
\cite{Kim:2009wb,Imamura:2009hc,Kim:2012uz,Hwang:2012jh,Hwang:2015wna,Nosaka:2018eip,Bachas:2019jaa}, 
however, general tests of mirror symmetry for 3d $\mathcal{N}=4$ gauge theories have not appeared in the literature 
except for the simplest Abelian mirror symmetry \cite{Razamat:2014pta}, 
in contrast to 3d $\mathcal{N}=2$ gauge theories \cite{Imamura:2011su, Kapustin:2011jm, Krattenthaler:2011da}. 
Additionally, one drawback of the 3d full-indices is that they are insensitive to the boundary conditions for 4d $\mathcal{N}=4$ gauge theory, 
which involve singular boundary conditions specified by the Nahm poles \cite{Gaiotto:2008sa}. 
In order to address these open issues, 
we evaluate supersymmetric 4d half-indices and 3d full-indices 
which count local operators both in four and three dimensions  
and present many identities of indices by checking several terms of series expansion. 
As discussed in \cite{Razamat:2014pta}, after the special limits, 
the full-indices of 3d $\mathcal{N}=4$ gauge theories reduce to the Hilbert series. 
As a result, the identities of indices provide promotions of identities of Hilbert series discussed in \cite{Cremonesi:2013lqa, Razamat:2014pta}. 
Furthermore, the full-indices of 3d $\mathcal{N}=4$ gauge theories can be used to generalize the 
analysis in \cite{Gaiotto:2019jvo} of the half-BPS boundary conditions and interfaces for 4d $\mathcal{N}=4$ SYM theory. 
We present general half-indices for enriched half-BPS boundary conditions and interfaces in 4d $\mathcal{N}=4$ gauge theory 
and check precise matching of the indices for dual pairs.

The organization of this paper is straightforward. 
In section \ref{sec_index} 
we briefly review the supersymmetric indices introduced in \cite{Gaiotto:2019jvo} 
and present formulae and some properties of half-indices for 4d $\mathcal{N}=4$ gauge theories and full-indices for 3d $\mathcal{N}=4$ gauge theories. 
In section \ref{sec_3dmirror1} 
we evaluate full-indices for 3d $\mathcal{N}=4$ Abelian gauge theories and check mirror symmetry. 
In section \ref{sec_3dmirror2} 
we further examine mirror symmetry by computing full-indices for 3d $\mathcal{N}=4$ non-Abelian gauge theories. 
We also briefly check in section \ref{sec_3dseiberg} Seiberg-like duality for 3d $\mathcal{N}=4$ gauge theories 
proposed in \cite{Gaiotto:2008ak} between ugly theory and good theory in terms of full-indices.  
In section \ref{sec_3dm4dS1} 
we discuss the half-BPS  enriched boundary conditions 
for 4d $\mathcal{N}=4$ SYM theory which involve 3d $\mathcal{N}=4$ gauge theories. 
We present strong evidence for dualities between them conjectured from string theory by calculating half-indices, 
which contain non-regular Nahm pole b.c. 
Finally in section \ref{sec_3dm4dS2} 
we study the half-BPS interfaces in 4d $\mathcal{N}=4$ $U(1)$ gauge theory 
including 3d $\mathcal{N}=4$ Abelian gauge theories. 
We test dualities between the interfaces by computing half-indices. 
In Appendix \ref{app_exp} 
we show 
the $q$-expansions of indices generated by Mathematica 
as well as the confirmed orders which the indices agree up to.

\section{Indices}
\label{sec_index}
We begin with a definition of the quarter-index introduced in \cite{Gaiotto:2019jvo}. 
It is a generalization of superconformal index in that 
it can count local operators living in different dimensions, i.e. in 4d bulk, 3d boundary and 2d junction.  
When the configuration has a trivial junction, it becomes the half-index that counts boundary local operators, 
while for the trivial interface, it becomes the full-index that counts bulk local operators. 
The quarter-index can be defined as the trace over the cohomology of the preserved supercharges
\begin{align}
\label{INDEX_def}
\mathbb{IV}(t,x;q)
&:={\Tr}_{\mathrm{Op}}(-1)^{F}q^{J+\frac{H+C}{4}}t^{H-C} x^{f}. 
\end{align}
Here $F$ is the Fermion number, 
$J$ is the generator of the $U(1)_{J}$ rotational symmetry in the space-time 
on which local operators are supported. 
$H$ and $C$ stands for the Cartan generators of the $SU(2)_{H}$ and $SU(2)_{C}$ R-symmetry groups respectively. 
$f$ is the Cartan generator of the global symmetry. 
The choice of fugacity in the index (\ref{INDEX_def}) 
is fixed in such a way that the power of $q$ is always strictly positive for a non-trivial local operator, by a unitarity bound. 
This ensures the convergence of the index. 
Consequently, the index can be a formal power series in $q$ 
whose coefficients are Laurent polynomials in the other fugacities. 

In this paper we focus on the configurations of 3d $\mathcal{N}=4$ gauge theories 
which may couple to 4d $\mathcal{N}=4$ gauge theories 
so that the indices (\ref{INDEX_def}) reduce to the full-indices $\mathbb{I}$ of 3d $\mathcal{N}=4$ gauge theories 
and/or the half-indices $\mathbb{II}$ of 4d $\mathcal{N}=4$ gauge theories. 
One may compute the indices for appropriate configurations by a localization procedure. 
However, we will not pursue that direction in this paper, 
instead we will count local operators seriously from physical consideration.  

In the description of indices we use the following notation by defining $q$-shifted factorial
\begin{align}
\label{qpoch_def}
(a;q)_{0}&:=1,\qquad
(a;q)_{n}:=\prod_{k=0}^{n-1}(1-aq^{k}),\qquad 
(q)_{n}:=\prod_{k=1}^{n}(1-q^{k}),\quad 
\quad  n\ge1,
\nonumber \\
(a;q)_{\infty}&:=\prod_{k=0}^{\infty}(1-aq^{k}),\qquad 
(q)_{\infty}:=\prod_{k=1}^{\infty} (1-q^k), 
\nonumber\\
(a^{\pm};q)_{\infty}&:=(a;q)_{\infty}(a^{-1};q)_{\infty}
\end{align}
where $a$ and $q$ are complex variables with $|q|<1$.

We compute the indices to test the 3d dualities and dualities of the half-BPS boundary conditions/interfaces 
which are conjectured from string theory \cite{Gaiotto:2008ak}. 
We consider five types of branes in Type IIB string theory whose world-volumes span the following directions:
\begin{itemize}
\item D3-branes extended along $x^{0}x^{1}x^{2}x^{6}$, 
\item NS5-branes extended along $x^{0}x^{1}x^{2}x^{3}x^{4}x^{5}$, 
\item D5-branes extended along $x^{0}x^{1}x^{2}x^{7}x^{8}x^{9}$, 
\item NS5$'$-branes extended along $x^{0}x^{1}x^{6}x^{7}x^{8}x^{9}$, 
\item D5$'$-branes extended along $x^{0}x^{1}x^{3}x^{4}x^{5}x^{6}$
\end{itemize}
In other words, the brane configuration is summarized as
\begin{align}
\label{brane1}
\begin{array}{ccccccccccc}
&0&1&2&3&4&5&6&7&8&9\\
\textrm{D3}
&\circ&\circ&\circ&-&-&-&\circ&-&-&- \\
\textrm{NS5}
&\circ&\circ&\circ&\circ&\circ&\circ&-&-&-&- \\
\textrm{D5}
&\circ&\circ&\circ&-&-&-&-&\circ&\circ&\circ \\
\textrm{NS5$'$}
&\circ&\circ&-&-&-&-&\circ&\circ&\circ&\circ \\
\textrm{D5$'$}
&\circ&\circ&-&\circ&\circ&\circ&\circ&-&-&- \\
\end{array}
\end{align}
The 3d $\mathcal{N}=4$ gauge theories are realized by considering D3-branes which are finite segments in the $x^6$ direction between NS5-branes 
and may intersect with D5-branes \cite{Hanany:1996ie}. 
The half-BPS boundaries and interfaces in 4d $\mathcal{N}=4$ gauge theories 
are realized by considering D3-branes which are (semi-)infinite D3-branes which end on or pass through 
a sequence of NS5- and D5-branes \cite{Gaiotto:2008sa}. 

Such brane setup is a nice tool for finding mirror pairs of 3d $\mathcal{N}=4$ gauge theories 
and dual pairs of half-BPS boundary conditions and interfaces in 4d $\mathcal{N}=4$ SYM theory 
by studying the action of S-duality \cite{Gaiotto:2008ak}.

\subsection{Indices of 4d $\mathcal{N}=4$ SYM theory}
\label{sec_index4d}

\subsubsection{Full-indices}
\label{sec_fullindex4d}
Four-dimensional $\mathcal{N}=4$ SYM theory has $SU(4)_{R}$ R-symmetry. 
It contains the adjoint scalar fields transforming as ${\bf 6}$ under the $SU(4)_{R}$. 
Let $X$ and $Y$ be the scalar fields transforming as  $({\bf 1},{\bf 3})$ and $({\bf 3},{\bf 1})$ under the $SU(2)_{C}\times SU(2)_{H}$ $\subset$ $SU(4)_{R}$. 
In the brane construction (\ref{brane1}), the scalar fields $X$ and $Y$ describe the positions of D3-branes along 
the $(x^{7}$, $x^{8}$, $x^{9})$ directions and $(x^{3}$, $x^{4}$, $x^{5})$ directions respectively. 
The theory also has the 4d gauginos $\lambda$ transforming as $({\bf 2},{\bf 2})$ 
under the $SU(2)_{C}$ $\times$ $SU(2)_{H}$. 

The local operators in 4d $\mathcal{N}=4$ gauge theory of gauge group $G$ which contribute to index have charges 
\begin{align}
\label{4dn4_ch}
\begin{array}{c|cccc}
&\partial^{n}X&\partial^{n}Y&\partial^{n}\lambda&\partial^{n}\overline{\lambda} \\ \hline
G&\textrm{adj}&\textrm{adj}&\textrm{adj}&\textrm{adj} \\
U(1)_{J}&n&n&n+\frac12&n+\frac12 \\
U(1)_{C}&0&2&+&+ \\
U(1)_{H}&2&0&+&+  \\
\textrm{fugacity}
&q^{n+\frac12}t^{2}s^{\alpha}
&q^{n+\frac12}t^{-2}s^{\alpha}
&-q^{n+1}s^{\alpha}
&-q^{n+1}s^{\alpha} \\
\end{array}
\end{align}

From (\ref{4dn4_ch}) one can express the index for 4d $\mathcal{N}=4$ $U(N)$ gauge theory as
\begin{align}
\label{4duN_INDEX}
\mathbb{I}^{\textrm{4d $U(N)$}}(t;q)
&=\frac{1}{N!} \frac{(q)_{\infty}^{2N}}
{(q^{\frac12} t^{2};q)_{\infty}^N(q^{\frac12} t^{-2};q)_{\infty}^N}
\oint 
\prod_{i=1}^{N}
\frac{ds_{i}}{2\pi is_{i}}
\prod_{i\neq j}
\frac{\left(\frac{s_{i}}{s_{j}};q\right)_{\infty}
\left(q\frac{s_{i}}{s_{j}};q\right)_{\infty}
}
{
\left(q^{\frac12} t^{2}\frac{s_{i}}{s_{j}};q\right)_{\infty}
\left(q^{\frac12} t^{-2}\frac{s_{i}}{s_{j}};q\right)_{\infty}
}. 
\end{align}
Here the denominator comes from the scalar fields $X$ and $Y$ 
while the numerator captures the 4d gauginos. 
The integration contour for gauge fugacities $s_{i}$ is taken as a unit torus $\mathbb{T}^{N}$.

\subsubsection{Half-indices}
\label{sec_hindex4d}
Four-dimensional $\mathcal{N}=4$ SYM theory admits half-BPS boundary conditions 
which preserve three-dimensional $\mathcal{N}=4$ supersymmetry 
with the R-symmetry group $SU(4)_{R}$ broken down to $SU(2)_{C}$ $\times$ $SU(2)_{H}$. 
In the brane setup (\ref{brane1}), they arise when parallel D3-branes end on a single fivebrane.  
There are two types of three-dimensional boundaries/interfaces at $x^2=0$ realized by NS5$'$- and D5$'$-branes 
and those at $x^6=0$ realized by NS5- and D5-branes.

Let us consider the half-BPS boundary conditions for 4d $\mathcal{N}=4$ $U(1)$ gauge theory. 
When a single D3-brane ends on the NS5-brane and D5-brane, 
one finds the Neumann b.c. $\mathcal{N}$ and Dirichlet b.c. $\mathcal{D}$ at $x^6=0$ for $U(1)$ gauge theory respectively:
\begin{align}
\label{4d_bc1}
\begin{array}{cccc}
\mathcal{N}:& F_{6\mu}|_{\partial}=0,& \partial_{\mu}X|_{\partial}=0,&\partial_{6}Y|_{\partial}=0 \\
\mathcal{D}:& F_{\mu\nu}|_{\partial}=0,& \partial_{6}X|_{\partial}=0,&\partial_{\mu}Y|_{\partial}=0 \\
\end{array}
\qquad \mu,\nu=0,1,2
\end{align}
On the other hand, when the NS5$'$-brane and D5$'$-brane end on a single D3-brane 
one obtains the Neumann b.c. $\mathcal{N}'$ and Dirichlet b.c. $\mathcal{D}'$ at $x^2=0$ for $U(1)$ gauge theory respectively:
\begin{align}
\label{4d_bc2}
\begin{array}{cccc}
\mathcal{N}':& F_{2\mu}|_{\partial}=0,& \partial_{2}X|_{\partial}=0,&\partial_{\mu}Y|_{\partial}=0 \\
\mathcal{D}':& F_{\mu\nu}|_{\partial}=0,& \partial_{\mu}X|_{\partial}=0,&\partial_{2}Y|_{\partial}=0 \\
\end{array}
\qquad \mu,\nu=0,1,6
\end{align}

The half-indices of the Neumann b.c. $\mathcal{N}$ and Dirichlet b.c. $\mathcal{D}'$ 
for 4d $\mathcal{N}=4$ $U(1)$ gauge theory takes the form
\begin{align}
\label{4du1_hINDEX_N1}
\mathbb{II}^{\textrm{4d $U(1)$}}_{\mathcal{N}}(t;q)=
\mathbb{II}^{\textrm{4d $U(1)$}}_{\mathcal{D}'}(t;q)
&=\frac{(q)_{\infty}}{(q^{\frac12}t^{-2};q)_{\infty}}. 
\end{align}
The denominator is associated to the scalar fields $Y$ charged under $U(1)_{C}$ 
while the numerator correspond to a half of the 4d gauginos. 
Likewise, the half-indices of Neumann b.c. $\mathcal{N}'$ and Dirichlet b.c. $\mathcal{D}$ is 
\begin{align}
\label{4du1_hINDEX_N2}
\mathbb{II}^{\textrm{4d $U(1)$}}_{\mathcal{D}}(t;q)=
\mathbb{II}^{\textrm{4d $U(1)$}}_{\mathcal{N}'}(t;q)
&=\frac{(q)_{\infty}}{(q^{\frac12}t^{2};q)_{\infty}}. 
\end{align}
The denominator captures the scalar fields $X$ charged under $U(1)_{H}$ 
whereas the numerator is associated to a half of the 4d gauginos.

The half-BPS boundary conditions corresponding to $N$ D3-branes ending on a single NS5-brane (or NS5')
are also Neumann b.c. for the $U(N)$ gauge theory. We can denote them as $\mathcal{N}$ and $\mathcal{N}'$
as in the  Abelian case. 
By contrast, when $N$ multiple D3-branes end on a single D5-brane, one finds a singular boundary condition 
associated to a regular Nahm pole \cite{Diaconescu:1996rk,Gaiotto:2008sa}. 
A single D5-brane or D5$'$-brane on which $N$ D3-branes end give rise to the Nahm or Nahm$'$ pole boundary conditions: 
\begin{align}
\label{4d_nahmbc}
\begin{array}{ccccc}
\textrm{Nahm}:& F_{\mu\nu}|_{\partial}=0,& D_{6}\vec{X}+\vec{X}\times \vec{X}|_{\partial}=0,&D_{\mu}\vec{Y}|_{\partial}=0&\qquad \mu,\nu=0,1,2 \\
\textrm{Nahm}':& F_{\mu\nu}|_{\partial}=0,& D_{\mu}\vec{X}|_{\partial}=0,&D_{2}\vec{Y}+\vec{Y}\times \vec{Y}|_{\partial}=0&\qquad  \mu,\nu=0,1,6 \\
\end{array}
\end{align}
where we denote the scalar fields by $\vec{X}$ and $\vec{Y}$ as they are the $SU(2)_{H}$ triplet and the $SU(2)_{C}$ triplet respectively. 
The Nahm  equations for the scalar fields $\vec{X}$ and $\vec{Y}$ have singular solutions
\begin{align}
\label{4d_nahmbc2}
\vec{X}(x^6)&=\frac{\vec{\mathfrak{t}}}{x^6},& 
\vec{Y}(x^2)&=\frac{\vec{\mathfrak{t}}}{x^2}
\end{align}
where $\vec{\mathfrak{t}}$ $=$ $(\mathfrak{t}_{1},\mathfrak{t}_{2},\mathfrak{t}_{3})$ is a triplet of elements of 
the Lie algebra $\mathfrak{g}$ $=$ $\mathfrak{u}(N)$ obeying the commutation relation $[\mathfrak{t}_{1},\mathfrak{t}_{2}]=\mathfrak{t}_{3}$ and cyclic permutation thereof. 
The choice of $\vec{\mathfrak{t}}$ specifies a homomorphism of Lie algebras $\rho:$ $\mathfrak{su}(2)$ $\rightarrow$ $\mathfrak{g}$ 
which maps the fundamental representation of $U(N)$ to the dimension $N$ irreducible representation of $\mathfrak{su}(2)$. 
When $N$ D3-branes end on multiple D5-branes, one finds other Nahm poles, 
including the Dirichlet b.c. as the trivial Nahm pole corresponding to the case with $N$ D5-branes.

The half-index of Neumann b.c. $\mathcal{N}$ for 4d $\mathcal{N}=4$ $U(N)$ gauge theory takes the form
\begin{align}
\label{4duN_hINDEX_N1}
\mathbb{II}_{\mathcal{N}}^{\textrm{4d $U(N)$}}(t;q)
&=
\frac{1}{N!}
\frac{(q)_{\infty}^{N}}
{(q^{\frac12} t^{-2};q)_{\infty}^{N}}
\oint 
\prod_{i=1}^{N}
\frac{ds_{i}}{2\pi is_{i}}
\prod_{i\neq j}
\frac{\left(
\frac{s_{i}}{s_{j}};q
\right)_{\infty}
}
{
\left(
q^{\frac12} t^{-2}\frac{s_{i}}{s_{j}};q
\right)_{\infty}
}.
\end{align}
Again the integration contour for gauge fugacities $s_{i}$ is a unit torus $\mathbb{T}^{N}$. 
The half-index of Dirichlet b.c. $\mathcal{D}$ for 4d $\mathcal{N}=4$ $U(N)$ gauge theory is given by
\begin{align}
\label{4duN_hINDEX_D1}
\mathbb{II}_{\mathcal{D}}^{\textrm{4d $U(N)$}}(t,z_{i};q)
&=\frac{(q)_{\infty}^{N} }{(q^{\frac12} t^2;q)_{\infty}^{N}}
\prod_{i\neq j}
\frac{
\left(q\frac{z_{i}}{z_{j}};q\right)_{\infty}}
{\left(q^{\frac12} t^2 \frac{z_{i}}{z_{j}};q\right)_{\infty}}
\end{align}
where $z_{i}$ is the fugacities associated to the boundary $U(N)$ global symmetry. 
The half-index for Nahm pole boundary conditions 
in 4d $\mathcal{N}=4$ $U(N)$ gauge theory is
\begin{align}
\label{4duN_nahm}
\mathbb{II}_{\textrm{Nahm}}^{\textrm{4d $U(N)$}}(t;q)
&=
\prod_{k=1}^{N}
\frac{(q^{\frac{k+1}{2}}t^{2(k-1)};q)_{\infty}}
{(q^{\frac{k}{2}}t^{2k};q)_{\infty}}. 
\end{align}
As discussed in \cite{Gaiotto:2019jvo}, 
the duality between the Neumann b.c. and the regular Nahm pole b.c. 
implies equality between the half-indices (\ref{4duN_hINDEX_N1}) and (\ref{4duN_nahm}). 
In section \ref{sec_3dm4dS1} we discuss more general dual descriptions of the half-indices, 
including half-index (\ref{4duN_hINDEX_D1}) for Dirichlet b.c. 

We can get 
similar expressions for the mirror boundary conditions, 
i.e. $\mathcal{N}'$, $\mathcal{D}'$ and Nahm$'$ by setting $t\rightarrow t^{-1}$.

\subsection{Indices of 3d $\mathcal{N}=4$ gauge theory}
\label{sec_index4d}

The 3d $\mathcal{N}=4$ hypermultiplet consists of 
a pair of complex scalars $\mathbb{H}$, $\widetilde{\mathbb{H}}$ 
forming a doublet of $SU(2)_{H}$ 
and a pair of complex fermions $\psi_{+}^{\mathbb{H}}$, $\psi_{+}^{\widetilde{\mathbb{H}}}$ 
forming a doublet of $SU(2)_{C}$. 
The charges of 3d $\mathcal{N}=4$ hypermultiplet is
\begin{align}
\label{3dn4_hm_ch}
\begin{array}{c|cccccc}
&\mathbb{H}&\widetilde{\mathbb{H}}&\psi_{+}^{\mathbb{H}}&\psi_{+}^{\widetilde{\mathbb{H}}}&\overline{\psi}_{-}^{\mathbb{H}}&\overline{\psi}_{-}^{\widetilde{\mathbb{H}}} \\ \hline
U(1)_{C}&0&0&-&-&+&+ \\
U(1)_{H}&+&+&0&0&0&0 \\
\end{array}
\end{align}

The 3d $\mathcal{N}=4$ Abelian vector multiplet consists of 
a 3d $U(1)$ gauge field $A_{\mu}^{\textrm{3d}}$, 
three scalars, which we denote by real and complex scalars $\sigma$, $\varphi$ forming the $SU(2)_{C}$ triplet, 
and two complex fermions $(\lambda_{\alpha}^{\textrm{3d}}, \eta_{\alpha}^{\textrm{3d}})$. 
The charges of 3d $\mathcal{N}=4$ vector multiplet is 
\begin{align}
\label{3dn4_vm_ch}
\begin{array}{c|ccccccc}
&A_{\mu}^{\textrm{3d}}&\sigma&\varphi&\lambda_{\pm}^{\textrm{3d}}&\overline{\lambda}_{\pm}^{\textrm{3d}}
&\eta_{\pm}^{\textrm{3d}}&\overline{\eta}_{\pm}^{\textrm{3d}} \\ \hline
U(1)_{C}&0&0&2&+&-&+&- \\
U(1)_{H}&0&0&0&+&-&-&+ \\
\end{array}
\end{align}

The 3d $\mathcal{N}=4$ superalgebra has an outer automorphism that interchanges $SU(2)_{C}$ and $SU(2)_{H}$. 
This automorphism makes the ordinary supermultiplets into twisted supermultiplets. 
The twisted hyper and vector multiplets can be obtained by exchanging the $U(1)_{H}$ and $U(1)_{C}$ charges 
 of the hypermultiplet and vector multiplet respectively. 
 
In three dimensions, a photon is electric magnetic dual to a scalar field, which we call dual photon. 
The dual photon is periodic when the gauge group is compact 
and the shift symmetry of the dual photon is a classical topological symmetry 
whose conserved current is $*F$ where $*$ is the Hodge star. 
The conservation of $*F$ follows from the Bianchi identity $dF=0$. 

\subsubsection{Matter multiplets}
\label{sec_index_gauge}

The index of a 3d $\mathcal{N}=2$ chiral multiplet of charge $+1$ under a $U(1)_{f}$ flavor symmetry with fugacity $x$ is
\begin{align}
\label{3dcm}
\mathbb{I}^{\textrm{3d CM}}(x;q)&=
\frac{(qx^{-1};q)_{\infty}}
{(x;q)_{\infty}}. 
\end{align}
Its denominator counts complex scalar and its $\partial$ derivatives 
while its numerator counts fermions and its $\partial$ derivatives included in the 3d $\mathcal{N}=2$ chiral multiplet. 

3d $\mathcal{N}=4$ hypermultiplet has the following operators which contribute to index:
\begin{align}
\label{3dn4_hm_ch2}
\begin{array}{c|cc|cc}
&\partial^{n}\mathbb{H}&\partial^{n}\widetilde{\mathbb{H}}&
\partial^{n}\overline{\psi}_{-}^{\mathbb{H}}&\partial^{n}\overline{\psi}_{-}^{\widetilde{\mathbb{H}}} \\ \hline
U(1)_{f}&+&-&+&- \\
U(1)_{J}&n&n&n+\frac12&n+\frac12 \\ 
U(1)_{C}&0&0&+&+ \\
U(1)_{H}&+&+&0&0  \\
\textrm{fugacity}&q^{n+\frac14}tx&q^{n+\frac14}tx^{-1}
&-q^{n+\frac34}t^{-1}x&-q^{n+\frac34}t^{-1}x^{-1} \\
\end{array}
\end{align}
The index for 3d $\mathcal{N}=4$ hypermultiplet is 
\begin{align}
\label{3dhm}
\mathbb{I}^{\textrm{3d HM}}(t,x;q)
&=\frac{(q^{\frac34}t^{-1}x;q)_{\infty}(q^{\frac34}t^{-1}x^{-1};q)_{\infty}}
{(q^{\frac14}t x;q)_{\infty}(q^{\frac14}tx^{-1};q)_{\infty}}
\nonumber\\
&=\mathbb{I}^{\textrm{3d CM}}(q^{\frac14}tx) \times \mathbb{I}^{\textrm{3d CM}}(q^{\frac14}tx^{-1}). 
\end{align}
It can be expanded as
\begin{align}
\label{3dhm2}
\mathbb{I}^{\textrm{3d HM}}(t,x;q)&=
\sum_{n=0}^{\infty} \sum_{k=0}^{n} \frac{(q^{\frac12}t^{-2};q)_{k} (q^{\frac12}t^{-2};q)_{n-k}}
{(q)_{k} (q)_{n-k}} x^{n-2k} q^{\frac{n}{4}} t^{n}. 
\end{align}

The free 3d $\mathcal{N}=4$ hypermultiplet has no Coulomb branch local operators surviving in the H-twist. 
Therefore in the H-twist limit $t\rightarrow q^{\frac14}$, the indices (\ref{3dhm}) and (\ref{3dhm2}) become trivial
\begin{align}
\label{3dhm_H}
\mathbb{I}^{\textrm{3d HM}}(t=q^{\frac14},x;q)&=1.
\end{align} 
On the other hand, in the C-twist limit $t\rightarrow q^{-\frac14}$, the indices reduce to 
\begin{align}
\label{3dhm_H}
\mathbb{I}^{\textrm{3d HM}}(t=q^{-\frac14},x;q)&=
\frac{1}{(1-x)(1-x^{-1})}.
\end{align}
This counts two bosonic generators in the algebra of Higgs branch local operators. 

The operators in 3d $\mathcal{N}=4$ twisted hypermultiplet which contribute to index are
\begin{align}
\label{3dn4_thm_ch}
\begin{array}{c|cc|cc}
&\partial^{n}\mathbb{T}&\partial^{n}\widetilde{\mathbb{T}}&
\partial^{n}\overline{\widetilde{\psi}}_{-}^{\mathbb{T}}&\partial^{n}\overline{\widetilde{\psi}}_{-}^{\widetilde{\mathbb{T}}} \\ \hline
U(1)_{f}&+&-&+&- \\
U(1)_{J}&n&n&n+\frac12&n+\frac12 \\ 
U(1)_{C}&+&+&0&0 \\
U(1)_{H}&0&0&+&+  \\
\textrm{fugacity}&q^{n+\frac14}t^{-1}x&q^{n+\frac14}t^{-1}x^{-1}&-q^{n+\frac34}tx&-q^{n+\frac34}tx^{-1} \\
\end{array}
\end{align}
The index for 3d $\mathcal{N}=4$ twisted hypermultiplet can be obtained from the index (\ref{3dhm}) 
by setting $t\rightarrow t^{-1}$
\begin{align}
\label{3dthm}
\mathbb{I}^{\textrm{3d tHM}}(t,x;q)
&=\frac{(q^{\frac34}tx;q)_{\infty}(q^{\frac34}tx^{-1};q)_{\infty}}
{(q^{\frac14}t^{-1}x;q)_{\infty}(q^{\frac14}t^{-1}x^{-1};q)_{\infty}}
\nonumber\\
&=\mathbb{I}^{\textrm{3d CM}}(q^{\frac14}t^{-1}x) \cdot \mathbb{I}^{\textrm{3d CM}}(q^{\frac14}t^{-1}x^{-1}). 
\end{align}
Again it has an expansion
\begin{align}
\label{3dthm2}
\mathbb{I}^{\textrm{3d tHM}}(t,x;q)&=
\sum_{n=0}^{\infty} \sum_{k=0}^{n} \frac{(q^{\frac12}t^2;q)_{\infty} (q^{\frac12} t^2;q)_{\infty}}
{(q)_{k} (q)_{n-k}} x^{n-2k} q^{\frac{n}{4}} t^{-n}. 
\end{align}

\subsubsection{Gauge multiplets}
\label{sec_index_gauge}
While in four-dimensional case the index only involves integration over the gauge group \cite{Kinney:2005ej,Romelsberger:2005eg}, 
in three-dimensional case the index would have non-perturbative contributions of monopole operators 
and contain the sum over the magnetic fluxes of monopole operators for all backgrounds \cite{Imamura:2011su,Kim:2009wb}. 

Let us firstly consider the perturbative contributions to the index. 
The charges of operators in 3d $\mathcal{N}=4$ vector multiplet contributing to the index are 
\begin{align}
\label{3dn4_vm_ch2}
\begin{array}{c|cc|cc}
&D^{n}(\sigma+i\rho)&D^{n}\varphi&
D^{n}\overline{\lambda}_{-}^{\textrm{3d}}&D^{n}\overline{\eta}_{-}^{\textrm{3d}} \\ \hline
G&\textrm{adj}&\textrm{adj}&\textrm{adj}&\textrm{adj} \\
U(1)_{J}&n&n&n+\frac12&n+\frac12 \\ 
U(1)_{C}&0&2&-&- \\
U(1)_{H}&0&0&-&+  \\
\textrm{fugacity}&q^{n}s^{\alpha}&q^{n+\frac12}t^{-2}s^{\alpha}
&-q^{n}s^{\alpha}&-q^{n+\frac12}t^{2}s^{\alpha} \\
\end{array}
\end{align}

The perturbative index contributed from the local operators in (\ref{3dn4_vm_ch2}) of 3d $\mathcal{N}=4$ $U(1)$ vector multiplet takes the form
\begin{align}
\label{3dn4u1}
\mathbb{I}^{\textrm{3d pert $U(1)$}}(t;q)&=
\frac{(q^{\frac12} t^2;q)_{\infty}}{(q^{\frac12}t^{-2};q)_{\infty}}\oint \frac{ds}{2\pi is}
\end{align}
where the integration contour of gauge fugacity $s$ is a unit circle. 
Similarly, the perturbative index for 3d $\mathcal{N}=4$ $U(N)$ vector multiplet takes the form
\begin{align}
\label{3dn4un}
\mathbb{I}^{\textrm{3d pert $U(N)$}}(t;q)&=
\frac{1}{N!}\frac{(q^{\frac12}t^2;q)_{\infty}^N}
{(q^{\frac12}t^{-2};q)_{\infty}^N}
\oint \prod_{i=1}^{N}\frac{ds_{i}}{2\pi is_{i}}
\prod_{i\neq j}\left(1-\frac{s_{i}}{s_{j}}\right)
\frac{\left(q^{\frac12}t^2\frac{s_{i}}{s_{j}}\right)}{\left(q^{\frac12}t^{-2}\frac{s_{i}}{s_{j}}\right)}
\end{align}
where the integration contour of gauge fugacities $s_{i}$ is a unit torus $\mathbb{T}^{N}$. 

Likewise, charges of operators in 3d $\mathcal{N}=4$ twisted vector multiplet are
\begin{align}
\label{3dn4_tvm_ch}
\begin{array}{c|cc|cc}
&D^{n}(\widetilde{\sigma}+i\widetilde{\rho})&D^{n}\widetilde{\varphi}&
D^{n}\overline{\widetilde{\lambda}}_{-}^{\textrm{3d}}&D^{n}\overline{\widetilde{\eta}}_{-}^{\textrm{3d}} \\ \hline
G&\textrm{adj}&\textrm{adj}&\textrm{adj}&\textrm{adj} \\
U(1)_{J}&n&n&n+\frac12&n+\frac12 \\ 
U(1)_{C}&0&0&+&+ \\
U(1)_{H}&0&2&+&-  \\
\textrm{fugacity}&q^{n}s_{\alpha}&q^{n+\frac12}t^{2}s_{\alpha}
&-q^{n}s_{\alpha}&-q^{n+\frac12}t^{-2}s_{\alpha} \\
\end{array}
\end{align}
We can obtain the index for 3d $\mathcal{N}=4$ twisted vector multiplet 
by setting $t\rightarrow t^{-1}$ for the index of 3d $\mathcal{N}=4$ vector multiplet. 

As 3d $\mathcal{N}=4$ $U(1)$ gauge theory appears from 
4d $\mathcal{N}=4$ $U(1)$ gauge theory on a segment with Neumann b.c. $\mathcal{N}$ at each end, 
we have a schematic relation
\begin{align}
\label{3dvm_prop1}
\mathbb{I}^{\textrm{3d pert $U(1)$}}
&=\frac{\mathbb{II}_{\mathcal{N}}^{\textrm{4d $U(1)$}} \times 
\mathbb{II}_{\mathcal{N}}^{\textrm{4d $U(1)$}} 
}
{\mathbb{I}^{\textrm{4d $U(1)$}}}.
\end{align}
When a 4d $\mathcal{N}=4$ $U(1)$ gauge theory is put on a slab with Neumann b.c. $\mathcal{N}$ and Dirichlet b.c. $\mathcal{D}$ at each end, 
we have 
\begin{align}
\label{3dvm_prop2}
1&=\frac{\mathbb{II}_{\mathcal{N}}^{\textrm{4d $U(1)$}} \times  
\mathbb{II}_{\mathcal{D}}^{\textrm{4d $U(1)$}}}{\mathbb{I}^{\textrm{4d $U(1)$}}}. 
\end{align}
This indicates that the resulting 3d theory is a trivial theory. 
For a 4d $\mathcal{N}=4$ $U(1)$ gauge theory on a segment obeying Dirichlet b.c. at both ends, 
we find that 
\begin{align}
\label{3dvm_prop3}
\mathbb{I}^{\textrm{3d pert $\widetilde{U(1)}$}}&
=\frac{\mathbb{II}_{\mathcal{N}'}^{\textrm{4d $U(1)$}} \times  
\mathbb{II}_{\mathcal{N}'}^{\textrm{4d $U(1)$}}}{\mathbb{I}^{\textrm{4d $U(1)$}}}. 
\end{align}
This reflects the fact that the resulting 3d theory is mirror to 3d $\mathcal{N}=4$ twisted $\widetilde{U(1)}$ gauge theory 
appearing from a 4d theory on a segment with Neumann b.c. $\mathcal{N}'$ at both ends.

Now consider the non-perturbative contributions to indices from monopole operators. 
Classically monopole operators are charged under the topological symmetry. 
In addition, quantum mechanically they can acquire non-trivial quantum numbers. 
Let us consider the canonical $U(1)$ R-charge as $\frac12(C-H)$ 
so that the complex scalar $\mathbb{H}$ in the hypermultiplet carries charge $-1/2$ 
and the adjoint complex scalar $\varphi$ in the vector multiplet have charge $+1$.  
Then the R-charge of a BPS bare monopole operator of magnetic charge $m$ in the IR CFT is given by 
\begin{align}
\label{monopole_dim}
\Delta(m)&=
\frac12 \sum_{i=1}^{N_{f}}\sum_{\lambda_{i} \in {\bf R}_{i}} |\lambda_{i}(m)|
-\sum_{\alpha\in \Delta_{+}}|\alpha(m)|. 
\end{align}
This formula was firstly proposed in \cite{Gaiotto:2008ak} and later verified in \cite{Bashkirov:2010kz, Benna:2009xd}. 
The first term in (\ref{monopole_dim}) is the contribution from $N_{f}$ hypermultiplets labeled by 
$i=1,\cdots, N_{f}$ transforming as representations $\{ {\bf R}_{i} \}_{i=1,\cdots, N_{f}}$ under the gauge group. 
The sum is taken over the weights $\lambda_{i}$ of ${\bf R}_{i}$. 
The second term is the contribution from vector multiplet. 
The sum is taken over the positive roots $\alpha\in \Delta_{+}$.

The R-charge (\ref{monopole_dim}) of bare monopole 
can specify its energy, or equivalently conformal dimension since the bare monopole is a BPS state. 
If all BPS monopole operators carry $\Delta(m)$ $>$ $\frac12$, the theory is called \textit{good}. 
If all BPS monopole operators have $\Delta(m)$ $\ge$ $\frac12$ and 
some saturate the unitarity bound $\Delta(m)=\frac12$, the theory is called \textit{ugly}. 
Otherwise, the theory is called \textit{bad}. 
In this paper, we focus on the good or ugly theories 
in which $\Delta(m)$ is identified with the conformal dimension 
and the quantum numbers $J+\frac{H+C}{4}$ of a bare monopole operator with magnetic charge $m$ is fixed to $\frac{\Delta(m)}{2}$ in the IR SCFT.

Taking into account the above, 
we can write the full-index of 3d $\mathcal{N}=4$ gauge theory 
with gauge group $G$ and $N_{f}$ hypermultiplets which takes the form 
\begin{align}
\label{full3dindex_0}
&
\mathbb{I}^{\textrm{3d $G$}}(t,x_{i},z_{i};q)
\nonumber\\
&=
\frac{1}{| \textrm{Weyl} (G)|}
\frac{
(q^{\frac12}t^2;q)_{\infty}^{\mathrm{rank}(G)}
}
{
(q^{\frac12}t^{-2};q)_{\infty}^{\mathrm{rank}(G)}
}
\sum_{m\in \mathrm{cochar}(G)}
\oint \prod_{\alpha\in \mathrm{roots} (G) }
\frac{ds}{2\pi is}
\frac{
\left(1-q^{\frac{|m\cdot \alpha|}{2}}s^{\alpha} \right)
\left(q^{\frac{1+|m\cdot \alpha|}{2}} t^2 s^{\alpha};q\right)_{\infty}
}
{
\left(q^{\frac{1+|m\cdot \alpha|}{2}} t^{-2} s^{\alpha};q\right)_{\infty}
}
\nonumber\\
&\times 
\prod_{i=1}^{N_{f}}
\prod_{\lambda_{i} \in {\bf R}_{i}}
\frac{
\left( q^{\frac34+\frac{|m\cdot \lambda_{i}|}{2}} t^{-1} s^{\pm\lambda_{i}} x_{i}^{\pm};q \right)_{\infty}
}
{
\left( q^{\frac14+\frac{|m\cdot \lambda_{i}|}{2}} t s^{\pm\lambda_{i}} x_{i}^{\pm};q \right)_{\infty}
}
q^{\frac{\Delta(m)}{2}}
\cdot 
t^{-2\Delta(m)}
\cdot 
z^{m}. 
\end{align}
Here the second line is the contribution from vector multiplet of gauge group $G$. 
The third line is the contribution from $N_{f}$ hypermultiplets transforming as representation $\{{\bf R}_{i}\}_{i=1,\cdots, N_{f}}$ 
of gauge group $G$. 
The fugacities $x$ are associated to the flavor symmetry that rotates $N_{f}$ hypermultiplets. 
For non-zero magnetic flux $m$, the expression is shifted from the index (\ref{3dhm}) for 3d hypermultiplet. 
This reflects the fact that in the presence of magnetic flux, the electrically charged states get an effective quantum numbers. 
The third line is the contribution from bare monopole operators. 
The fugacities $z$ are associated to the topological symmetry.

In order to check mirror symmetry and dual boundary conditions, 
one also needs to evaluate the mirror version of the full-index for 3d $\mathcal{N}=4$ gauge theory. 
The mirror index of 3d $\mathcal{N}=4$ gauge theory consisting of 
twisted vector multiplet of gauge group $\widetilde{G}$ 
and $\widetilde{N_{f}}$ twisted hypermultiplets transforming as representation $\{ {\bf R}_{i}\}_{i=1,\cdots, \widetilde{N_{f}}} $ is given by
\begin{align}
\label{full3dindex_1}
\widetilde{\mathbb{I}}^{\textrm{3d $\widetilde{G}$}}(t,x_{i},z_{i};q)
&=
\mathbb{I}^{\textrm{3d $\widetilde{G}$}}(t^{-1},z_{i},x_{i};q)
\end{align}
%
where the fugacities $z_{i}$ and $x_{i}$ are associated to the flavor symmetry and the topological symmetry 
in the twisted 3d $\mathcal{N}=4$ theories as they are exchanged under mirror symmetry. 
In the following sections, we show various identities 
between indices (\ref{full3dindex_0}) and (\ref{full3dindex_1}) for mirror pairs.

\section{Abelian mirror symmetry}
\label{sec_3dmirror1}
\begin{figure}
\begin{center}
\includegraphics[width=10cm]{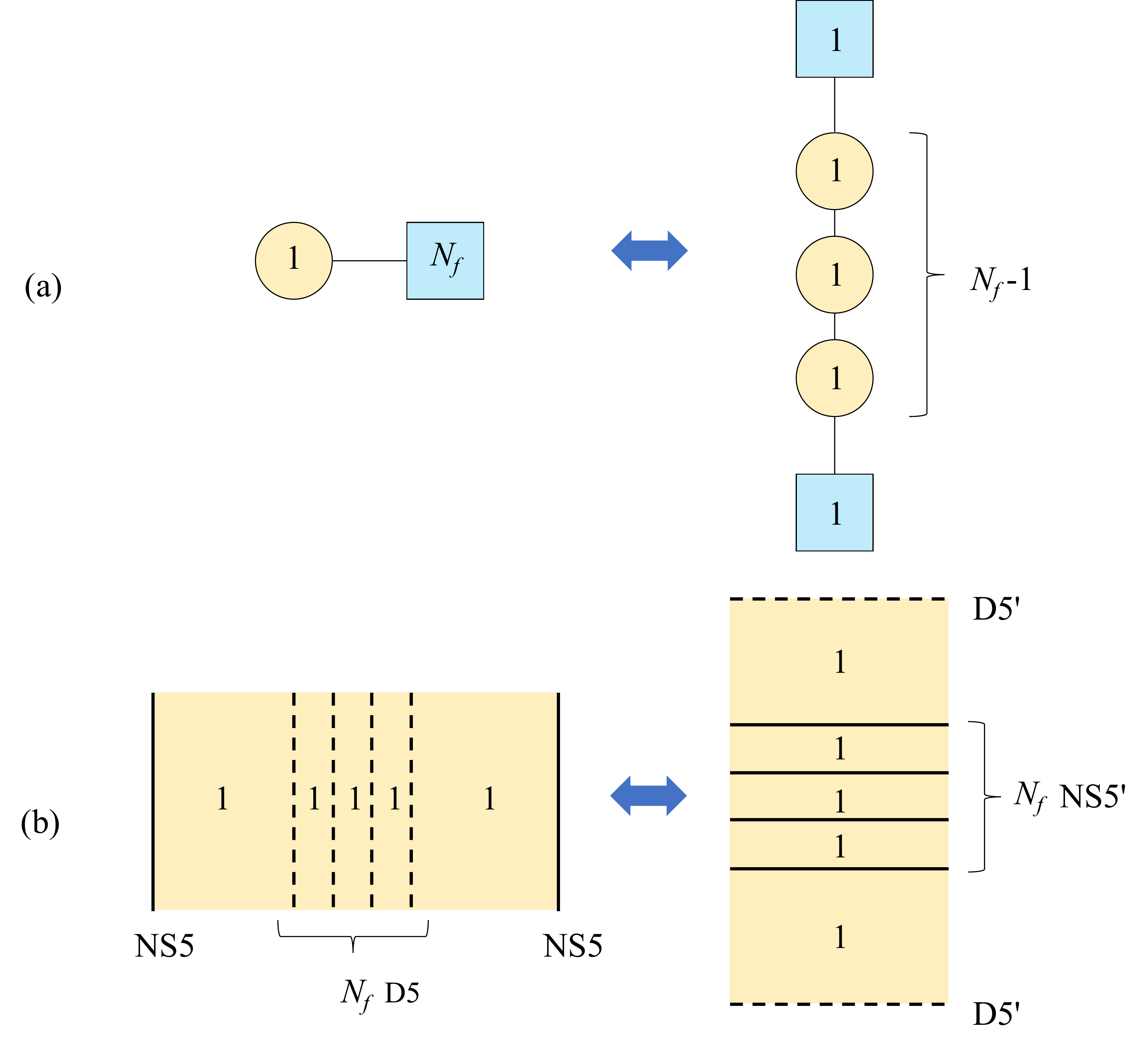}
\caption{
(a) The quiver diagrams of SQED$_{N_{f}}$ and its mirror $U(1)^{N_{f}-1}$ quiver gauge theory. 
(b) The brane configurations of SQED$_{N_{f}}$ and its mirror $U(1)^{N_{f}-1}$ quiver gauge theory. 
}
\label{figsqed1}
\end{center}
\end{figure}

In this section, we consider the 3d $\mathcal{N}=4$ Abelian gauge theory and its mirror. 
The mirror of 3d $\mathcal{N}=4$ $U(1)$ gauge theory with $N_{f}$ hypermultiplets 
is a twisted $U(1)^{N_{f}-1}$ quiver gauge theory with $N_{f}$ twisted hypermultiplets \cite{Kapustin:1999ha}. 
The quiver diagram and the corresponding brane configuration are depicted in Figure \ref{figsqed1}. 
We check that two indices for mirror pairs coincide with each other.

\subsection{SQED$_{1}$}
\label{sec_3dsqed1}
We begin with a 3d $\mathcal{N}=4$ Abelian gauge theory with a single charged hypermultiplet, which we call SQED$_{1}$. 
The 3d $\mathcal{N}=4$ SQED$_{1}$ is mirror to a 3d $\mathcal{N}=4$ twisted hypermultiplet \cite{Kapustin:1999ha}. 
The flavor symmetry of the twisted hypermultiplet is mapped to the topological symmetry in SQED$_{1}$. 
This is the simplest 3d $\mathcal{N}=4$ Abelian mirror symmetry 
and index computation has been already computed in \cite{Razamat:2014pta}. 
We further extract a Higgsing interpretation of the indices by picking up residues at poles in the indices. 

The full index of SQED$_{1}$ is given by 
\begin{align}
\label{sqed1full}
\mathbb{I}^{\textrm{3d SQED}_{1}}(t,x;q)&=
\frac{(q^{\frac12} t^2;q)_{\infty}}{(q^{\frac12} t^{-2};q)_{\infty}}
\sum_{m\in \mathbb{Z}}
\oint \frac{ds}{2\pi is} 
\frac{(q^{\frac34+\frac{|m|}{2}} t^{-1}s;q)_{\infty} (q^{\frac34+\frac{|m|}{2}} t^{-1}s^{-1};q)_{\infty}}
{(q^{\frac14+\frac{|m|}{2}} t s;q)_{\infty} (q^{\frac14+\frac{|m|}{2}} ts^{-1};q)_{\infty}}
q^{\frac{|m|}{4}}t^{-|m|} x^{m} 
\end{align}
where $m$ is the magnetic monopole charge and the fugacity $x$ is associated to a $U(1)_{t}$ topological symmetry. 
The factors $q^{\frac{|m|}{4}}$ $t^{-|m|}$ $x^m$ are associated to contributions from bare monopole operators 
with the R-charge $\Delta(m)$ $=$ $\frac{|m|}{2}$. 
The remaining factors in the integrand involve shifts of quantum numbers due to the background magnetic flux. 
The index for SQED$_{1}$ already appeared in \cite{Kapustin:2011jm,Razamat:2014pta} and 
the index (\ref{sqed1full}) coincides with the index (\ref{3dthm}) of a 3d $\mathcal{N}=4$ twisted hypermultiplet. 

\begin{figure}
\begin{center}
\includegraphics[width=9.5cm]{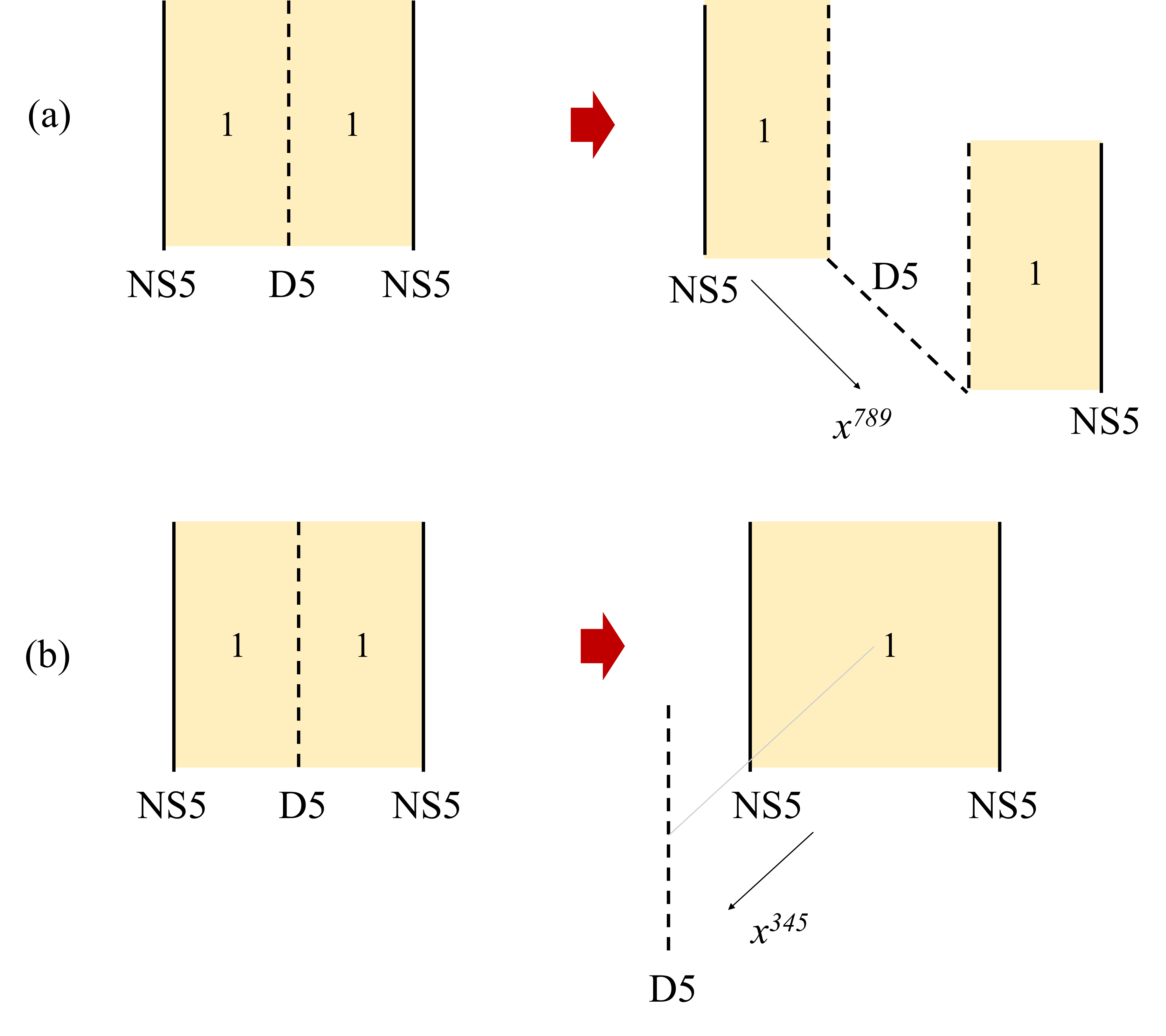}
\caption{
(a) The Higgsing procedure of 3d $\mathcal{N}=4$ SQED$_{1}$ 
splitting a D3-brane along the D5-brane 
and separating one of the NS5-brane in the $x^{7,8,9}$ directions. 
(b) The Higgsing procedure of 3d $\mathcal{N}=4$ SQED$_{1}$ 
separating the D5-brane from 
the stretched D3-brane between the NS5-branes. 
}
\label{fighiggs1}
\end{center}
\end{figure}

As discussed in \cite{Gaiotto:2019jvo}, 
poles in the integrand and their residues may have a physical interpretation. 
When FI-like parameters are turned on, 
elementary fields with gauge charges get non-trivial vevs, which leads to a Higgsing the gauge group. 
Consequently, the index is written as a sum over residues 
which are associated to the index of the Higgsed theory. 
In the brane setup, this corresponds to a certain deformation of the brane configuration. 
In order to extract a Higgsing interpretation by expanding the index (\ref{sqed1full}), 
it is enough to consider the perturbative SQED$_{1}$ index corresponding to $m=0$ that takes the form
\begin{align}
\label{sqed1index}
\mathbb{I}^{\textrm{3d pert SQED}_{1}}(t;q)
&=
\underbrace{
\frac{(q^{\frac12}t^2;q)_{\infty}}{(q^{\frac12}t^{-2};q)_{\infty}}
\oint \frac{ds}{2\pi is}
}_{\mathbb{I}^{\textrm{3d pert $U(1)$}}}
\underbrace{
\frac{(q^{\frac34}t^{-1}s;q)_{\infty} (q^{\frac34}t^{-1}s^{-1};q)_{\infty}}
{(q^{\frac14}ts;q)_{\infty} (q^{\frac14}ts^{-1};q)_{\infty}}
}_{\mathbb{I}^{\textrm{3d HM}}(s)}. 
\end{align}
This can be evaluated by considering the residues at poles of charged hypermultiplet 
$s=q^{\frac14+m}t$ as
\begin{align}
\label{sqed1index1}
\mathbb{I}^{\textrm{3d pert SQED}_{1}}(t;q)
&=\frac{(q^{\frac12}t^2;q)_{\infty}^2}
{(q)_{\infty}^2} \sum_{m=0}^{\infty} 
\frac{(q^{1+m};q)_{\infty}^2}{(q^{\frac12+m}t^2;q)_{\infty}^2}q^{\frac{m}{2}} t^{-2m}. 
\end{align}
As the residue sum begins with $1$, the Higgsed theory is a trivial theory. 
This would imply that the expansion (\ref{sqed1index1}) is associated to a Higgsing process 
which splits a D3-brane along the D5-brane 
and separates one of the NS5-brane in the $x^{7,8,9}$ directions 
(see Figure \ref{fighiggs1}).

Correspondingly, we get the mirror index by selecting out the zero-charge sector from 
the index (\ref{3dthm}) of 3d $\mathcal{N}=4$ twisted hypermultiplet
\begin{align}
\label{3dthmdual}
\oint \frac{dx}{2\pi ix} 
\mathbb{I}^{\textrm{3d tHM}}(t,x;q)&=
\oint \frac{dx}{2\pi ix} 
\underbrace{
\frac{\left(q^{\frac34}t x;q\right)_{\infty} \left(q^{\frac34}t x^{-1};q\right)_{\infty}}
{\left(q^{\frac14}t^{-1} x;q\right)_{\infty} \left(q^{\frac14}t^{-1} x^{-1};q\right)_{\infty}}
}_{\mathbb{I}^{\textrm{3d tHM}}\left(x\right)}. 
\end{align}
The indices (\ref{sqed1index1}) and (\ref{3dthmdual}) can be shown to be equal. 
To see the equivalence, we can firstly calculate the index (\ref{3dthmdual}) 
by taking the sum over residues at poles of bi-fundamental twisted hypermultiplet 
$x=q^{\frac14+m}t^{-1}$ as
\begin{align}
\label{3dthmdual1}
\oint \frac{dx}{2\pi ix} 
\mathbb{I}^{\textrm{3d tHM}}(t,x;q)&=
\frac{(q^{\frac12}t^2;q)_{\infty} (q^{\frac12}t^{-2};q)_{\infty}}{(q)_{\infty}^2} 
\sum_{m=0}^{\infty} \frac{(q^{1+m};q)_{\infty}^2}{(q^{\frac12+m}t^{-2};q)_{\infty}^2}q^{\frac{m}{2}} t^{2m}. 
\end{align}
On the other hand, we can expand the index (\ref{sqed1index}) of SQED$_{1}$ 
in terms of the expansion (\ref{3dhm2}) of the hypermultiplet index as 
\begin{align}
\label{sqed1index2}
\mathbb{I}^{\textrm{3d pert SQED}_{1}}(t;q)
&=
\frac{(q^{\frac12}t^2;q)_{\infty}}{(q^{\frac12}t^{-2};q)_{\infty}}
\oint \frac{ds}{2\pi is}\sum_{n=0}^{\infty} \sum_{k=0}^{n} 
\frac{(q^{\frac12} t^{-2};q)_{k} (q^{\frac12}t^{-2};q)_{n-k}}
{(q)_{k}(q)_{n-k}} s^{n-2k} q^{\frac{n}{4}} t^n
\nonumber\\
&=\frac{(q^{\frac12}t^2;q)_{\infty} (q^{\frac12}t^{-2};q)_{\infty}}{(q)_{\infty}^2} 
\sum_{m=0}^{\infty} \frac{(q^{1+m};q)_{\infty}^2}{(q^{\frac12+m}t^{-2};q)_{\infty}^2}q^{\frac{m}{2}} t^{2m}.
\end{align}
This agrees with the expression (\ref{3dthmdual1}). 
Also we observe that 
the expansion in the sum (\ref{sqed1index2}) starts from 
the perturbative index $\mathbb{I}^{\textrm{3d pert $U(1)$}}$ of 3d $\mathcal{N}=4$ $U(1)$ vector multiplet. 
This would be associated to the Higgsing process of separating the D5-brane from 
the stretched D3-brane between the NS5-branes 
(see Figure \ref{fighiggs1}). 

Introducing a Wilson line operator $\mathcal{W}_{n}$ of charge $n$, 
we obtain the index
\begin{align}
\label{sqed1fullw1}
\mathbb{I}^{\textrm{3d SQED}_{1}}_{\mathcal{W}_{n}}(t,x;q)&=
\frac{(q^{\frac12}t^2;q)_{\infty}}{(q^{\frac12}t^{-2};q)_{\infty}}
\sum_{m\in \mathbb{Z}}
\oint \frac{ds}{2\pi is} 
\frac{(q^{\frac34+\frac{|m|}{2}} t^{-1}s;q)_{\infty} 
(q^{\frac34+\frac{|m|}{2}} t^{-1}s^{-1};q)_{\infty}
}{
(q^{\frac14+\frac{|m|}{2}} ts;q)_{\infty} 
(q^{\frac14+\frac{|m|}{2}} ts^{-1};q)_{\infty}
}
s^{n} 
q^{\frac{|m|}{4}} t^{-|m|} x^{m}. 
\end{align}
As shown in Appendix \ref{app_wil}, we have checked that this agrees with 
\begin{align}
\label{thmv1}
\mathbb{I}^{\textrm{3d tHM}}_{\mathcal{V}_{n}}(t,x;q)&=
\frac{(q^{\frac34+\frac{|n|}{2}} t x;q)_{\infty} (q^{\frac34+\frac{|n|}{2}}tx^{-1};q)_{\infty}}
{(q^{\frac{1}{4}+\frac{|n|}{2}} t x;q)_{\infty} (q^{\frac{1}{4}+\frac{|n|}{2}} t^{-1}x^{-1};q)_{\infty}} 
q^{\frac{|n|}{4}} t^{|n|}. 
\end{align}
We see that the index (\ref{thmv1}) has the contributions from bare monopole operator associated with flavor symmetry 
and that the quantum numbers of twisted hypermultiplets are affected by the flux.

\subsection{$T[SU(2)]$}
\label{sec_3dtsu2}
The next example is a 3d $\mathcal{N}=4$ Abelian gauge theory with two charged hypermultiplets, which we call $T[SU(2)]$. 
The dimension of the Higgs branch is $\dim_{\mathbb{C}}\mathcal{M}_{H}^{T[SU(2)]}$ $=$ $2\cdot (1\times 2-1^2)$ $=2$ 
and the dimension of the Coulomb branch is $\dim_{\mathbb{C}}\mathcal{M}_{C}^{T[SU(2)]}$ $=$ $2\cdot 1=2$. 
This is self-mirror with two global symmetries, 
a $SU(2)_{f}$ flavor symmetry on $\mathcal{M}_{H}$ 
and enhanced $SU(2)_{t}$ topological symmetry  on $\mathcal{M}_{C}$ 
which are exchanged under mirror symmetry. 

The index of $T[SU(2)]$ reads
\begin{align}
\label{tsu2full1}
&
\mathbb{I}^{T[SU(2)]}(t,x_{\alpha},z_{\alpha};q)
\nonumber\\
&=
\frac{(q^{\frac12}t^2;q)_{\infty}}{(q^{\frac12}t^{-2};q)_{\infty}} 
\sum_{m\in \mathbb{Z}} \oint \frac{ds}{2\pi is} 
\frac{
(q^{\frac34+\frac{|m|}{2}} t^{-1}s^{\pm}x_{1}^{\pm};q)_{\infty}
}
{(q^{\frac14+\frac{|m|}{2}} t s^{\pm}x_{1}^{\pm};q)_{\infty}
}
\frac{
(q^{\frac34+\frac{|m|}{2}} t^{-1}s^{\pm}x_{2}^{\pm};q)_{\infty}
}
{(q^{\frac14+\frac{|m|}{2}} t s^{\pm}x_{2}^{\pm};q)_{\infty} 
}
q^{\frac{|m|}{2}} t^{-2|m|} z_{1}^{m} z_{2}^{-m}
\end{align}
where the fugacities $x_{\alpha}$ are associated to the $SU(2)$ flavor symmetry with $x_{1}x_{2}=1$ 
and $z_{\alpha}$ are the fugacities for the topological symmetry. 

The index (\ref{tsu2full1}) of $T[SU(2)]$ coincides with the index of $\widetilde{T[SU(2)]}$ 
which is obtained from $T[SU(2)]$ by replacing their supermultiplets with the twisted counterparts:
\begin{align}
\label{tsu2full2}
&
\mathbb{I}^{\widetilde{T[SU(2)]}}(t,x_{\alpha},z_{\alpha};q)
\nonumber\\
&=
\frac{(q^{\frac12}t^{-2};q)_{\infty}}{(q^{\frac12} t^2;q)_{\infty}} 
\sum_{m\in \mathbb{Z}} \oint \frac{ds}{2\pi is}
\frac{
(q^{\frac34+\frac{|m|}{2}} ts^{\pm}z_{1}^{\pm};q)_{\infty} 
}
{
(q^{\frac14+\frac{|m|}{2}} t^{-1}s^{\pm}z_{1}^{\pm};q)_{\infty} 
}
\frac{
(q^{\frac34+\frac{|m|}{2}} ts^{\pm}z_{2}^{\pm};q)_{\infty} 
}
{
(q^{\frac14+\frac{|m|}{2}} t^{-1}s^{\pm}z_{2}^{\pm};q)_{\infty} 
}
 q^{\frac{|m|}{2}} t^{2|m|} x_{1}^{m}x_{2}^{-m}. 
\end{align}
Here the role of fugacities $x_{\alpha}$ and $z_{\alpha}$ are exchanged.

We can draw a lesson from the expansion of the index. 
To get a Higgsing procedure, 
we simply look at the sector with zero flavor charge of the perturbative index of $T[SU(2)]$
\begin{align}
\label{tsu2index1}
&
\oint \frac{dx_{1}}{2\pi ix_{1}} \frac{dx_{2}}{2\pi ix_{2}}
\mathbb{I}^{\textrm{3d pert $T[SU(2)]$}}(t;q)
\nonumber\\
&=
\underbrace{
\frac{(q^{\frac12}t^2;q)_{\infty}}{(q^{\frac12}t^{-2};q)_{\infty}}
\oint \frac{ds_{1}}{2\pi is_{1}}}_{\mathbb{I}^{\textrm{3d pert $U(1)$}}} 
\oint \frac{dx_{1}}{2\pi ix_{1}} 
\oint \frac{dx_{2}}{2\pi ix_{2}}
\underbrace{
\frac{
(q^{\frac34}t^{-1}s^{\pm}x_{1}^{\pm};q)_{\infty} 
}
{
(q^{\frac14}ts^{\pm}x_{1}^{\pm};q)_{\infty} 
}
}_{\mathbb{I}^{\textrm{3d HM}} (sx_{1})}
\cdot 
\underbrace{
\frac{
(q^{\frac34}t^{-1}s^{\pm}x_{2}^{\pm};q)_{\infty} 
}
{
(q^{\frac14}ts^{\pm}x_{2}^{\pm};q)_{\infty} 
}
}_{\mathbb{I}^{\textrm{3d HM}}(sx_{2})}.
\end{align}
This can be evaluated as 
\begin{align}
\label{tsu2index2}
&
\oint \frac{dx_{1}}{2\pi ix_{1}} \frac{dx_{2}}{2\pi ix_{2}}
\mathbb{I}^{\textrm{pert $T[SU(2)]$}}(t;q)
\nonumber\\
&=
\frac{(q^{\frac12}t^2;q)_{\infty} (q^{\frac12}t^{-2};q)_{\infty}^3}
{(q)_{\infty}^4} 
\sum_{n,m=0}^{\infty} \frac{(q^{1+n};q)_{\infty}^2 (q^{1+m};q)_{\infty}^2}{(q^{\frac12+n}t^{-2};q)_{\infty}^2 (q^{\frac12+m}t^{-2};q)_{\infty}^2} 
q^{\frac{n+m}{2}} t^{2 (n+m)}.
\end{align}
Again the expansion (\ref{tsu2index2}) has a Higgsing interpretation. 
As its first term is identified with the index $\mathbb{I}^{\textrm{3d pert $U(1)$}}$ of 3d $\mathcal{N}=4$ $U(1)$ vector multiplet. 
It is associated to a Higgsing manipulation of 
separation of two D5-branes from a suspended D3-brane between two NS5-branes.

\subsection{SQED$_{N_{f}}$}
\label{sec_3dsqednf}
Now we would like to discuss the generalization of full-indices for 3d $\mathcal{N}=4$ Abelian gauge theories.  
Consider a 3d $\mathcal{N}=4$ Abelian gauge theory with $N_{f}$ charged hypermultiplets, which we call SQED$_{N_{f}}$. 

For SQED$_{N_{f}}$ the R-charge of a bare monopole is $\Delta(m)$ $=$ $\frac{N_{f}|m|}{2}$  
and the index takes the form
\begin{align}
\label{sqednf_full}
\mathbb{I}^{\textrm{3d SQED}_{N_{f}}}(t,x_{\alpha},z_{\alpha};q)&=
\frac{(q^{\frac12} t^2;q)_{\infty}}{(q^{\frac12} t^{-2};q)_{\infty}} 
\sum_{m\in \mathbb{Z}}\oint \frac{ds}{2\pi is}  
\prod_{\alpha=1}^{N_{f}} 
\frac{(q^{\frac34+\frac{|m|}{2}} t^{-1}s^{\pm} x_{\alpha}^{\mp};q)_{\infty}}
{(q^{\frac14+\frac{|m|}{2}} t s^{\pm} x_{\alpha}^{\pm};q)_{\infty}}
q^{\frac{N_{f}|m|}{4}} t^{-N_{f}|m|} z_{1}^{m} z_{2}^{-m}
\end{align}
where $x_{\alpha}$ are the fugacities for the $SU(N_{f})$ flavor symmetry with $\prod_{\alpha=1}^{N_{f}} x_{\alpha}=1$  
and $z_{\alpha}$ are the fugacities for the topological symmetry. 

As discussed in \cite{Razamat:2014pta}, one can obtain the Hilbert series for branches of vacua in the special fugacity limit of the indices. 
We briefly check this in our notation as follows. 
The Coulomb branch of SQED$_{N_{f}}$ is the $A_{N_{f}-1}$ singularity $\mathbb{C}^2/\mathbb{Z}_{N_{f}}$ \cite{Intriligator:1996ex}. 
Taking the Coulomb limit $q$ $\rightarrow 0$ while keeping $\widetilde{\mathfrak{t}}^{\frac12}$ $=$ $q^{\frac14}t^{-1}$ constant, 
and setting $z=z_{1}z_{2}^{-1}$
the index (\ref{sqednf_full}) reduces to
\begin{align}
\label{sqednf_C}
\mathbb{I}^{\textrm{3d SQED}_{N_{f}}}_{\textrm{Coulomb}}(\widetilde{\mathfrak{t}},z)
&=\frac{1}{(1-\widetilde{\mathfrak{t}})}
\sum_{m\in \mathbb{Z}}\widetilde{\mathfrak{t}}^{\frac{N_{f}|m|}{2}}
z^m
\nonumber\\
&=
\frac{1-\widetilde{\mathfrak{t}}^{N_{f}}}
{\left(1-\widetilde{\mathfrak{t}}\right) 
\left(1-\widetilde{\mathfrak{t}}^{\frac{N_{f}}{2}} z\right)
\left(1-\widetilde{\mathfrak{t}}^{\frac{N_{f}}{2}} z^{-1}\right)
}. 
\end{align}
This is identified with the refined Hilbert series of the Coulomb branch of SQED$_{N_{f}}$ \cite{Cremonesi:2013lqa}. 
The factors $1/(1-\widetilde{\mathfrak{t}})$, $1/(1-\widetilde{\mathfrak{t}}^{\frac{N_{f}}{2}}z)$ 
and $1/(1-\widetilde{\mathfrak{t}}^{\frac{N_{f}}{2}}z^{-1})$ would correspond to 
the scalar field $\varphi$, the monopole operator $V_{+}$ of magnetic flux $+1$ 
and the monopole operator $V_{-}$ of magnetic flux $-1$ respectively. 
They obey the relation $V_{+}V_{-}=\varphi^{N_{f}}$ at dimension $N_{f}$ and topological charge $0$ \cite{Borokhov:2002cg} 
which is encoded by the numerator. 

In fact, by setting $z=1$ in (\ref{sqednf_C}), we get the Hilbert series of the $A_{N_{f}-1}$ singularity $\mathbb{C}^2/\mathbb{Z}_{N_{f}}$ \cite{Benvenuti:2006qr}
\begin{align}
\label{sqednf_C1}
\mathbb{I}^{\textrm{3d SQED}_{N_{f}}}_{\textrm{Coulomb}}(\widetilde{\mathfrak{t}},z=1)
&=
\frac{1-\widetilde{\mathfrak{t}}^{N_{f}}}
{\left(1-\widetilde{\mathfrak{t}}\right) 
\left(1-\widetilde{\mathfrak{t}}^{\frac{N_{f}}{2}}\right)^2
}
\nonumber\\
&=\mathrm{Hilb}[\mathbb{C}^2/\mathbb{Z}_{N_{f}}]
. 
\end{align}

The mirror of SQED$_{N_{f}}$, 
which we denote by $\widetilde{[1]-(1)^{N_{f}-1}-[1]}$ has a gauge group $U(1)^{N_{f}}/U(1)$ 
and the twisted hypermultiplets that are associated to the links of the extended Dynkin diagram of $A_{N_{f}-1}$. 
The twisted hypers carry charges $(+,-,0,\cdots,0)$, $(0,+,-,\cdots,0)$, $\cdots$, $(-,0,\cdots,0,+)$ under the $U(1)^{N_{f}}$. 
For $\widetilde{[1]-(1)^{N_{f}-1}-[1]}$ theory, the magnetic fluxes are labeled by $N_{f}-1$ integers 
$(m_{1},\cdots, m_{N_{f}-1})$ corresponding to $(N_{f}-1)$ $U(1)$ topological symmetries. 

The conformal dimension of bare monopole is $\Delta(m)$ $=$ 
$\frac12(|m_{1}|+|m_{N_{f}-1}|+\sum_{i=1}^{N_{f}-2}|m_{i}-m_{i+1}|)$. 
Then the index of the mirror quiver gauge theory $\widetilde{[1]-(1)^{N_{f}-1}-[1]}$ reads
\begin{align}
\label{msqednf_full}
&\mathbb{I}^{\textrm{3d $\widetilde{[1]-(1)^{N_{f}-1}-[1]}$}}(t,x_{\alpha},z_{\alpha};q)
\nonumber\\
&=
\frac{(q^{\frac12}t^{-2};q)_{\infty}^{N_{f}-1}}
{(q^{\frac12}t^2;q)_{\infty}^{N_{f}-1}} \sum_{m_{1},\cdots,m_{N_{f}-1}\in \mathbb{Z}} 
\oint \prod_{i=1}^{N_{f}-1}\frac{ds_{i}}{2\pi is_{i}}
\nonumber\\
&\times 
\frac{(q^{\frac34+\frac{|m_{1}|}{2}} t s_{1}^{\pm} z_{1}^{\mp};q)_{\infty}}
{(q^{\frac14+\frac{|m_{1}|}{2}} t^{-1}s_{1}^{\pm}z_{1}^{\mp};q)_{\infty}} 
\cdot 
\prod_{i=1}^{N_{f}-2} 
\frac{
(q^{\frac34+\frac{|m_{i}-m_{i+1}|}{2}}ts_{i}^{\pm}s_{i+1}^{\mp};q)_{\infty}
}
{(q^{\frac34+\frac{|m_{i}-m_{i+1}|}{2}}t^{-1}s_{i}^{\pm} s_{i+1}^{\mp};q)_{\infty}}
\cdot 
\frac{(q^{\frac34+\frac{|m_{N_{f}-1}|}{2}} t s_{N_{f}-1}^{\pm} z_{2}^{\mp};q)_{\infty}}
{(q^{\frac14+\frac{|m_{N_{f}-1}|}{2}} t^{-1}s_{N_{f}-1}^{\pm}z_{2}^{\mp};q)_{\infty}}
\nonumber\\
&\times 
q^{\frac{|m_{1}|}{4}+\frac{|m_{N_{f}-1}|}{4}+\sum_{i=1}^{N_{f}-2} \frac{|m_{i}-m_{i+1}|}{4}} 
t^{|m_{1}|+|m_{N_{f}-1}|+\sum_{i=1}^{N_{f}-2}|m_{i}-m_{i+1}|} 
\prod_{\alpha=1}^{N_{f}-1} \left(\frac{x_{\alpha}}{x_{\alpha+1}}\right)^{m_{\alpha}}
\left(\frac{x_{N_{f}}}{x_{1}}\right)^{m_{1}+m_{N_{f}-1}}.
\end{align}
Here the fugacities $z_{\alpha}$ are used for the flavor symmetry 
while the fugacities $x_{\alpha}$ are associated to the topological symmetry. 
The index (\ref{msqednf_full}) would be equal to the index (\ref{sqednf_full}) of SQED$_{N_{f}}$. 
As shown in Appendix \ref{app_abe}, we have confirmed that they agree with each other for $N=1,2,3,4$ up to order $q^5$. 

The Coulomb branch of $[1]-(1)^{N_{f}-1}-[1]$ is the reduced moduli space of one instanton of $SU(N_{f})$. 
Taking the Higgs limit $q$ $\rightarrow$ $0$ while keeping $\mathfrak{t}^{\frac12}$ $=$ $q^{\frac14}t$ constant, 
setting $z_{\alpha}$ $=$ $x_{\alpha}x_{\alpha+1}^{-1}$ and $m_{0}$ $=$ $m_{N_{f}}$ $\equiv$ $0$, 
the index (\ref{msqednf_full}) becomes 
\begin{align}
\label{msqednf_H}
\mathbb{I}^{\textrm{3d $\widetilde{[1]-(1)^{N_{f}-1}-[1]}$}}_{\textrm{Higgs}}(\mathfrak{t},z_{i})
&=\frac{1}{(1-\mathfrak{t})^{N_{f}-1}}\sum_{m_{1},\cdots,m_{N_{f}-1}} \mathfrak{t}^{\frac12 \sum_{i=0}^{N_{f}-1} |m_{i}-m_{i+1}|}
\prod_{i=1}^{N_{f}-1}z_{i}^{m_{i}}
\nonumber\\
&=\mathrm{Hilb}^{{\textrm{3d $\widetilde{[1]-(1)^{N_{f}-1}-[1]}$}}}(\mathfrak{t},z_{i})
\end{align}
where we have removed decoupled $U(1)$ by gauge fixing to set $m_{1}+m_{N_{f}-1}=0$. 
This is the refined Hilbert series of the Coulomb branch of $[1]-(1)^{N_{f}-1}-[1]$.

\section{Non-Abelian mirror symmetry}
\label{sec_3dmirror2}
In this section we test 3d $\mathcal{N}=4$ mirror symmetry for non-Abelian gauge theories by computing 3d full-indices. 
We confirm that two indices nicely agree to each other.

\subsection{$(N)-[2N]$}
\label{sec_3dNw2N}
Let us argue for the balanced $U(N)$ gauge theory with $2N$ hypers  
which we denote by $(N)-[2N]$. 
The magnetic charges for $U(N)$ gauge theory are given by $N$-tuples of integers $(m_{1},\cdots,m_{N})$. 
The dimension of Coulomb branch is 
\begin{align}
\label{Nw2Ndim_C}
\dim_{\mathbb{C}}\mathcal{M}_{C}^{(N)-[2N]}&=2 N
\end{align}
while the dimension of Higgs branch is 
\begin{align}
\label{Nw2Ndim_H}
\dim_{\mathbb{C}}\mathcal{M}_{H}^{(N)-[2N]}
&=2\cdot (N\times 2N-N^2)=2N^2.
\end{align}

The R-charge of bare monopole is 
\begin{align}
\label{Nw2N_monodim}
\Delta(m)&=N\sum_{i=1}^{N}|m_{i}|-\sum_{i<j}|m_{i}-m_{j}|. 
\end{align}
Here the first terms are contributed from the $2N$ fundamental hypers 
while the second terms are the contributions from the $U(N)$ vector multiplet. 

The index of $(N)-[2N]$ takes the form 
\begin{align}
\label{3dNw2Na}
&\mathbb{I}^{(N)-[2N]}(t,x_{\alpha},z_{\alpha};q)
\nonumber\\
&=
\frac{1}{N!} 
\frac{(q^{\frac12}t^2;q)_{\infty}^{N}}
{(q^{\frac12}t^{-2};q)_{\infty}^{N}} 
\sum_{m_{1},\cdots,m_{N}\in \mathbb{Z}} 
\oint 
\prod_{i=1}^{N} \frac{ds_{i}}{2\pi is_{i}} \prod_{i<j}
\frac{
\left(1-q^{\frac{|m_{i}-m_{j}|}{2}} s_{i}^{\pm}s_{j}^{\mp} \right)
\left(q^{\frac{1+|m_{i}-m_{j}|}{2}} t^2 s_{i}^{\pm} s_{j}^{\mp};q \right)_{\infty}
}
{
\left(q^{\frac{1+|m_{i}-m_{j}|}{2}} t^{-2} s_{i}^{\pm} s_{j}^{\mp};q \right)_{\infty}
}
\nonumber\\
&\times
\prod_{i=1}^{N} \prod_{\alpha=1}^{2N}
\frac{
\left(q^{\frac34+\frac{|m_{i}|}{2}}t^{-1}s_{i}^{\pm}x_{\alpha}^{\pm};q \right)_{\infty}
}
{
\left(q^{\frac14+\frac{|m_{i}|}{2}}t s_{i}^{\pm}x_{\alpha}^{\pm};q \right)_{\infty}
}
\nonumber\\
&\times 
q^{\frac{N}{2}\sum_{i=1}^{N} |m_{i}|-\sum_{i<j}\frac{|m_{i}-m_{j}|}{2}}
\cdot 
t^{-2N\sum_{i=1}^{N}|m_{i}|+2\sum_{i<j}|m_{i}-m_{j}|}
\cdot 
\left(\frac{z_{1}}{z_{2}} \right)^{\sum_{i=1}^{N}m_{i}}
\end{align}
where $x_{\alpha}$ are the fugacities for the $SU(2N)$ flavor symmetry with $\prod_{i=1}^{2N} x_{\alpha}=1$  
and $z_{\alpha}$ are the fugacities for the topological symmetry. 

The mirror of $(N)-[2N]$ is the quiver gauge theory 
$\widetilde{\begin{smallmatrix}
(1)-(2)-&\cdots&-(N)-&\cdots&-(2)-(1)\\
&&|&&\\
&&[2]&&\\
\end{smallmatrix}}$. 
The dimension of the Coulomb branch is 
\begin{align}
\label{mNw2Ndim_C}
&\dim_{\mathbb{C}}\mathcal{M}_{C}^{\widetilde{\begin{smallmatrix}
(1)-(2)-&\cdots&-(N)-&\cdots&-(2)-(1)\\
&&|&&\\
&&[2]&&\\
\end{smallmatrix}}}
\nonumber\\
&=2\cdot \left(2 \sum_{k=1}^{N-1}k + N\right)
=2N^2
\end{align}
and the dimension of Higgs branch is 
\begin{align}
\label{mNw2Ndim_H}
&\dim_{\mathbb{C}}\mathcal{M}_{H}^{\widetilde{\begin{smallmatrix}
(1)-(2)-&\cdots&-(N)-&\cdots&-(2)-(1)\\
&&|&&\\
&&[2]&&\\
\end{smallmatrix}}}
\nonumber\\
&=2\cdot 
\left(2\times \sum_{k=1}^{N-1}k(k+1)+2\times N
-2\sum_{k=1}^{N-1}k^2-N^2\right)
=2N.
\end{align}
As predicted by mirror symmetry, 
the dimensions (\ref{mNw2Ndim_C}) and (\ref{mNw2Ndim_H}) 
agree with the dimensions (\ref{Nw2Ndim_H}) and (\ref{Nw2Ndim_C}) respectively. 

Classically there is  a $U(1)^{N^2}$ topological symmetry 
and we label the magnetic fluxes by $N$ sets of $k$-tuple of integers  
 $\{m_{i}^{(k)}\}_{i=1,\cdots, k}$ with $k=1,\cdots, N$ 
 and 
 $(N-1)$ sets of $(2N-k)$-tuple of integers 
 $\{m_{i}^{(k)}\}_{i=1,\cdots, 2N-k}$ with $k=N+1,\cdots, 2N-1$, 
 which are in total $N^2$ integers. 
The conformal dimension of bare monopole is 
\begin{align}
\label{mNw2N_monodim}
\Delta(m)&=
\frac12 \sum_{k=1}^{2N-2} \sum_{i} \sum_{j} |m_{i}^{(k)}-m_{j}^{(k+1)}|
+\sum_{i=1}^{N}|m_{i}^{(N)}|-\sum_{k=1}^{2N-1}\sum_{i<j} |m_{i}^{(k)} -m_{j}^{(k)}|.
\end{align}
The first terms are the contributions from 
the bi-fundamental twisted hypermultiplets. 
The sum over $i$ runs from $1$ to $k$ for $k\le N$ 
and from $1$ to $2N-k$ for $N<k$ 
while the sum over $j$ runs from $1$ to $k+1$ for $k\le N-1$ 
and from $1$ to $2N-k-1$ for $N-1<k$.  
The second terms are the contributions from the two twisted hypermultiplets transforming as 
fundamental representation under the $U(N)$ gauge symmetry, 
and the third terms are the contributions from 
the twisted vector multiplets.

Then the index of the quiver gauge theory 
$\widetilde{
\begin{smallmatrix}
(1)-(2)-&\cdots&-(N)-&\cdots&-(2)-(1)\\
&&|&&\\
&&[2]&&\\
\end{smallmatrix}}$ reads
\begin{align}
\label{3dNw2Nb}
&\mathbb{I}^{
\textrm{3d }
\widetilde{
\begin{smallmatrix}
(1)-(2)-&\cdots&-(N)-&\cdots&-(2)-(1)\\
&&|&&\\
&&[2]&&\\
\end{smallmatrix}}}
(t,x_{\alpha},z_{\alpha};q)
\nonumber\\
&=
\prod_{k=1}^{N}\frac{1}{k!} 
\frac{(q^{\frac12}t^{-2};q)_{\infty}^{k}}
{(q^{\frac12}t^2;q)_{\infty}^{k}}
\sum_{m_{1}^{(k)}, \cdots, m_{k}^{(k)}\in \mathbb{Z}}
\oint \prod_{i=1}^{k} 
\frac{ds_{i}^{(k)}}
{2\pi is_{i}^{(k)}}
\nonumber\\
&\times 
\prod_{i< j}
\frac{
\left(1-q^{\frac{|m_{i}^{(k)}-m_{j}^{(k)}|}{2}} s_{i}^{(k)\pm} s_{j}^{(k)\mp} \right)
\left(q^{\frac{1+|m_{i}^{(k)}-m_{j}^{(k)}|}{2}} t^{-2} s_{i}^{(k)\pm}  s_{j}^{(k)\mp} ; q\right)_{\infty}
}
{
\left(q^{\frac{1+|m_{i}^{(k)}-m_{j}^{(k)}|}{2}} t^{2} s_{i}^{(k)\pm}  s_{j}^{(k)\mp} ; q\right)_{\infty}
}
\nonumber\\
&\times 
\prod_{k=1}^{N-1}\frac{1}{(N-k)!} 
\frac{(q^{\frac12}t^{-2};q)_{\infty}^{N-k}}
{(q^{\frac12}t^2;q)_{\infty}^{N-k}}
\sum_{m_{1}^{(N+k)}, \cdots, m_{N-k}^{(N+k)}\in \mathbb{Z}}
\oint \prod_{i=1}^{N-k} 
\frac{ds_{i}^{(N+k)}}
{2\pi is_{i}^{(N+k)}}
\nonumber\\
&\times 
\prod_{i< j}
\frac{
\left(1-q^{\frac{|m_{i}^{(N+k)}-m_{j}^{(N+k)}|}{2}} s_{i}^{(N+k)\pm} s_{j}^{(N+k)\mp} \right)
\left(q^{\frac{1+|m_{i}^{(N+k)}-m_{j}^{(N+k)}|}{2}} t^{-2} s_{i}^{(N+k)\pm}  s_{j}^{(N+k)\mp} ; q\right)_{\infty}
}
{
\left(q^{\frac{1+|m_{i}^{(N+k)}-m_{j}^{(N+k)}|}{2}} t^{2} s_{i}^{(N+k)\pm}  s_{j}^{(N+k)\mp} ; q\right)_{\infty}
}
\nonumber\\
&\times 
\prod_{k=1}^{2N-2}
\prod_{i} \prod_{j}
\frac{
\left(q^{\frac34+\frac{|m_{i}^{(k)} -m_{j}^{(k+1)}|}{2}} t s_{i}^{(k)\pm} s_{j}^{(k+1)\mp};q \right)_{\infty}
}
{
\left(q^{\frac14+\frac{|m_{i}^{(k)} -m_{j}^{(k+1)}|}{2}} t^{-1} s_{i}^{(k)\pm} s_{j}^{(k+1)\mp};q \right)_{\infty}
}
\cdot 
\prod_{i=1}^{N}\prod_{\alpha=1}^{2} 
\frac{
\left(q^{\frac34+\frac{|m_{i}^{(N)}|}{2}} t s_{i}^{(N)\pm} z_{\alpha}^{\pm};q \right)_{\infty}
}
{
\left(q^{\frac14+\frac{|m_{i}^{(N)}|}{2}} t^{-1} s_{i}^{(N)\pm} z_{\alpha}^{\pm};q \right)_{\infty}
}
\nonumber\\
&\times 
q^{\sum_{k=1}^{2N-2} \sum_{i} \sum_{j} \frac{|m_{i}^{(k)} -m_{j}^{(k+1)} |}{4}
+\sum_{i=1}^{N} \frac{|m_{i}^{(N)}|}{2}
-\sum_{k=1}^{2N-1}\sum_{i<j} \frac{|m_{i}^{(k)} -m_{j}^{(k)}|}{2}
}
\nonumber\\
&\times 
t^{\sum_{k=1}^{2N-2} \sum_{i} \sum_{j} |m_{i}^{(k)}-m_{j}^{(k+1)}|
+2\sum_{i=1}^{N}|m_{i}^{(N)}|-2\sum_{k=1}^{2N-1}\sum_{i<j} |m_{i}^{(k)} -m_{j}^{(k)}|}
\nonumber\\
&\times 
\prod_{k=1}^{2N-1} 
\left(\frac{x_{k}}{x_{k+1}} \right)^{\sum_{i=1}^{k}m_{i}^{(k)}} 
\cdot 
\left(\frac{x_{2N}}{x_{1}} \right)^{m_{1}^{(1)}+\sum_{i=1}^{2N-1}m_{i}^{(2N-1)}}
\end{align}
where $z_{\alpha}$ are the fugacities for the flavor symmetry 
and $x_{\alpha}$ are the fugacities for the topological symmetry. 
We expect that 
the index (\ref{3dNw2Na}) coincides with the index (\ref{3dNw2Nb}).

\begin{figure}
\begin{center}
\includegraphics[width=13cm]{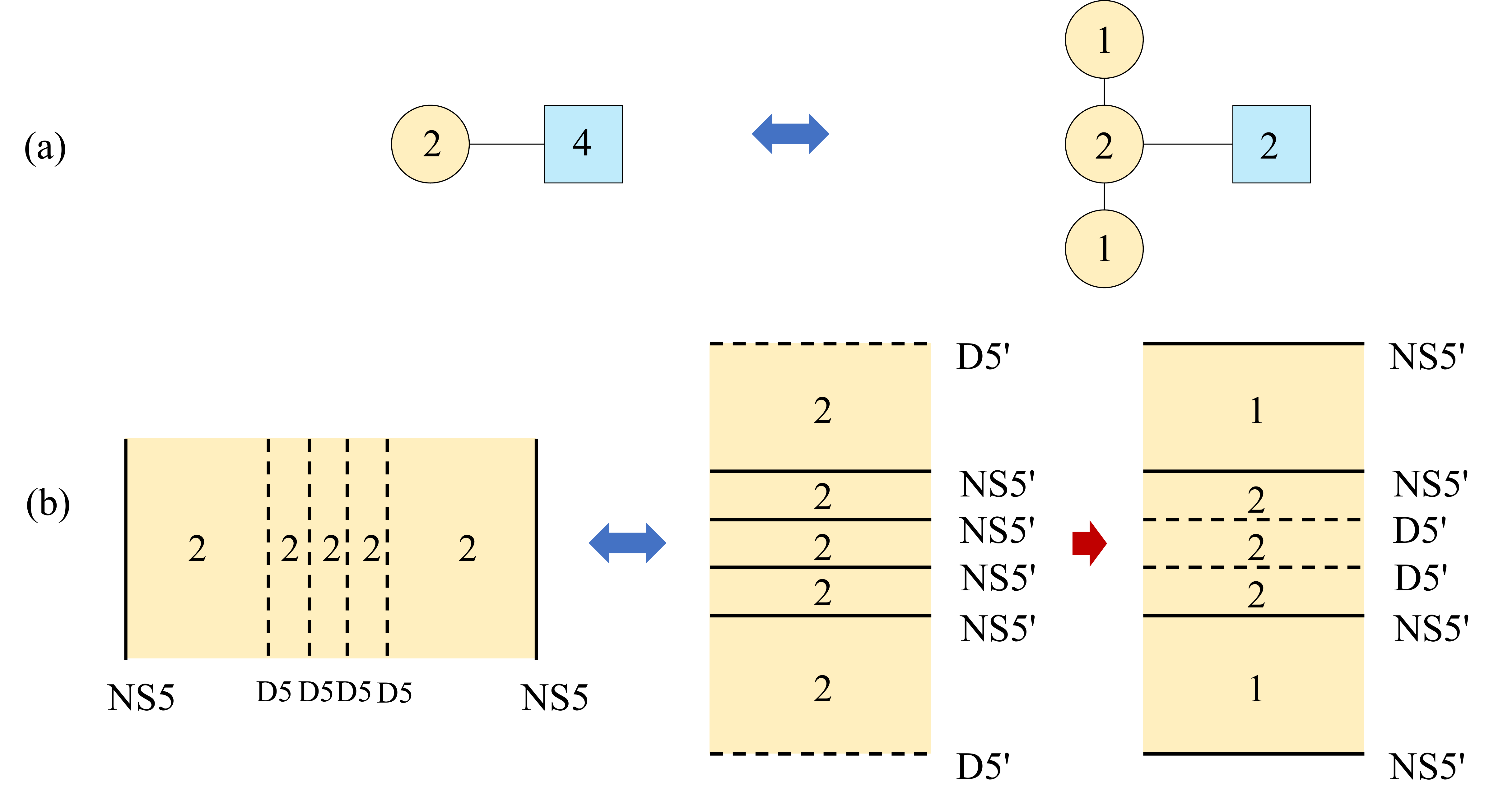}
\caption{
(a) The quiver diagrams of $(2)-[4]$ and its mirror quiver gauge theory. 
(b) The brane configurations of $(2)-[4]$ and its mirror quiver gauge theory. 
}
\label{fig24}
\end{center}
\end{figure}

For example, 
consider 3d $\mathcal{N}=4$ $U(2)$ gauge theory with four fundamental hypermultiplets, which we denote by $(2)-[4]$. 
This is the simplest balanced non-Abelian gauge theory. 
The  quiver diagram and the brane construction are shown in Figure \ref{fig24}. 

The index of $(2)-[4]$ is 
\begin{align}
\label{3d24a}
&\mathbb{I}^{\textrm{3d $(2)-[4]$}}(t,x_{\alpha},z_{\alpha};q)
\nonumber\\
&=
\frac12 \frac{(q^{\frac12}t^2;q)_{\infty}^2}{(q^{\frac12}t^{-2};q)_{\infty}^2} 
\sum_{m_{1},m_{2}\in \mathbb{Z}}
\oint \frac{ds_{1}}{2\pi is_{1}}\frac{ds_{2}}{2\pi is_{2}} 
\nonumber\\
&\times 
\frac{
\left(1-q^{\frac{|m_{1}-m_{2}|}{2}}\frac{s_{1}}{s_{2}} \right)
\left(1-q^{\frac{|m_{1}-m_{2}|}{2}}\frac{s_{2}}{s_{1}} \right)
\left( q^{\frac{1+|m_{1}-m_{2}|}{2}} t^2 \frac{s_{1}}{s_{2}};q \right)_{\infty}
\left( q^{\frac{1+|m_{1}-m_{2}|}{2}} t^2 \frac{s_{2}}{s_{1}};q \right)_{\infty}
}
{
\left( q^{\frac{1+|m_{1}-m_{2}|}{2}} t^{-2} \frac{s_{1}}{s_{2}};q \right)_{\infty}
\left( q^{\frac{1+|m_{1}-m_{2}|}{2}} t^{-2} \frac{s_{2}}{s_{1}};q \right)_{\infty}
}
\nonumber\\
&\times 
\prod_{i=1}^{2} \prod_{\alpha=1}^{4} 
\frac{
\left(q^{\frac34+\frac{|m_{i}|}{2}} t^{-1}s_{i}^{\pm} x_{\alpha}^{\pm};q \right)_{\infty}
}
{
\left(q^{\frac14+\frac{|m_{i}|}{2}} t s_{i}^{\pm} x_{\alpha}^{\pm};q \right)_{\infty}
}
\nonumber\\
&\times 
q^{|m_{1}|+|m_{2}|-\frac{|m_{1}-m_{2}|}{2}} 
t^{-4|m_{1}|-4|m_{2}|+2|m_{1}-m_{2}|} 
z_{1}^{m_{1}+m_{2}} z_{2}^{-m_{1}-m_{2}}
\end{align}
where $x_{\alpha}$ are the fugacities for the $SU(4)$ flavor symmetry with $\prod_{\alpha=1}^{4} x_{\alpha}=1$  
and $z_{\alpha}$ are the fugacities for the topological symmetry. 

In the Coulomb limit where 
one keeps $\widetilde{\mathfrak{t}}^{\frac12}$ $=$ $q^{\frac14}t^{-1}$ constant and sends $q\rightarrow 0$ and $t\rightarrow \infty$, 
the index (\ref{3d24a}) reduces to 
\begin{align}
\label{3d24_Clim}
&\mathbb{I}_{\textrm{Coulomb}}^{\textrm{3d $(2)-[4]$}}(\widetilde{\mathfrak{t}},z)
\frac12 \frac{1}{(1-\widetilde{\mathfrak{t}})^2}
\sum_{m_{1},m_{2}\in \mathbb{Z}} 
\oint \frac{ds_{1}}{2\pi is_{1}} \frac{ds_{2}}{2\pi is_{2}} 
\delta_{m_{1},m_{2}} \frac{(1-s_{1}^{\pm}s_{2}^{\mp})}
{(1-\widetilde{\mathfrak{t}}s_{1}^{\pm}s_{2}^{\mp})} 
\widetilde{\mathfrak{t}}^{\Delta(m)} z^{m_{1}+m_{2}}
\end{align}
where we have defined $z:=$ $z_{1}z_{2}^{-1}$. 
Let $\lambda(m_{i})$ be a partition of $2$ which is associated to the magnetic flux $m_{i}$ 
obeying $\sum_{i=1}^{2}\lambda_{j}(m_{i})$ $=2$ and $\lambda_{j}(m_{i})\ge \lambda_{j+1}(m_{i})$. 
Making use of the formula 
\begin{align}
\label{HLint_id}
\frac{1}{N!}\frac{1}{(1-x)^{N}} \oint \prod_{i=}^{N} 
\frac{ds_{i}}{2\pi is_{i}} \prod_{i\neq j} \frac{\left(1-\frac{s_{i}}{s_{j}} \right)}
{\left(1-x\frac{s_{i}}{s_{j}} \right)}&=\prod_{k=1}^{N}\frac{1}{1-x^{k}},
\end{align}
we get 
\begin{align}
\label{3d24_Clim2}
\mathbb{I}_{\textrm{Coulomb}}^{\textrm{3d $(2)-[4]$}}(\widetilde{\mathfrak{t}},z)
&=\sum_{m_{1},m_{2}\in \mathbb{Z}} 
\widetilde{\mathfrak{t}}^{2|m_{1}|+2|m_{2}|-|m_{1}-m_{2}|}
\cdot 
z^{m_{1}+m_{2}}
\cdot 
P_{U(2)}(\widetilde{\mathfrak{t}},m_{i}).
\end{align}
Here 
\begin{align}
\label{cfac_u2}
P_{U(2)}(\widetilde{\mathfrak{t}},m_{i})&=
\prod_{k=1}^2 \frac{1}{(1-\widetilde{\mathfrak{t}}^k)^{\lambda^{T}_{k}(m_{i})}}
\end{align}
is the factor which counts the number of Casimir 
where  $\lambda_{k}^{T}(m_{i})$ is the length of the $k$-th row of the transposed Young tableau $\lambda^{T}(m_{i})$. 
The expression (\ref{3d24_Clim2}) is identified with the refined Hilbert series for the Coulomb branch of $(2)-[4]$ 
which can be further written as \cite{Cremonesi:2013lqa}
\begin{align}
\label{3d24_Clim3}
\mathbb{I}_{\textrm{Coulomb}}^{\textrm{3d $(2)-[4]$}}(\widetilde{\mathfrak{t}},z)
&=\mathrm{Hilb}^{\textrm{3d $(2)-[4]$}}(\widetilde{\mathfrak{t}},z)
\nonumber\\
&=\prod_{i=1}^{2} \frac{1-\widetilde{\mathfrak{t}}^{5-i}}
{\left(1-\widetilde{\mathfrak{t}}^{i}\right) 
\left(1-z\widetilde{\mathfrak{t}}^{3-i}\right)
\left(1-z^{-1}\widetilde{\mathfrak{t}}^{3-i}\right)}.
\end{align}
The factors $1/(1-\widetilde{\mathfrak{t}}^i)$ with $i=1,2$ correspond to 
the 2 generators $\Tr \varphi^i$ where $\varphi$ is the adjoint scalar field. 
The factors $1/(1-z\widetilde{\mathfrak{t}}^{3-i})$ 
and $1/(1-z^{-1}\widetilde{\mathfrak{t}}^{3-i})$ with $i=1,2$ describe 
the monopole operators $V_{+}$ with magnetic flux $(+,0)$ 
and $V_{-}$ with magnetic flux $(-,0)$ dressed by the adjoint complex scalar field.

The index of the mirror quiver gauge theory $\widetilde{
\begin{smallmatrix}
(1)-(2)-(1)\\
|\\
[2]\\
\end{smallmatrix}}$ 
reads
\begin{align}
\label{3d24b}
&\mathbb{I}^{\textrm{3d $\widetilde{
\begin{smallmatrix}
(1)-(2)-(1)\\
|\\
[2]\\
\end{smallmatrix}}$}}(t,x_{\alpha},z_{\alpha};q)
\nonumber\\
&=
\frac{(q^{\frac12}t^{-2};q)_{\infty}}
{(q^{\frac12}t^2;q)_{\infty}}
\sum_{m_{1}\in \mathbb{Z}}
\oint \frac{ds_{1}}{2\pi is_{1}} 
\cdot 
\frac12 \frac{(q^{\frac12}t^{-2};q)_{\infty}^2}
{(q^{\frac12}t^2;q)_{\infty}^2}
\sum_{m_{2},m_{3}\in \mathbb{Z}}
\oint \frac{ds_{2}}{2\pi is_{2}} \frac{ds_{3}}{2\pi is_{3}}
\nonumber\\
&\times 
\frac{
\left(1-q^{\frac{|m_{2}-m_{3}|}{2}}\frac{s_{2}}{s_{3}} \right)
\left(1-q^{\frac{|m_{2}-m_{3}|}{2}}\frac{s_{3}}{s_{2}} \right)
\left( q^{\frac{1+|m_{2}-m_{3}|}{2}} t^{-2} \frac{s_{2}}{s_{3}};q \right)_{\infty}
\left( q^{\frac{1+|m_{2}-m_{3}|}{2}} t^{-2} \frac{s_{3}}{s_{2}};q \right)_{\infty}
}
{
\left( q^{\frac{1+|m_{2}-m_{3}|}{2}} t^{2} \frac{s_{2}}{s_{3}};q \right)_{\infty}
\left( q^{\frac{1+|m_{2}-m_{3}|}{2}} t^{2} \frac{s_{3}}{s_{2}};q \right)_{\infty}
}
\nonumber\\
&\times 
\frac{(q^{\frac12} t^{-2};q)_{\infty}}{(q^{\frac12} t^2;q)_{\infty}} 
\sum_{m_{4}\in \mathbb{Z}} 
\oint \frac{ds_{4}}{2\pi is_{4}} 
\nonumber\\
&\times 
\prod_{i=2}^{3} 
\frac{
\left(q^{\frac34+\frac{|m_{1}-m_{i}|}{2}} t s_{1}^{\pm} s_{i}^{\mp};q \right)_{\infty}
}
{
\left(q^{\frac14+\frac{|m_{1}-m_{i}|}{2}} t^{-1} s_{1}^{\pm} s_{i}^{\mp};q \right)_{\infty}
}
\cdot 
\prod_{i=2}^{3}\prod_{\alpha=1}^{2} 
\frac{
\left(q^{\frac34+\frac{|m_{i}|}{2}} t s_{i}^{\pm} z_{\alpha}^{\pm};q \right)_{\infty}
}
{
\left(q^{\frac14+\frac{|m_{i}|}{2}} t^{-1} s_{i}^{\pm} z_{\alpha}^{\pm};q \right)_{\infty}
}
\cdot 
\prod_{i=2}^{3}
\frac{
\left(q^{\frac34+\frac{|m_{i}-m_{4}|}{2}} t s_{i}^{\pm} s_{4}^{\mp};q \right)_{\infty}
}{
\left(q^{\frac14+\frac{|m_{i}-m_{4}|}{2}} t^{-1} s_{i}^{\pm} s_{4}^{\mp};q \right)_{\infty}
}
\nonumber\\
&\times 
q^{\frac{|m_{1}-m_{2}|+|m_{1}-m_{3}|}{4} + \frac{|m_{2}|}{2}+\frac{|m_{3}|}{2} + \frac{|m_{2}-m_{4}|+|m_{3}-m_{4}|}{4}-\frac{|m_{2}-m_{3}|}{2}}
\nonumber\\
&\times 
t^{|m_{1}-m_{2}|+|m_{1}-m_{3}|+2|m_{2}|+2|m_{3}|+|m_{2}-m_{4}|+|m_{3}-m_{4}|-2|m_{2}-m_{3}|}
\nonumber\\
&\times 
\left(\frac{x_{1}}{x_{2}} \right)^{m_{1}} 
\left(\frac{x_{2}}{x_{3}} \right)^{m_{2}+m_{3}}
\left(\frac{x_{3}}{x_{4}} \right)^{m_{4}}
\left(\frac{x_{4}}{x_{1}} \right)^{m_{1}+m_{4}}
\end{align}
where $x_{\alpha}$ are the fugacities for the topological symmetry  
while $z_{\alpha}$ are the fugacities for the flavor symmetry. 
As expected, we find that the indices (\ref{3d24a}) and (\ref{3d24b}) coincide with each other 
up to order $q^3$ (see Appendix \ref{app_nonabe}).

\subsection{$T[SU(N)]$}
\label{sec_3dtsuN}
\begin{figure}
\begin{center}
\includegraphics[width=10cm]{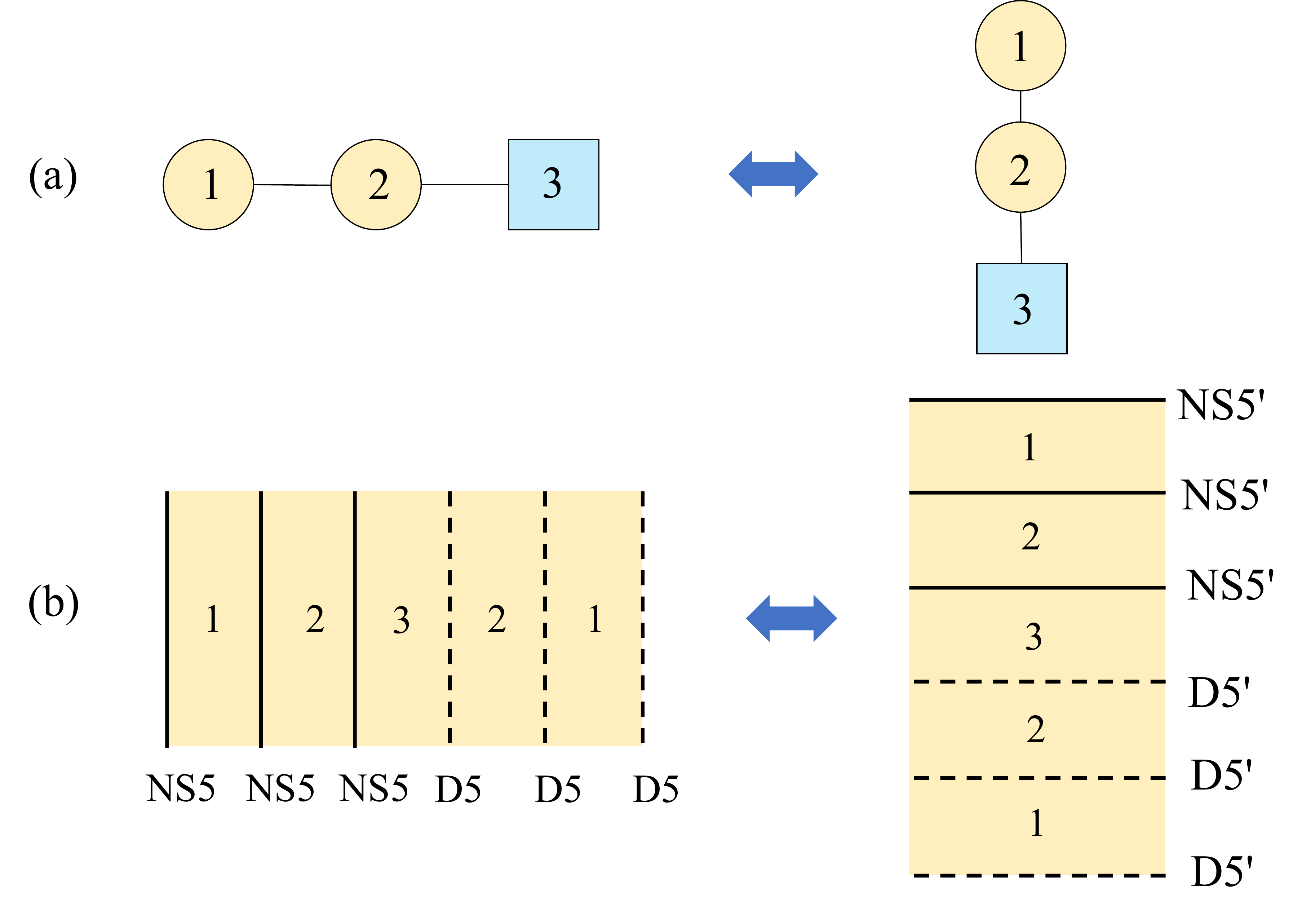}
\caption{
(a) The quiver diagrams of $T[SU(3)]$. 
(b) The brane configurations of $T[SU(3)]$. 
}
\label{figtsu3}
\end{center}
\end{figure}

Consider the quiver gauge theory $(1)-(2)-$ $\cdots$ $-(N-1)-[N]$, which we call $T[SU(N)]$. 
For example, for $T[SU(3)]$ theory the quiver diagram and the brane construction are illustrated in Figure \ref{figtsu3}. 

The dimension of the Coulomb branch of $T[SU(N)]$ is 
\begin{align}
\label{tsundim_C}
\dim_{\mathbb{C}} \mathcal{M}_{C}^{T[SU(N)]}
&=2\sum_{k=1}^{N-1}k=N(N-1)
\end{align}
and the dimension of the Higgs branch of $T[SU(N)]$ is 
\begin{align}
\label{tsundim_H}
\dim_{\mathbb{C}}\mathcal{M}_{H}^{T[SU(N)]}
=2\left[\sum_{k=1}^{N-1}k(k+1)-\sum_{k=1}^{N-1}k^2 \right]
=N(N-1).
\end{align}
The quiver gauge theory $T[SU(N)]$ is a self-mirror theory 
whose Coulomb branch and Higgs branch are identical 
as the dimensions (\ref{tsundim_C}) and (\ref{tsundim_H}) are equal. 

From brane construction, 
a D3-brane ending on the NS5-brane can be viewed as a magnetic monopole. 
Hence the Coulomb branch or Higgs branch of $T[SU(N)]$ is interpreted as 
the moduli space of $SU(N)$ monopoles 
which are formed by 
one with magnetic charge $(+,-,0,\cdots, 0)$, 
two with magnetic charge $(0,+,-,\cdots, 0)$, 
$\cdots$, 
$(N-1)$ with magnetic charge $(0,0,\cdots,+,-)$ 
together with $N$ fixed Dirac monopoles with magnetic charge $(0,0,\cdots,0,+)$. 

The magnetic fluxes for $T[SU(N)]$ is labeled by $N-1$ sets of $k$-tuple of integers, 
that is $\frac{N(N-1)}{2}$ integers $\{m_{i}^{(k)}\}_{i=1,\cdots, k}$ with $k$ $=$ $1,\cdots, N-1$. 
The canonical R-charge of bare monopole is 
\begin{align}
\label{tsuN_monodim}
\Delta(m)&=
\frac12 
\sum_{k=1}^{N-2}\sum_{i=1}^{k}\sum_{j=1}^{k+1} |m_{i}^{(k)}-m_{j}^{(k+1)}|
+\frac{N}{2}\sum_{i=1}^{N-1} |m_{i}^{(N-1)}|
-\sum_{k=1}^{N-1}\sum_{i<j}|m_{i}^{(k)}-m_{j}^{(k)}|
\end{align}
where the first terms are the contributions from the bi-fundamental hypers, 
the second terms are those from the $N$ fundamental hypers, 
and the third terms are those from the vector multiplets.

The index of $T[SU(N)]$ takes the form
\begin{align}
\label{tsuNa}
&\mathbb{I}^{T[SU(N)]}(t,x_{\alpha},z_{\alpha};q)
\nonumber\\
&=
\prod_{k=1}^{N-1} 
\left[
\frac{1}{k!}
\frac{(q^{\frac12}t^2;q)_{\infty}^{k}}
{(q^{\frac12}t^{-2};q)_{\infty}^{k}}
\sum_{m_{1}^{(k)},\cdots, m_{k}^{(k)}\in \mathbb{Z}} 
\oint \prod_{i=1}^{k} 
\frac{ds_{i}^{(k)}}{2\pi is_{i}^{(k)}} 
\right.
\nonumber\\
&\times 
\left.
\prod_{i<j}
\frac{
\left( 1-q^{\frac{|m_{i}^{(k)}-m_{j}^{(k)}|}{2}} s_{i}^{(k)\pm} s_{j}^{(k)\mp} \right)
\left( q^{\frac{1+|m_{i}^{(k)}-m_{j}^{(k)}|}{2}} t^2 s_{i}^{(k)\pm} s_{j}^{(k)\mp};q \right)_{\infty}
}{
\left( q^{\frac{1+|m_{i}^{(k)}-m_{j}^{(k)}|}{2}} t^{-2} s_{i}^{(k)\pm} s_{j}^{(k)\mp};q \right)_{\infty}
}
\right]
\nonumber\\
&\times 
\prod_{k=1}^{N-2}\prod_{i=1}^{k}\prod_{j=1}^{k+1}
\frac{
\left(q^{\frac34+\frac{|m_{i}^{(k)}-m_{j}^{(k+1)}|}{2}} t^{-1} s_{i}^{(k)\pm} s_{j}^{(k)\mp};q \right)_{\infty}
}
{
\left(q^{\frac14+\frac{|m_{i}^{(k)}-m_{j}^{(k+1)}|}{2}} t s_{i}^{(k)\pm} s_{j}^{(k)\mp};q \right)_{\infty}
}
\prod_{i=1}^{N-1}\prod_{\alpha=1}^{N}
\frac{
\left(q^{\frac34+\frac{|m_{i}^{(N-1)}|}{2}} t^{-1} s_{i}^{(N-1)\pm} x_{\alpha}^{\pm};q \right)_{\infty}
}
{
\left(q^{\frac14+\frac{|m_{i}^{(N-1)}|}{2}} t s_{i}^{(N-1)\pm} x_{\alpha}^{\pm};q \right)_{\infty}
}
\nonumber\\
&\times 
q^{\frac14 \sum_{k=1}^{N-2}\sum_{i=1}^{k}\sum_{j=1}^{k+1} |m_{i}^{(k)}-m_{j}^{(k+1)}|
+ \frac{N}{4}\sum_{i=1}^{N-1} |m_{i}^{(N-1)}|
- \frac12\sum_{k=1}^{N-1}\sum_{i<j}|m_{i}^{(k)}-m_{j}^{(k)}|
}
\nonumber\\
&\times 
t^{
-\sum_{k=1}^{N-2}\sum_{i=1}^{k}\sum_{j=1}^{k+1} |m_{i}^{(k)}-m_{j}^{(k+1)}|
-N\sum_{i=1}^{N-1} |m_{i}^{(N-1)}|
+ 2 \sum_{k=1}^{N-1}\sum_{i<j}|m_{i}^{(k)}-m_{j}^{(k)}|
}
\nonumber\\
&\times 
\prod_{k=1}^{N-1}
\left(\frac{z_{k}}{z_{k+1}} \right)^{\sum_{i=1}^{k}m_{i}^{(k)}} 
\cdot 
\left( 
\frac{z_{N}}{z_{1}}
\right)^{m_{1}^{(1)} +\sum_{i=1}^{N-1} m_{i}^{(N-1)}}
\end{align}
where $x_{\alpha}$ are the fugacities for the $SU(N)$ flavor symmetry 
with $\prod_{\alpha=1}^{N}x_{\alpha}=1$ 
and $z_{\alpha}$ are the fugacities for the topological symmetry. 

We expect that the index (\ref{tsuNa}) is equal to 
the index of $\widetilde{T[SU(N)]}$
\begin{align}
\label{mtsuNa}
&\mathbb{I}^{\widetilde{T[SU(N)]}}(t,x_{\alpha},z_{\alpha};q)
\nonumber\\
&=
\prod_{k=1}^{N-1} 
\left[
\frac{1}{k!}
\frac{(q^{\frac12}t^{-2};q)_{\infty}^{k}}
{(q^{\frac12}t^{2};q)_{\infty}^{k}}
\sum_{m_{1}^{(k)},\cdots, m_{k}^{(k)}\in \mathbb{Z}} 
\oint \prod_{i=1}^{k} 
\frac{ds_{i}^{(k)}}{2\pi is_{i}^{(k)}} 
\right.
\nonumber\\
&\times 
\left.
\prod_{i<j}
\frac{
\left( 1-q^{\frac{|m_{i}^{(k)}-m_{j}^{(k)}|}{2}} s_{i}^{(k)\pm} s_{j}^{(k)\mp} \right)
\left( q^{\frac{1+|m_{i}^{(k)}-m_{j}^{(k)}|}{2}} t^{-2} s_{i}^{(k)\pm} s_{j}^{(k)\mp};q \right)_{\infty}
}{
\left( q^{\frac{1+|m_{i}^{(k)}-m_{j}^{(k)}|}{2}} t^{2} s_{i}^{(k)\pm} s_{j}^{(k)\mp};q \right)_{\infty}
}
\right]
\nonumber\\
&\times 
\prod_{k=1}^{N-2}\prod_{i=1}^{k}\prod_{j=1}^{k+1}
\frac{
\left(q^{\frac34+\frac{|m_{i}^{(k)}-m_{j}^{(k+1)}|}{2}} t s_{i}^{(k)\pm} s_{j}^{(k)\mp};q \right)_{\infty}
}
{
\left(q^{\frac14+\frac{|m_{i}^{(k)}-m_{j}^{(k+1)}|}{2}} t^{-1} s_{i}^{(k)\pm} s_{j}^{(k)\mp};q \right)_{\infty}
}
\prod_{i=1}^{N-1}\prod_{\alpha=1}^{N}
\frac{
\left(q^{\frac34+\frac{|m_{i}^{(N-1)}|}{2}} t s_{i}^{(N-1)\pm} z_{\alpha}^{\pm};q \right)_{\infty}
}
{
\left(q^{\frac14+\frac{|m_{i}^{(N-1)}|}{2}} t^{-1} s_{i}^{(N-1)\pm} z_{\alpha}^{\pm};q \right)_{\infty}
}
\nonumber\\
&\times 
q^{\frac14 \sum_{k=1}^{N-2}\sum_{i=1}^{k}\sum_{j=1}^{k+1} |m_{i}^{(k)}-m_{j}^{(k+1)}|
+ \frac{N}{4}\sum_{i=1}^{N-1} |m_{i}^{(N-1)}|
- \frac12\sum_{k=1}^{N-1}\sum_{i<j}|m_{i}^{(k)}-m_{j}^{(k)}|
}
\nonumber\\
&\times 
t^{
\sum_{k=1}^{N-2}\sum_{i=1}^{k}\sum_{j=1}^{k+1} |m_{i}^{(k)}-m_{j}^{(k+1)}|
+N\sum_{i=1}^{N-1} |m_{i}^{(N-1)}|
- 2 \sum_{k=1}^{N-1}\sum_{i<j}|m_{i}^{(k)}-m_{j}^{(k)}|
}
\nonumber\\
&\times 
\prod_{k=1}^{N-1}
\left(\frac{x_{k}}{x_{k+1}} \right)^{\sum_{i=1}^{k}m_{i}^{(k)}} 
\cdot 
\left( 
\frac{x_{N}}{x_{1}}
\right)^{m_{1}^{(1)} +\sum_{i=1}^{N-1} m_{i}^{(N-1)}}
\end{align}
with $z_{\alpha}$ being the fugacities for the flavor symmetry 
satisfying $\prod_{\alpha=1}^{N} z_{\alpha}=1$ 
and $x_{\alpha}$ being the fugacities for the topological symmetry. 
In fact, we have checked that 
the indices (\ref{tsuNa}) and (\ref{mtsuNa}) agree up to order $q^3$ (see Appendix \ref{app_nonabe}).

\subsection{$(1)-(2)-(1)$}
\label{sec_3d121}
\begin{figure}
\begin{center}
\includegraphics[width=13cm]{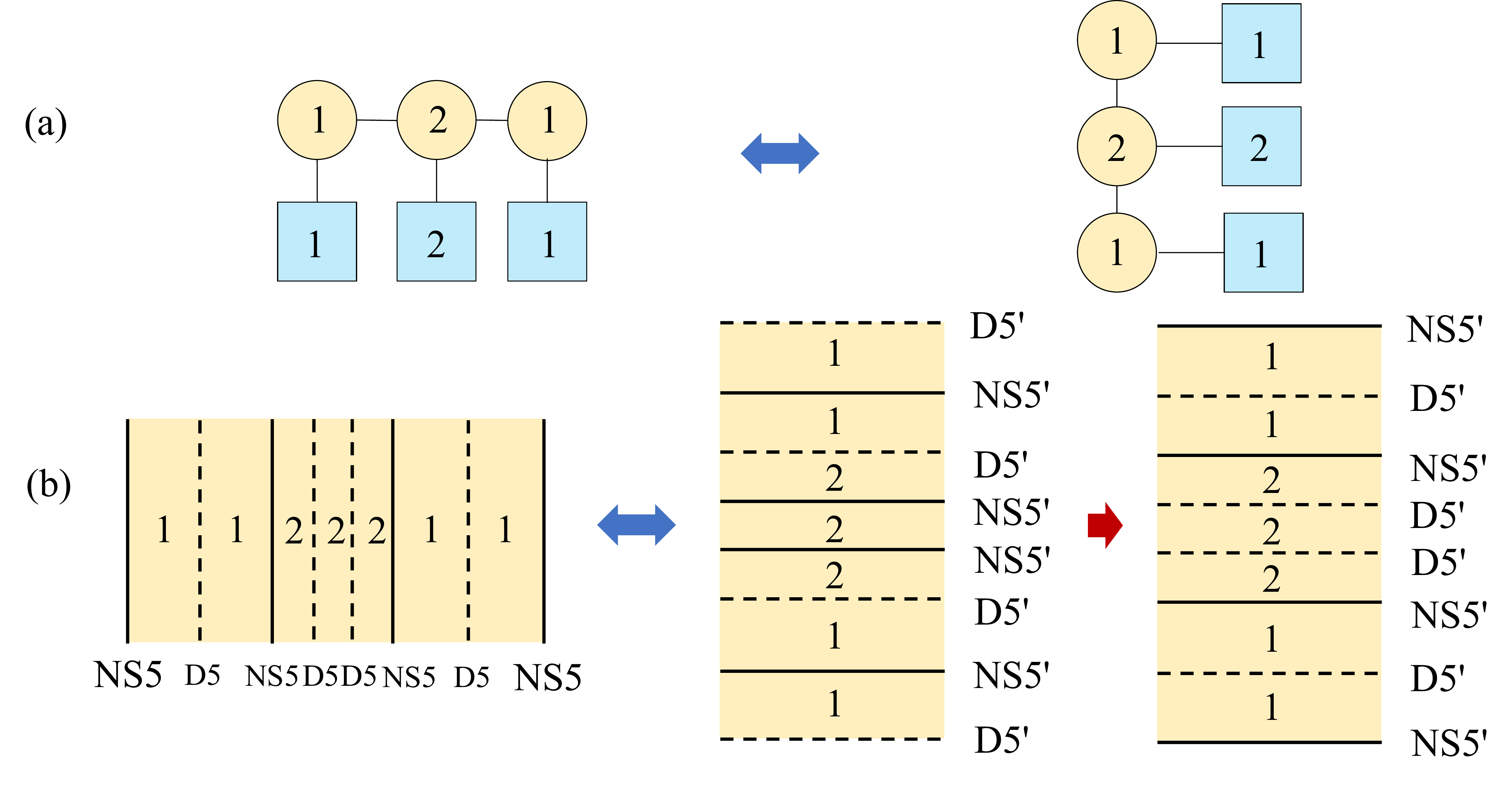}
\caption{
(a) The quiver diagrams of the self-mirror $U(1)\times U(2)\times U(1)$ quiver gauge theory. 
(b) The brane configurations of the self-mirror $U(1)\times U(2)\times U(1)$ quiver gauge theory. 
}
\label{fig121sd}
\end{center}
\end{figure}

Consider a quiver gauge theory $\begin{smallmatrix}
(1)&-&(2)&-&(1)\\
|&&|&&|\\
[1]&&[2]&&[1]\\
\end{smallmatrix}$ 
whose quiver diagram and the brane construction are illustrated in Figure \ref{fig121sd}. 
This is a self-mirror quiver theory. 

The dimension of the Coulomb branch is 
\begin{align}
\label{121sd_dim_C}
\dim_{\mathbb{C}}\mathcal{M}_{C}^{\begin{smallmatrix}
(1)&-&(2)&-&(1)\\
|&&|&&|\\
[1]&&[2]&&[1]\\
\end{smallmatrix}}&=2\cdot (1+2+1)=8
\end{align}
and the dimension of the Higgs branch is 
\begin{align}
\label{121sd_dim_H}
\dim_{\mathbb{C}}\mathcal{M}_{H}^{\begin{smallmatrix}
(1)&-&(2)&-&(1)\\
|&&|&&|\\
[1]&&[2]&&[1]\\
\end{smallmatrix}}&=2\cdot (1\times 1+1\times2+2\times 2+2\times 1+1\times 1-1^2-2^2-1^2)=8.
\end{align}

Let us label the magnetic fluxes for $\begin{smallmatrix}
(1)&-&(2)&-&(1)\\
|&&|&&|\\
[1]&&[2]&&[1]\\
\end{smallmatrix}$ by four integers $m_{1},m_{2},m_{3}, m_{4}$. 
The R-charge of bare monopole operator is given by
\begin{align}
\label{121_monodim}
\Delta(m)&=
\frac{|m_{1}|}{2}
+\frac{|m_{1}-m_{2}|+|m_{1}-m_{3}|}{2}
+|m_{2}|+|m_{3}|
\nonumber\\
&
+\frac{|m_{2}-m_{4}|+|m_{3}-m_{4}|}{2}
+\frac{|m_{4}|}{2}
-|m_{2}-m_{3}|. 
\end{align}
The terms $\frac{|m_{1}|}{2}$ $+$ $\frac{|m_{4}|}{2}$ are contributed from 
the charged hypers under the $U(1)$ gauge factors 
while the terms $|m_{2}|+|m_{3}|$ are the contributions from the fundamental hypers of $U(2)$ gauge symmetry. 
The terms $\prod_{i<j}|m_{i}-m_{j}|$ are the contributions from the bi-fundamental hypers. 
The term $-|m_{2}-m_{3}|$ are the contribution from the $U(2)$ vector multiplet. 

The index of $\begin{smallmatrix}
(1)&-&(2)&-&(1)\\
|&&|&&|\\
[1]&&[2]&&[1]\\
\end{smallmatrix}$ is evaluated as 
\begin{align}
\label{121sd_a}
&
\mathbb{I}^{\textrm{3d 
$\begin{smallmatrix}
(1)&-&(2)&-&(1)\\
|&&|&&|\\
[1]&&[2]&&[1]\\
\end{smallmatrix}$
}}(t,x_{\alpha},z_{\alpha};q)
\nonumber\\
&=
\frac{(q^{\frac12}t^2;q)_{\infty}}
{(q^{\frac12}t^{-2};q)_{\infty}}
\sum_{m_{1}\in \mathbb{Z}}
\oint \frac{ds_{1}}{2\pi is_{1}}
\nonumber\\
&\times 
\frac12 \frac{(q^{\frac12}t^2;q)_{\infty}^2}{(q^{\frac12}t^{-2};q)_{\infty}^2}
\sum_{m_{2},m_{3}\in \mathbb{Z}} \oint \frac{ds_{2}}{2\pi is_{2}}\frac{ds_{3}}{2\pi is_{3}} 
\frac{
\left(1-q^{\frac{|m_{2}-m_{3}|}{2}} s_{2}^{\pm} s_{3}^{\mp} \right)
\left( q^{\frac{1+|m_{2}-m_{3}|}{2}} t^2 s_{2}^{\pm} s_{3}^{\mp};q \right)_{\infty}
}
{
\left( q^{\frac{1+|m_{2}-m_{3}|}{2}} t^{-2} s_{2}^{\pm} s_{3}^{\mp};q \right)_{\infty}
}
\nonumber\\
&\times 
\frac{(q^{\frac12}t^2;q)_{\infty}}
{(q^{\frac12}t^{-2};q)_{\infty}}
\sum_{m_{4}\in \mathbb{Z}}
\oint \frac{ds_{4}}{2\pi is_{4}}
\nonumber\\
&\times 
\frac{
\left( q^{\frac34+\frac{|m_{1}|}{2}} t^{-1} s_{1}^{\pm} x_{1}^{\pm};q \right)_{\infty}
}
{
\left( q^{\frac14+\frac{|m_{1}|}{2}} t s_{1}^{\pm} x_{1}^{\pm};q \right)_{\infty}
}
\cdot 
\prod_{i=2}^{3}
\frac{
\left( q^{\frac34+\frac{|m_{1}-m_{i}|}{2}} t^{-1} s_{1}^{\pm} s_{i}^{\mp};q \right)_{\infty}
}
{
\left( q^{\frac14+\frac{|m_{1}-m_{i}|}{2}} t s_{1}^{\pm} s_{i}^{\mp};q \right)_{\infty}
}
\prod_{i=2}^{3}\prod_{\alpha=2}^{3} 
\frac{
\left( q^{\frac34+\frac{|m_{i}|}{2}} t^{-1} s_{i}^{\pm} x_{\alpha}^{\pm};q \right)_{\infty}
}
{
\left( q^{\frac14+\frac{|m_{i}|}{2}} t s_{i}^{\pm} x_{\alpha}^{\pm};q \right)_{\infty}
}
\nonumber\\
&\times 
\prod_{i=2}^3
\frac{
\left( q^{\frac34+\frac{|m_{i}-m_{4}|}{2}} t^{-1} s_{i}^{\pm} s_{4}^{\mp};q \right)_{\infty}
}
{
\left( q^{\frac14+\frac{|m_{i}-m_{4}|}{2}} t s_{i}^{\pm} s_{4}^{\mp};q \right)_{\infty}
}
\cdot 
\frac{
\left( q^{\frac34+\frac{|m_{4}|}{2}} t^{-1} s_{4}^{\pm} x_{4}^{\pm};q \right)_{\infty}
}
{
\left( q^{\frac14+\frac{|m_{4}|}{2}} t s_{4}^{\pm} x_{4}^{\pm};q \right)_{\infty}
}
\nonumber\\
&\times 
q^{\frac{|m_{1}|}{4}+\frac{|m_{1}-m_{2}|+|m_{1}-m_{3}|}{4}+\frac{|m_{2}|+|m_{3}|}{2}
+\frac{|m_{2}-m_{4}|+|m_{3}-m_{4}|}{4} +\frac{|m_{4}|}{4} -\frac{|m_{2}-m_{3}|}{2}}
\nonumber\\
&\times 
t^{-|m_{1}|-|m_{1}-m_{2}|-|m_{1}-m_{3}|-2|m_{2}| -2|m_{3}|-|m_{2}-m_{4}| -|m_{3}-m_{4}|-|m_{4}|+2|m_{2}-m_{3}| }
\nonumber\\
&\times 
\left( \frac{z_{1}}{z_{2}} \right)^{m_{1}}
\left( \frac{z_{2}}{z_{3}} \right)^{m_{2}+m_{3}}
\left( \frac{z_{3}}{z_{4}} \right)^{m_{4}}
\left( \frac{z_{4}}{z_{1}} \right)^{m_{1}+m_{4}}
\end{align}
where $x_{\alpha}$ are the fugacities for the flavor symmetry 
and $z_{\alpha}$ are the fugavcities for the topological symmetry. 

As expected, we have confirmed that 
the index (\ref{121sd_a}) coincides with the index of $\widetilde{\begin{smallmatrix}
(1)&-&(2)&-&(1)\\
|&&|&&|\\
[1]&&[2]&&[1]\\
\end{smallmatrix}}$ 
which consists of twisted supermultiplets
\begin{align}
\label{121sd_b}
&
\mathbb{I}^{\textrm{3d 
$\widetilde{\begin{smallmatrix}
(1)&-&(2)&-&(1)\\
|&&|&&|\\
[1]&&[2]&&[1]\\
\end{smallmatrix}}$
}}(t,x_{\alpha},z_{\alpha};q)
\nonumber\\
&=
\frac{(q^{\frac12}t^{-2};q)_{\infty}}
{(q^{\frac12}t^{2};q)_{\infty}}
\sum_{m_{1}\in \mathbb{Z}}
\oint \frac{ds_{1}}{2\pi is_{1}}
\nonumber\\
&\times 
\frac12 \frac{(q^{\frac12}t^{-2};q)_{\infty}^2}{(q^{\frac12}t^{2};q)_{\infty}^2}
\sum_{m_{2},m_{3}\in \mathbb{Z}} \oint \frac{ds_{2}}{2\pi is_{2}}\frac{ds_{3}}{2\pi is_{3}} 
\frac{
\left(1-q^{\frac{|m_{2}-m_{3}|}{2}} s_{2}^{\pm} s_{3}^{\mp} \right)
\left( q^{\frac{1+|m_{2}-m_{3}|}{2}} t^{-2} s_{2}^{\pm} s_{3}^{\mp};q \right)_{\infty}
}
{
\left( q^{\frac{1+|m_{2}-m_{3}|}{2}} t^{2} s_{2}^{\pm} s_{3}^{\mp};q \right)_{\infty}
}
\nonumber\\
&\times 
\frac{(q^{\frac12}t^{-2};q)_{\infty}}
{(q^{\frac12}t^{2};q)_{\infty}}
\sum_{m_{4}\in \mathbb{Z}}
\oint \frac{ds_{4}}{2\pi is_{4}}
\nonumber\\
&\times 
\frac{
\left( q^{\frac34+\frac{|m_{1}|}{2}} t s_{1}^{\pm} z_{1}^{\pm};q \right)_{\infty}
}
{
\left( q^{\frac14+\frac{|m_{1}|}{2}} t^{-1} s_{1}^{\pm} z_{1}^{\pm};q \right)_{\infty}
}
\cdot 
\prod_{i=2}^{3}
\frac{
\left( q^{\frac34+\frac{|m_{1}-m_{i}|}{2}} t s_{1}^{\pm} s_{i}^{\mp};q \right)_{\infty}
}
{
\left( q^{\frac14+\frac{|m_{1}-m_{i}|}{2}} t^{-1} s_{1}^{\pm} s_{i}^{\mp};q \right)_{\infty}
}
\prod_{i=2}^{3}\prod_{\alpha=2}^{3} 
\frac{
\left( q^{\frac34+\frac{|m_{i}|}{2}} t s_{i}^{\pm} z_{\alpha}^{\pm};q \right)_{\infty}
}
{
\left( q^{\frac14+\frac{|m_{i}|}{2}} t^{-1} s_{i}^{\pm} z_{\alpha}^{\pm};q \right)_{\infty}
}
\nonumber\\
&\times 
\prod_{i=2}^3
\frac{
\left( q^{\frac34+\frac{|m_{i}-m_{4}|}{2}} t s_{i}^{\pm} s_{4}^{\mp};q \right)_{\infty}
}
{
\left( q^{\frac14+\frac{|m_{i}-m_{4}|}{2}} t^{-1} s_{i}^{\pm} s_{4}^{\mp};q \right)_{\infty}
}
\cdot 
\frac{
\left( q^{\frac34+\frac{|m_{4}|}{2}} t s_{4}^{\pm} z_{4}^{\pm};q \right)_{\infty}
}
{
\left( q^{\frac14+\frac{|m_{4}|}{2}} t^{-1} s_{4}^{\pm} z_{4}^{\pm};q \right)_{\infty}
}
\nonumber\\
&\times 
q^{\frac{|m_{1}|}{4}+\frac{|m_{1}-m_{2}|+|m_{1}-m_{3}|}{4}+\frac{|m_{2}|+|m_{3}|}{2}
+\frac{|m_{2}-m_{4}|+|m_{3}-m_{4}|}{4} +\frac{|m_{4}|}{4} -\frac{|m_{2}-m_{3}|}{2}}
\nonumber\\
&\times 
t^{|m_{1}|+|m_{1}-m_{2}|+|m_{1}-m_{3}|+2|m_{2}|+2|m_{3}|+|m_{2}-m_{4}| +|m_{3}-m_{4}|+|m_{4}|-2|m_{2}-m_{3}| }
\nonumber\\
&\times 
\left( \frac{x_{1}}{x_{2}} \right)^{m_{1}}
\left( \frac{x_{2}}{x_{3}} \right)^{m_{2}+m_{3}}
\left( \frac{x_{3}}{x_{4}} \right)^{m_{4}}
\left( \frac{x_{4}}{x_{1}} \right)^{m_{1}+m_{4}}
\end{align}
where $z_{\alpha}$ are the fugacities for the flavor symmetry 
and $x_{\alpha}$ are the fugacities for the topological symmetry 
(see Appendix \ref{app_nonabe}).

\begin{figure}
\begin{center}
\includegraphics[width=13cm]{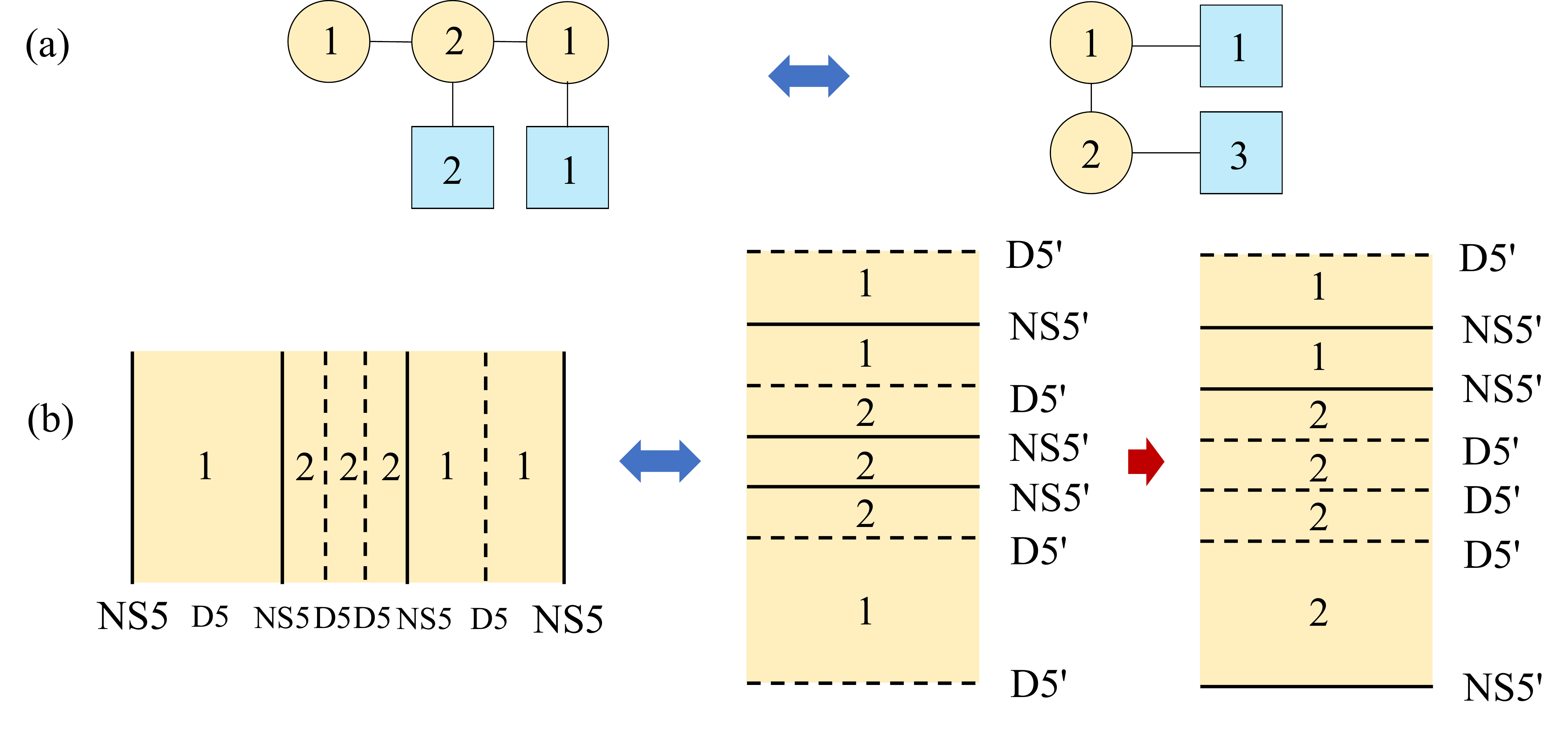}
\caption{
(a) The quiver diagrams of the $U(1)\times U(2)\times U(1)$ quiver gauge theory with two fundamental hypers for the centered $U(2)$ gauge node  
and one charged hyper for the end of the $U(1)$ node and its mirror theory. 
(b) The corresponding brane configurations. 
}
\label{fig121notsd}
\end{center}
\end{figure}

As a next example, let us consider a quiver gauge theory $\begin{smallmatrix}
(1)&-&(2)&-&(1)\\
&&|&&|\\
&&[2]&&[1]\\
\end{smallmatrix}$. 
This is not self-mirror and 
the quiver diagram and the brane construction are drawn in Figure \ref{fig121notsd}. 
The dimension of the Coulomb branch is 
\begin{align}
\label{121021_dimC}
\dim_{\mathbb{C}}\mathcal{M}_{C}^{\begin{smallmatrix}
(1)&-&(2)&-&(1)\\
&&|&&|\\
&&[2]&&[1]\\
\end{smallmatrix}}
&=2\cdot 
(1+2+1)=8
\end{align}
while the dimension of the Higgs branch is 
\begin{align}
\label{121021_dimH}
\dim_{\mathbb{C}}\mathcal{M}_{H}^{\begin{smallmatrix}
(1)&-&(2)&-&(1)\\
&&|&&|\\
&&[2]&&[1]\\
\end{smallmatrix}}
&=2\cdot (1\times 2+2\times 2+2\times 1+1\times 1-1^2-2^2-1^2)
=6.
\end{align}

Again the magnetic fluxes are labeled by integers $m_{1},m_{2},m_{3}, m_{4}$. 
The canonical R-charge of bare monopole is given by 
\begin{align}
\label{121021_monodim}
\Delta(m)&=
\frac{|m_{1}-m_{2}|+|m_{1}-m_{3}|}{2}
+|m_{2}|+|m_{3}|
\nonumber\\
&
+\frac{|m_{2}-m_{4}|+|m_{3}-m_{4}|}{2}
+\frac{|m_{4}|}{2}
-|m_{2}-m_{3}|,
\end{align}
which is obtained by eliminating the contribution $\frac{|m_{1}|}{2}$ 
of the charged hyper from the R-charge (\ref{121_monodim}) for self-mirror quiver gauge theory 
 $\begin{smallmatrix}
(1)&-&(2)&-&(1)\\
|&&|&&|\\
[1]&&[2]&&[1]\\
\end{smallmatrix}$.

We can evaluate the index of $\begin{smallmatrix}
(1)&-&(2)&-&(1)\\
&&|&&|\\
&&[2]&&[1]\\
\end{smallmatrix}$ as
\begin{align}
\label{121_021a}
&
\mathbb{I}^{\textrm{3d $\begin{smallmatrix}
(1)&-&(2)&-&(1)\\
&&|&&|\\
&&[2]&&[1]\\
\end{smallmatrix}$}}(t,x_{\alpha},z_{\alpha};q)
\nonumber\\
&=
\frac{(q^{\frac12}t^2;q)_{\infty}}
{(q^{\frac12} t^{-2};q)_{\infty}}
\sum_{m_{1},\mathbb{Z}} \oint \frac{ds_{1}}{2\pi is_{1}} 
\nonumber\\
&\times 
\frac12 \frac{(q^{\frac12} t^2;q)_{\infty}^2}
{(q^{\frac12} t^{-2};q)_{\infty}^2}
\sum_{m_{2},m_{3}\in \mathbb{Z}} \oint \frac{ds_{2}}{2\pi is_{2}} \frac{ds_{3}}{2\pi is_{3}} 
\frac{
\left(1-q^{\frac{|m_{2}-m_{3}|}{2}} s_{2}^{\pm} s_{3}^{\mp} \right)
\left(q^{\frac{1+|m_{2}-m_{3}|}{2}} t^2 s_{2}^{\pm} s_{3}^{\mp};q\right)_{\infty}
}
{
\left(q^{\frac{1+|m_{2}-m_{3}|}{2}} t^{-2} s_{2}^{\pm} s_{3}^{\mp};q\right)_{\infty}
}
\nonumber\\
&\times 
\frac{
(q^{\frac12}t^{2};q)_{\infty}
}
{
(q^{\frac12}t^{-2};q)_{\infty}
}
\sum_{m_{4}\in \mathbb{Z}} \oint \frac{ds_{4}}{2\pi is_{4}}
\nonumber\\
&\times 
\prod_{i=2}^{3} 
\frac{
\left(q^{\frac34+\frac{|m_{1}-m_{i}|}{2}} t^{-1} s_{1}^{\pm} s_{i}^{\mp};q \right)_{\infty}
}
{
\left(q^{\frac14+\frac{|m_{1}-m_{i}|}{2}} t s_{1}^{\pm} s_{i}^{\mp};q \right)_{\infty}
}
\cdot 
\prod_{i=2}^{3} \prod_{\alpha=1}^{2}
\frac{
\left( q^{\frac34+\frac{|m_{i}|}{2}} t^{-1} s_{i}^{\pm} x_{\alpha}^{\pm};q \right)_{\infty}
}
{
\left( q^{\frac14+\frac{|m_{i}|}{2}} t s_{i}^{\pm} x_{\alpha}^{\pm};q \right)_{\infty}
}
\nonumber\\
&\times 
\prod_{i=2}^{3} 
\frac{
\left( q^{\frac34+\frac{|m_{i}-m_{4}|}{2}} t^{-1} s_{i}^{\pm} s_{4}^{\mp};q \right)_{\infty}
}
{
\left( q^{\frac14+\frac{|m_{i}-m_{4}|}{2}} t s_{i}^{\pm} s_{4}^{\mp};q \right)_{\infty}
}
\cdot 
\frac{
\left( q^{\frac34+\frac{|m_{4}|}{2}} t^{-1} s_{4}^{\pm} x_{3}^{\pm};q \right)_{\infty}
}
{
\left( q^{\frac14+\frac{|m_{4}|}{2}} t s_{4}^{\pm} x_{3}^{\pm};q \right)_{\infty}
}
\nonumber\\
&\times 
q^{\frac{|m_{1}-m_{2}|}{4} +\frac{|m_{1}-m_{3}|}{4} +\frac{|m_{2}|+|m_{3}|}{2} 
+\frac{|m_{2}-m_{4}|}{4}+\frac{|m_{3}-m_{4}|}{4} +\frac{|m_{4}|}{4}-\frac{|m_{2}-m_{3}|}{2} }
\nonumber\\
&\times 
t^{-|m_{1}-m_{2}| -|m_{1}-m_{3}| -2|m_{2}|-2|m_{3}|-|m_{2}-m_{4}|-|m_{3}-m_{4}|-|m_{4}|+2|m_{2}-m_{3}| }
\nonumber\\
&\times 
\left(\frac{z_{4}}{z_{2}} \right)^{m_{1}}
\left(\frac{z_{2}}{z_{3}} \right)^{m_{2}+m_{3}}
\left(\frac{z_{3}}{z_{1}} \right)^{m_{4}}
\left(\frac{z_{1}}{z_{4}} \right)^{m_{1}+m_{4}}
\end{align}
where the fugacities $x_{\alpha}$ are associated to the flavor symmetry 
while the fugacities $z_{\alpha}$ are associated to the topological symmetry.

The mirror of $\begin{smallmatrix}
(1)&-&(2)&-&(1)\\
&&|&&|\\
&&[2]&&[1]\\
\end{smallmatrix}$ is the quiver gauge theory 
$\widetilde{\begin{smallmatrix}
(2)&-&(1)\\
|&&|\\
[3]&&[1]\\
\end{smallmatrix}}$ 
whose quiver diagram and brane construction are illustrated in Figure \ref{fig121notsd}. 

The dimension of the Coulomb branch is 
\begin{align}
\label{m121021_dimC}
\dim_{\mathbb{C}}\mathcal{M}_{C}^{\widetilde{\begin{smallmatrix}
(2)&-&(1)\\
|&&|\\
[3]&&[1]\\
\end{smallmatrix}}}
&=2\cdot (2+1)=6
\end{align}
and the dimension of the Higgs branch is 
\begin{align}
\label{m121021_dimH}
\dim_{\mathbb{C}}\mathcal{M}_{H}^{\widetilde{\begin{smallmatrix}
(2)&-&(1)\\
|&&|\\
[3]&&[1]\\
\end{smallmatrix}}}
&=2\cdot (2\times 3+2\times 1+1\times 1-2^2-1^2)=8. 
\end{align}

The magnetic fluxes are labeled by 
three integers $m_{1}, m_{2}, m_{3}$. 
We have the bare monopole operator with dimension  
\begin{align}
\label{m121021_monodim}
\Delta(m)&=
\frac32|m_{1}|+\frac32|m_{2}|
+\frac{|m_{1}-m_{3}|}{2}
+\frac{|m_{2}-m_{3}|}{2}
+\frac{|m_{3}|}{2}
-|m_{1}-m_{2}|. 
\end{align}
Here the first two terms are contributed from 
three fundamental twisted hypers for $U(2)$ gauge symmetry, 
the next two terms are the contributions from bi-fundamental twisted hypers, 
the second from the last is contributed from charged twisted hyper for $U(1)$ gauge symmetry, 
and the last comes from the $U(2)$ twisted vector multiplet. 

The index of $\widetilde{\begin{smallmatrix}
(2)&-&(1)\\
|&&|\\
[3]&&[1]\\
\end{smallmatrix}}$ is given by 
\begin{align}
\label{121_021b}
&
\mathbb{I}^{\textrm{3d $\widetilde{\begin{smallmatrix}
(2)&-&(1)\\
|&&|\\
[3]&&[1]\\
\end{smallmatrix}}$}}(t,x_{\alpha},z_{\alpha};q)
\nonumber\\
&=\frac12 \frac{(q^{\frac12}t^{-2};q)_{\infty}^2}
{(q^{\frac12}t^2;q)_{\infty}^2} 
\sum_{m_{1},m_{2}\in \mathbb{Z}}
\oint \frac{ds_{1}}{2\pi is_{1}} \frac{ds_{2}}{2\pi is_{2}} 
\frac{
\left( 1-q^{\frac{|m_{1}-m_{2}|}{2}} s_{1}^{\pm} s_{2}^{\mp} \right)
\left( q^{\frac{1+|m_{1}-m_{2}|}{2}} t^{-2} s_{1}^{\mp} s_{2}^{\mp};q \right)_{\infty}
}
{
\left( q^{\frac{1+|m_{1}-m_{2}|}{2}} t^{2} s_{1}^{\mp} s_{2}^{\mp};q \right)_{\infty}
}
\nonumber\\
&\times 
\frac{(q^{\frac12}t^{-2};q)_{\infty}}
{(q^{\frac12}t^2;q)_{\infty}} 
\sum_{m_{3}\in \mathbb{Z}} \oint \frac{ds_{3}}{2\pi is_{3}}
\nonumber\\
&\times 
\prod_{i=1}^{2}\prod_{\alpha=1}^{3}
\frac{
\left(q^{\frac34+\frac{|m_{i}|}{2}} t s_{i}^{\pm} z_{\alpha}^{\pm};q \right)_{\infty}
}
{
\left(q^{\frac14+\frac{|m_{i}|}{2}} t^{-1} s_{i}^{\pm} z_{\alpha}^{\pm};q \right)_{\infty}
}
\prod_{i=1}^{2} 
\frac{
\left(q^{\frac34+\frac{|m_{i}-m_{3}|}{2}} t s_{i}^{\pm} s_{3}^{\mp};q \right)_{\infty}
}
{
\left(q^{\frac14+\frac{|m_{i}-m_{3}|}{2}} t^{-1} s_{i}^{\pm} s_{3}^{\mp};q \right)_{\infty}
}
\cdot 
\frac{
\left(q^{\frac34+\frac{|m_{3}|}{2}} t s_{3}^{\pm} z_{4}^{\pm};q \right)_{\infty}
}
{
\left(q^{\frac14+\frac{|m_{3}|}{2}} t^{-1} s_{3}^{\pm} z_{4}^{\pm};q \right)_{\infty}
}
\nonumber\\
&\times 
q^{\frac34|m_{1}|+\frac34|m_{2}|+\frac{|m_{1}-m_{3}|}{4}+\frac{|m_{2}-m_{3}|}{4}+\frac{|m_{3}|}{4}-\frac{|m_{1}-m_{2}|}{2}}
\nonumber\\
&\times 
t^{3|m_{1}|+3|m_{2}|+|m_{1}-m_{3}|+|m_{2}-m_{3}|+|m_{3}|-2|m_{1}-m_{2}|}
\nonumber\\
&\times 
\left( \frac{x_{1}}{x_{2}} \right)^{m_{1}+m_{2}}
\left( \frac{x_{2}}{x_{3}} \right)^{m_{3}}
\left( \frac{x_{3}}{x_{1}} \right)^{m_{1}+m_{2}+m_{3}}.
\end{align}
As predicted from mirror symmetry, 
the associated global symmetry of fugacities $x_{\alpha}$ and $z_{\alpha}$ are swapped 
and the indices (\ref{121_021a}) and (\ref{121_021b}) coincide with each other (see Appendix \ref{app_nonabe}).

\subsection{$(1)-(2)-[4]$}
\label{sec_3d124}
\begin{figure}
\begin{center}
\includegraphics[width=12.5cm]{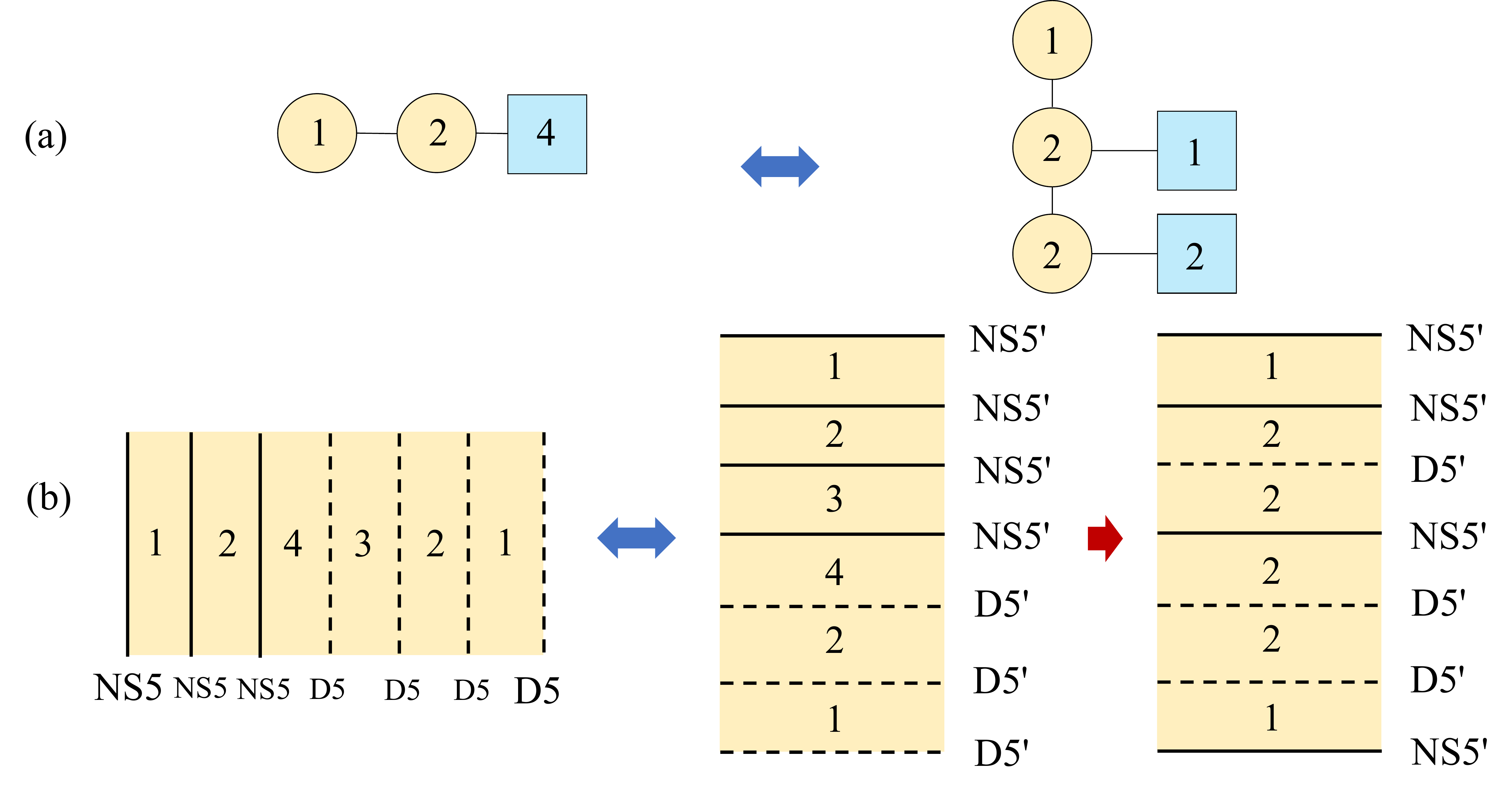}
\caption{
(a) The quiver diagrams of the $(1)-(2)-[4]$ quiver gauge theory and its mirror theory. 
(b) The corresponding brane configurations.  
}
\label{fig124}
\end{center}
\end{figure}

Consider the 3d $\mathcal{N}=4$ quiver gauge theory $(1)-(2)-[4]$. 
The quiver diagram and the brane construction are illustrated in Figure \ref{fig124}. 

The dimension of the Coulomb branch is 
\begin{align}
\label{124dim_c}
\dim_{\mathbb{C}}\mathcal{M}_{C}^{(1)-(2)-[4]}
&=2\cdot (1+2)=6
\end{align}
and the dimension of the Higgs branch is 
\begin{align}
\label{124dim_h}
\dim_{\mathbb{C}}\mathcal{M}_{H}^{(1)-(2)-[4]}
&=2\cdot (1\times 2+2\times 4-1^2-2^2)=10.
\end{align}

For  $(1)-(2)-[4]$ we can label the magnetic fluxes in terms of three integers $m_{1}, m_{2}, m_{3}$. 
The R-charge of bare monopole operator is expressed as
\begin{align}
\label{124_monodim}
\Delta(m)&=
\frac{|m_{1}-m_{2}|}{2}+\frac{|m_{1}-m_{3}|}{2}+
2|m_{2}|+2|m_{3}|-|m_{2}-m_{3}|
\end{align}
where the first two terms are the contributions from bi-fundamental hypers, 
the next two terms are those from four fundamental hypers for the $U(2)$ gauge node 
and the last is contributed from the $U(2)$ vector multiplet.

The index of $(1)-(2)-[4]$ is 
\begin{align}
\label{3d124a}
&\mathbb{I}^{\textrm{3d $(1)-(2)-[4]$}}(t,x_{\alpha},z_{\alpha};q)
\nonumber\\
&=\frac{(q^{\frac12}t^2;q)_{\infty}}
{(q^{\frac12}t^{-2};q)_{\infty}}\sum_{m_{1}\in \mathbb{Z}}
\oint \frac{ds_{1}}{2\pi is_{1}}
\nonumber\\
&\times 
\frac12 \frac{(q^{\frac12}t^2;q)_{\infty}^2}
{(q^{\frac12}t^{-2};q)_{\infty}^2}\sum_{m_{2},m_{3}\in \mathbb{Z}}
\oint \frac{ds_{2}}{2\pi is_{2}}\frac{ds_{3}}{2\pi is_{3}} 
\frac{
\left( 1-q^{\frac{|m_{2}-m_{3}|}{2}}s_{2}^{\pm}s_{3}^{\mp} \right)
\left( q^{\frac{1+|m_{2}-m_{3}|}{2}} t^2 s_{2}^{\pm}s_{3}^{\mp};q\right)_{\infty}
}
{
\left(q^{\frac{1+|m_{2}-m_{3}|}{2}} t^{-2}s_{2}^{\pm} s_{3}^{\mp};q \right)_{\infty}
}
\nonumber\\
&\times 
\prod_{i=2}^{3} 
\frac{
\left(q^{\frac34+\frac{|m_{1}-m_{i}|}{2}} t^{-1} s_{1}^{\pm} s_{i}^{\mp};q \right)_{\infty}
}
{
\left( q^{\frac14+\frac{|m_{1}-m_{i}|}{2}} t s_{1}^{\pm} s_{i}^{\mp};q\right)_{\infty}
}
\cdot 
\prod_{i=2}^{3}\prod_{\alpha=1}^{4} 
\frac{
\left( q^{\frac34+\frac{|m_{i}|}{2}} t^{-1}s_{i}^{\pm}x_{\alpha}^{\pm};q \right)_{\infty}
}
{
\left(q^{\frac14+\frac{|m_{i}|}{2}} t s_{i}^{\pm} x_{\alpha}^{\pm};q \right)_{\infty}
}
\nonumber\\
&\times 
q^{\frac{|m_{1}-m_{2}|}{4}+\frac{|m_{1}-m_{3}|}{4}+|m_{2}|+|m_{3}|-\frac{|m_{2}-m_{3}|}{2}}
\nonumber\\
&\times 
t^{-|m_{1}-m_{2}|-|m_{1}-m_{3}|-4|m_{2}|-4|m_{3}|+2|m_{2}-m_{3}|}
\nonumber\\
&\times 
\left(\frac{z_{3}}{z_{1}} \right)^{m_{1}} 
\left(\frac{z_{1}}{z_{2}} \right)^{m_{2}+m_{3}}
\left(\frac{z_{2}}{z_{3}} \right)^{m_{1}+m_{2}+m_{3}}
\end{align}
where $x_{\alpha}$ and $z_{\alpha}$ are the fugacities for 
flavor symmetry and topological symmetry respectively. 

The mirror of $(1)-(2)-[4]$ is the quiver gauge theory $\widetilde{
\begin{smallmatrix}
(2)&-&(2)&-&(1)\\
|&&|&&\\
[2]&&[1]&&\\
\end{smallmatrix}}$. 
The corresponding quiver diagram and brane configuration 
are shown in Figure \ref{fig124}. 

The dimension of the Coulomb branch is 
\begin{align}
\label{m124_c}
\dim_{\mathbb{C}}\mathcal{M}_{C}^{
\widetilde{
\begin{smallmatrix}
(2)&-&(2)&-&(1)\\
|&&|&&\\
[2]&&[1]&&\\
\end{smallmatrix}}
}&=2\cdot (2+2+1)=10
\end{align}
and the dimension of the Higgs branch is 
\begin{align}
\label{m124_h}
\dim_{\mathbb{C}}\mathcal{M}_{H}^{
\widetilde{
\begin{smallmatrix}
(2)&-&(2)&-&(1)\\
|&&|&&\\
[2]&&[1]&&\\
\end{smallmatrix}}
}&=2\cdot 
(2\times 2+2\times 2+2\times 1+2\times 1-2^2-2^2-1^2)=6. 
\end{align}

The dimension of monopole operator is 
\begin{align}
\label{m124_monodim}
\Delta(m)&=
|m_{1}|+|m_{2}|
+\sum_{i=1}^{2}\sum_{j=3}^{4}\frac{|m_{i}-m_{j}|}{2}
+\frac{|m_{3}|+|m_{4}|}{2}
+\sum_{i=3}^{4}\frac{|m_{i}-m_{5}|}{2}
\nonumber\\
&-\frac{|m_{1}-m_{2}|}{2}-\frac{|m_{3}-m_{4}|}{2}
\end{align}
where the first line includes the contributions from the twisted hyper multiplets 
while the second line describes the contributions from two $U(2)$ twisted vector multiplets. 

As shown in Appendix \ref{app_nonabe}, 
we have checked that 
the index (\ref{3d124a}) beautifully coincides with 
the index of the mirror quiver gauge theory: 
\begin{align}
\label{3d124b}
&\mathbb{I}^{\textrm{3d $\widetilde{
\begin{smallmatrix}
(2)&-&(2)&-&(1)\\
|&&|&&\\
[2]&&[1]&&\\
\end{smallmatrix}}$}}(t,x_{\alpha},z_{\alpha};q)
\nonumber\\
&=
\frac12 \frac{(q^{\frac12}t^{-2};q)_{\infty}^2}
{(q^{\frac12}t^2;q)_{\infty}^2} \sum_{m_{1},m_{2}\in \mathbb{Z}}
\oint \frac{ds_{1}}{2\pi is_{1}} \frac{ds_{2}}{2\pi is_{2}}
\frac{
\left(1-q^{\frac{|m_{1}-m_{2}|}{2}}s_{1}^{\pm}s_{2}^{\mp} \right)
\left(q^{\frac{1+|m_{1}-m_{2}|}{2}} t^{-2}s_{1}^{\pm}s_{2}^{\mp};q \right)_{\infty}
}
{\left(q^{\frac{1+|m_{1}-m_{2}|}{2}} t^2 s_{2}^{\pm} s_{3}^{\mp};q \right)_{\infty}}
\nonumber\\
&\times 
\frac12 \frac{(q^{\frac12}t^{-2};q)_{\infty}^2}
{(q^{\frac12}t^2;q)_{\infty}^2} \sum_{m_{3},m_{4}\in \mathbb{Z}}
\oint \frac{ds_{3}}{2\pi is_{3}} \frac{ds_{4}}{2\pi is_{4}}
\frac{
\left(1-q^{\frac{|m_{1}-m_{2}|}{2}}s_{3}^{\pm}s_{4}^{\mp} \right)
\left(q^{\frac{1+|m_{1}-m_{2}|}{2}} t^{-2}s_{3}^{\pm}s_{4}^{\mp};q \right)_{\infty}
}
{\left(q^{\frac{1+|m_{1}-m_{2}|}{2}} t^2 s_{3}^{\pm} s_{4}^{\mp};q \right)_{\infty}}
\nonumber\\
&\times 
\frac{(q^{\frac12}t^{-2};q)_{\infty}}
{(q^{\frac12}t^2;q)_{\infty}}
\sum_{m_{5}\in \mathbb{Z}} \oint \frac{ds_{5}}{2\pi is_{5}}
\nonumber\\
&\times 
\prod_{i=1}^{2}\prod_{\alpha=1}^{2} 
\frac{
\left( q^{\frac34+\frac{|m_{i}|}{2}} t s_{i}^{\pm}z_{\alpha}^{\pm};q \right)_{\infty}
}
{
\left( q^{\frac14+\frac{|m_{i}|}{2}} t^{-1} s_{i}^{\pm}z_{\alpha}^{\pm};q \right)_{\infty}
}
\cdot 
\prod_{i=1}^{2} \prod_{j=3}^{4}
\frac{
\left( q^{\frac34+\frac{|m_{i}-m_{j}|}{2}} t s_{i}^{\pm} s_{j}^{\mp};q \right)_{\infty}
}
{
\left( q^{\frac14+\frac{|m_{i}-m_{j}|}{2}} t^{-1} s_{i}^{\pm} s_{j}^{\mp};q \right)_{\infty}
}
\nonumber\\
&\times 
\prod_{i=3}^{4} 
\frac{
\left(q^{\frac34+\frac{|m_{i}|}{2}} t s_{i}^{\pm} z_{3}^{\pm};q \right)_{\infty}
}
{
\left(q^{\frac14+\frac{|m_{i}|}{2}} t^{-1} s_{i}^{\pm} z_{3}^{\pm};q \right)_{\infty}
} 
\cdot 
\prod_{i=3}^{4} 
\frac{
\left( q^{\frac34+\frac{|m_{i}-m_{5}|}{2}} t s_{i}^{\pm} s_{5}^{\mp};q \right)_{\infty}
}
{
\left( q^{\frac14+\frac{|m_{i}-m_{5}|}{2}} t^{-1} s_{i}^{\pm} s_{5}^{\mp};q \right)_{\infty}
}
\nonumber\\
&\times 
q^{\frac{|m_{1}|+|m_{2}|}{2}+\sum_{i=1}^{2}\sum_{j=3}^{4}\frac{|m_{i}-m_{j}|}{4}+\frac{|m_{3}|+|m_{4}|}{4}
+\sum_{i=3}^{4}\frac{|m_{i}-m_{5}|}{4}-\frac{|m_{1}-m_{2}|}{2}-\frac{|m_{3}-m_{4}|}{2}}
\nonumber\\
&\times 
t^{2|m_{1}|+2|m_{2}|+\sum_{i=1}^{2}\sum_{j=3}^{4}|m_{i}-m_{j}|+|m_{3}|+|m_{4}|
+\sum_{i=3}^{4}|m_{i}-m_{5}|-2|m_{1}-m_{2}|-2|m_{3}-m_{4}|}
\nonumber\\
&\times 
\left(\frac{x_{1}}{x_{2}}\right)^{m_{1}+m_{2}} 
\left(\frac{x_{2}}{x_{3}} \right)^{m_{3}+m_{4}}
\left(\frac{x_{3}}{x_{4}} \right)^{m_{5}}
\left(\frac{x_{4}}{x_{1}} \right)^{m_{1}+m_{2}+m_{5}}
\end{align}
where the fugacities $z_{\alpha}$ and $x_{\alpha}$ are associated to 
the flavor and topological symmetries of the mirror theory respectively.

\section{Seiberg-like duality}
\label{sec_3dseiberg}

The 3d $\mathcal{N}=4$ $U(N)$ gauge theory with $(2N-1)$ hypermultiplets 
is expected to be equivalent in the IR to a 
$U(N-1)$ gauge theory with $(2N-1)$ hypermultiplets 
times a free twisted hypermultiplet \cite{Gaiotto:2008ak}. 
This is interpreted as Seiberg-like duality between the ugly and good theories. 
The equality of the Hilbert series for the Coulomb branch is shown in \cite{Cremonesi:2013lqa} 
and that for the Hilbert series for the Higgs branch is shown in \cite{Razamat:2014pta}. 
Here we would like to check the equalities of 3d full-indices. 

We conjecture the identity of indices: 
\begin{align}
\label{seiberglike1}
\mathbb{I}^{\textrm{3d $(N)-[2N-1]$}}(t,x_{\alpha},z_{\alpha};q)
&=
\mathbb{I}^{\textrm{3d $(N-1)-[2N-1]$}}(t,x_{\alpha},z_{\alpha};q)
\times 
\mathbb{I}^{\textrm{3d tHM}}\left(t,\frac{z_{1}}{z_{2}};q \right)
\end{align}
where 
\begin{align}
\label{seiberglike2}
&
\mathbb{I}^{\textrm{3d $(N)-[N_{f}]$}}(t,x_{\alpha},z_{\alpha};q)
\nonumber\\
&=
\frac{1}{N!}
\frac{(q^{\frac12}t^2;q)_{\infty}^N}
{(q^{\frac12}t^{-2};q)_{\infty}^{N}}
\sum_{m_{1},\cdots,m_{N}}
\oint \prod_{i=1}^{N} \frac{ds_{i}}{2\pi is_{i}}
\prod_{i<j}
\frac{
\left( 1-q^{\frac{|m_{i}-m_{j}|}{2}} s_{i}^{\pm} s_{j}^{\mp} \right)
\left( q^{\frac{1+|m_{i}-m_{j}|}{2}} t^{2} s_{i}^{\pm} s_{j}^{\mp};q \right)_{\infty}
}
{
\left( q^{\frac{1+|m_{i}-m_{j}|}{2}} t^{-2} s_{i}^{\pm} s_{j}^{\mp};q \right)_{\infty}
}
\nonumber\\
&\times 
\prod_{i=1}^{N}\prod_{\alpha=1}^{N_{f}}
\frac{(q^{\frac34+\frac{|m_{i}|}{2}} t^{-1} s_{i}^{\pm} x_{\alpha}^{\pm};q)_{\infty}}
{(q^{\frac14+\frac{|m_{i}|}{2}} t s_{i}^{\pm} x_{\alpha}^{\pm};q)_{\infty}}
\nonumber\\
&\times 
q^{\frac{N_{f}}{4}\sum_{i=1}^{N}|m_{i}|-\sum_{i<j}\frac{|m_{i}-m_{j}|}{2} }
\cdot 
t^{-N_{f}\sum_{i=1}^{N}|m_{i}|+2\sum_{i<j}|m_{i}-m_{j}|}
\cdot 
\left( \frac{z_{1}}{z_{2}} \right)^{\sum_{i=1}^{N}m_{i}}
\end{align}
is the index for 3d $\mathcal{N}=4$ $U(N)$ gauge theory with 
$N_{f}$ fundamental hypermultiplets. 
In fact, we have checked that 
(\ref{seiberglike1}) holds for $N=2,3$ up to order $q^3$ 
(see Appendix \ref{app_seiberg} for the $q$-expansions).

\section{Dualities of boundary conditions}
\label{sec_3dm4dS1}
Making use of the results in section \ref{sec_3dmirror1} and \ref{sec_3dmirror2}, 
we construct the dual half-BPS boundary conditions for 4d $\mathcal{N}=4$ gauge theories 
by including 3d $\mathcal{N}=4$ gauge theories. 
In order to check the dualities conjectured from the action of S-duality in string theory, 
we compute the half-indices which encode the half-BPS boundary conditions for 4d $\mathcal{N}=4$ gauge theories 
as well as the full-indices for 3d $\mathcal{N}=4$ gauge theories. 
The indices for some half-BPS interfaces in 4d $\mathcal{N}=4$ gauge theory 
were studied in \cite{Gang:2012ff, Gaiotto:2019jvo}. 
The analysis in this section and in section \ref{sec_3dm4dS2} provides more general examples 
which realize 4d-3d dualities.

\subsection{4d $U(N)|$3d $U(M)$}
\label{sec_4duN_3duM}
\begin{figure}
\begin{center}
\includegraphics[width=9cm]{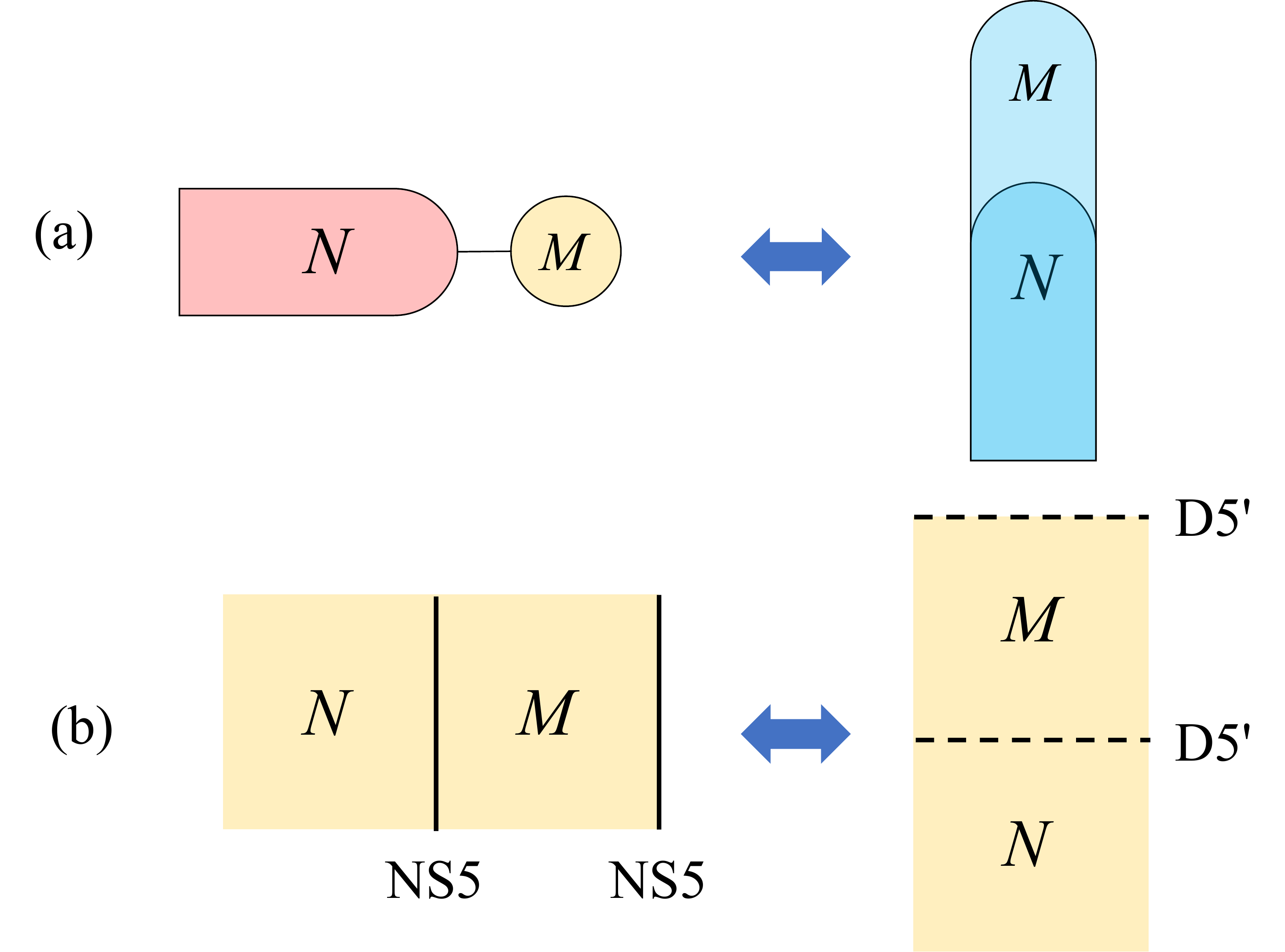}
\caption{
(a) The quiver diagrams of the enriched Neumann b.c. 4d $U(N)|$3d $U(M)$ and its mirror. 
(b) The corresponding brane constructions. 
While the enriched Neumann b.c. 4d $U(N)|$3d $U(M)$ involves 4d gauge and 3d gauge symmetries, 
its mirror b.c. breaks 4d gauge symmetry. 
}
\label{fig4duN3duM}
\end{center}
\end{figure}

Now let us consider the enriched Neumann b.c. $\mathcal{N}$ for 4d $\mathcal{N}=4$ $U(N)$ gauge theory 
which is coupled to 3d $\mathcal{N}=4$ $U(M)$ vector multiplet by the 3d bi-fundamental hypermultiplet. 
We denote this boundary condition by 4d $U(N)|$3d $U(M)$. 
It can be realized in the brane construction as in Figure \ref{fig4duN3duM}. 
There are finite $M$ D3-branes stretched between the NS5-branes 
and semi-infinite $N$ D3-branes extending along the one side of the NS5-brane in the $x^6$ direction. 
The computation of the half-indices for the enriched b.c. 4d $U(N)|$3d $U(M)$ involves 
two sets of contour integral corresponding to the 4d gauge and 3d gauge symmetries. 
On the other hand, the mirror boundary condition is rather simple as it has no gauge symmetry.

We can compute the half-index for 
the boundary condition 4d $U(N)|$3d $U(M)$ as 
\begin{align}
\label{uN_uMa}
&
\mathbb{II}^{\textrm{4d $U(N)|(M)|$}}_{\mathcal{N}}(t,z_{\alpha};q)
\nonumber\\
&=
\underbrace{
\frac{1}{N!}
\frac{(q)_{\infty}^{N}}{(q^{\frac12}t^{-2};q)_{\infty}^{N}} 
\oint \prod_{i=1}^{N} 
\frac{ds_{i}}{2\pi is_{i}} 
\prod_{i\neq j}
\frac{
\left( \frac{s_{i}}{s_{j}};q \right)_{\infty}
}
{
\left(q^{\frac12}t^{-2} \frac{s_{i}}{s_{j}};q \right)_{\infty}
}
}_{\mathbb{II}_{\mathcal{N}}^{\textrm{4d $U(N)$}}}
\nonumber\\
&\times 
\frac{1}{M!} \frac{(q^{\frac12}t^2;q)_{\infty}^M}{(q^{\frac12} t^{-2};q)_{\infty}^M}
\sum_{m_{1},\cdots, m_{M}\in \mathbb{Z}}
\oint \prod_{i=N+1}^{N+M}
\prod_{i<j}
\frac{
\left(1-q^{\frac{|m_{i-N}-m_{j-N}|}{2}}s_{i}^{\pm} s_{j}^{\mp} \right)
\left(q^{\frac{1+|m_{i-N}-m_{j-N}|}{2}} t^2 s_{i}^{\pm} s_{j}^{\mp};q \right)_{\infty}
}
{
\left(q^{\frac{1+|m_{i-N}-m_{j-N}|}{2}} t^{-2} s_{i}^{\pm} s_{j}^{\mp};q \right)_{\infty}
}
\nonumber\\
&\times 
\prod_{i=1}^{N}\prod_{j=N+1}^{N+M}
\frac{
(q^{\frac34+\frac{|m_{j-N}|}{2}} t^{-1} s_{i}^{\pm} s_{j}^{\mp};q)_{\infty}
}
{
(q^{\frac14+\frac{|m_{j-N}|}{2}} t s_{i}^{\pm} s_{j}^{\mp};q)_{\infty}
}
\nonumber\\
&\times 
q^{\sum_{i=1}^{M}\frac{N|m_{i}|}{4}-\sum_{i<j}\frac{|m_{i}-m_{j}|}{2} }
\cdot 
t^{-\sum_{i=1}^{M}N|m_{i}|+2\sum_{i<j}|m_{i}-m_{j}|}
\cdot 
\left( \frac{z_{1}}{z_{2}} \right)^{\sum_{i=1}^{M}m_{i}}.
\end{align}
The contributions in the second line and third line are the half-index for 
4d $U(N)$ gauge theory satisfying Neumann b.c. $\mathcal{N}$ 
and full-index for 3d $U(M)$ gauge theory respectively. 
The contributions in the fourth line count the 3d bi-fundamental hypermultiplet. 
The last line involves the contribution from the monopole operator 
whose canonical R-charge is 
\begin{align}
\label{uN_uM_monodim}
\Delta(m)&=\sum_{i=1}^{M}\frac{N|m_{i}|}{2}-\sum_{i<j}|m_{i}-m_{j}|. 
\end{align}
We assume the condition $N\ge 2M-1$, 
which guarantees that all monopole operators are above the unitarity bound 
so that the half-indices are convergent. 

The S-dual boundary condition 4d $U(N)|[M]|$ is 
associated with two D5$'$-branes. 
The D5$'$-brane defect would break the 4d gauge group $U(N)$ 
down to $U(M)$ block-diagonal subgroup. 
This may lead to the singular boundary condition specified by the Nahm pole of rank $(N-M)$. 
The another D5$'$-brane on which $M$ D3-branes terminate 
requires the singular boundary specified by the Nahm pole of rank $M$. 
There would be $M$ different types of contributions to half-index from broken gauge theory 
characterized by the Nahm pole of rank $(N-M)$, as discussed in \cite{Gaiotto:2019jvo}. 

The half-index for the boundary condition 4d $U(N)|[M]|$ takes the form
\begin{align}
\label{uN_uMb}
&
\mathbb{II}^{\textrm{4d $U(N)|[M]|$}}_{\mathcal{D}'}(t,z_{\alpha};q)
\nonumber\\
&=
\underbrace{
\prod_{k=1}^{N-M}
\frac{(q^{\frac{k+1}{2}}t^{-2(k-1)};q )_{\infty}}
{(q^{\frac{k}{2}} t^{-2k};q)_{\infty}}
}_{\mathbb{II}_{\textrm{Nahm}'}^{\textrm{4d $U(N-M)$}}}
\cdot 
\underbrace{
\prod_{l=1}^{M}
\frac{(q^{\frac{l+1}{2}} t^{-2(l-1)} ;q)_{\infty}}
{(q^{\frac{l}{2}} t^{-2l};q)_{\infty}}
}_{\mathbb{II}_{\textrm{Nahm}'}^{\textrm{4d $U(M)$}}}
\prod_{l=1}^{M}
\frac{
\left( q^{\frac{N-2M}{4}+\frac{l+1}{2}} t^{-(N-2M)-2(l-1)} z_{1}^{\pm} z_{2}^{\mp};q \right)_{\infty}
}
{
\left( q^{\frac{N-2M}{4}+\frac{l}{2}} t^{-(N-2M)-2l} z_{1}^{\pm} z_{2}^{\mp};q \right)_{\infty}
}.
\end{align}
We expect that the half-index (\ref{uN_uMa}) for the enriched Neumann boundary condition 4d $U(N)|$3d $U(M)$ is equal to 
the half-index (\ref{uN_uMb}) for its dual boundary condition 4d $U(N)|[M]|$. 
As shown in Appendix \ref{app_bc}, we have checked that 
they coincide for 
$(N,M)$ $=$ $(1,1)$, $(2,1)$, $(3,1)$, $(3,2)$, $(4,2)$ and $(5,2)$ 
up to certain orders of $q$.

\subsection{4d $U(N)|$3d $U(N-1)\times\cdots\times U(1)$}
\label{sec_4duN3dto1}
\begin{figure}
\begin{center}
\includegraphics[width=11cm]{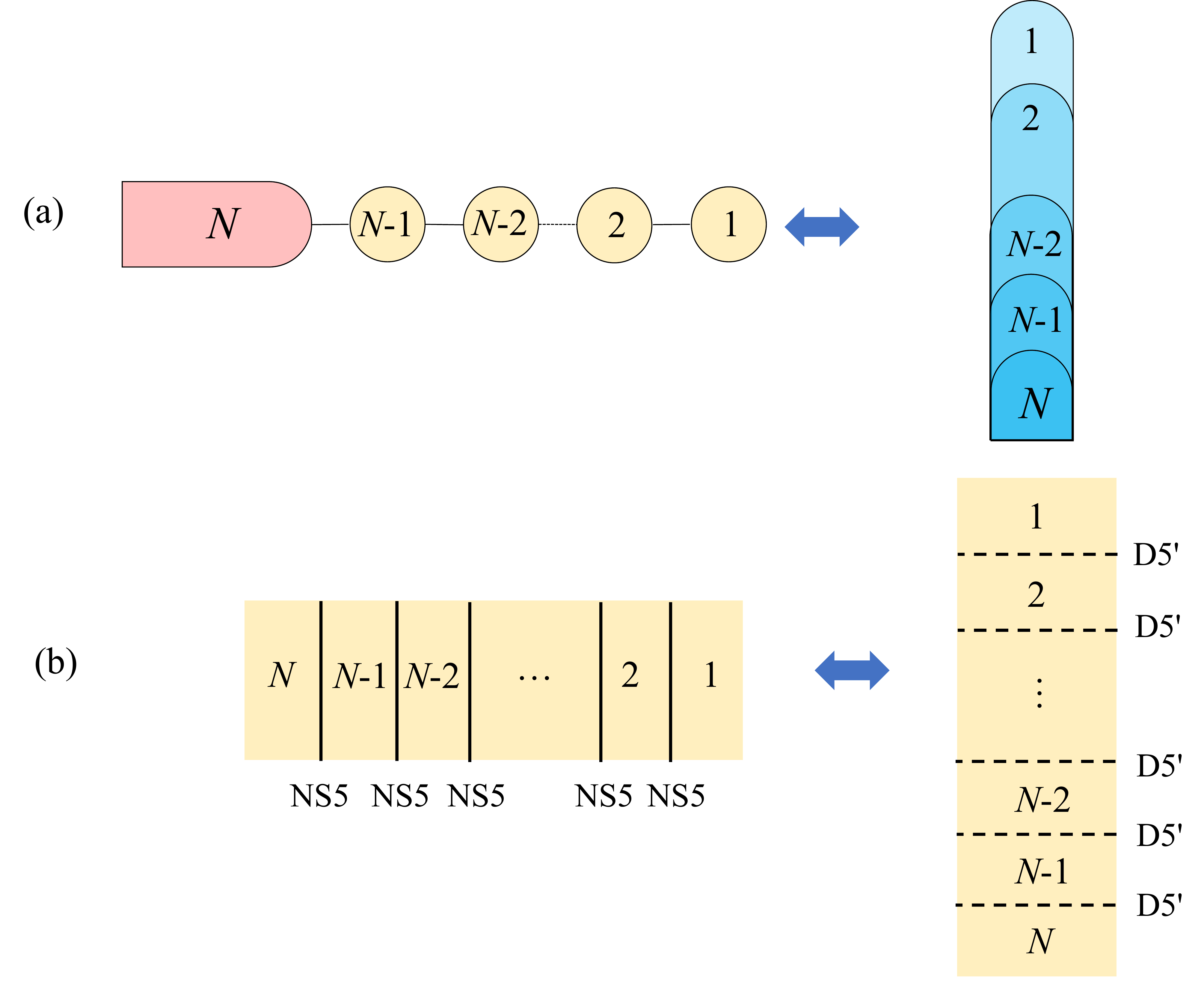}
\caption{
(a) The quiver diagrams of the enriched Neumann b.c. 4d $U(N)|$3d $U(N-1)\times U(N-2)\times \cdots \times U(1)$ 
and its mirror, that is Dirichlet b.c. $\mathcal{D}'$ for 4d $U(N)$ gauge theory.
(b) The corresponding brane constructions. 
}
\label{figdirichlet}
\end{center}
\end{figure}

Now we present the half-index of the enriched Neumann boundary condition 4d $U(N)|$3d $U(N-1)\times\cdots\times U(1)$, 
which is the Neumann b.c. $\mathcal{N}$ for 4d $\mathcal{N}=4$ $U(N)$ gauge theory 
coupled to 3d $\mathcal{N}=4$ $U(N-1)\times U(N-2)$ $\times \cdots \times$ $U(1)$ quiver gauge theory 
through the 3d hypermultiplet transforming as 
$({\bf N}, \overline{\bf N-1})$ $\oplus$ $(\overline{\bf N}, {\bf N-1})$ 
under the $U(N)\times U(N-1)$ gauge group. 
This is expected to be dual to the Dirichlet boundary condition $\mathcal{D}'$ 
for 4d $\mathcal{N}=4$ $U(N)$ SYM theory \cite{Gaiotto:2008sa} 
(see Figure \ref{figdirichlet}). 

One can compute the half-index of the enriched Neumann boundary condition 4d $U(N)|$3d $U(N-1)\times\cdots\times U(1)$ as 
\begin{align}
\label{uN_to_u1a}
&
\mathbb{II}^{\textrm{4d $U(N)|(N-1)-(N-2)-\cdots-(1)$}}_{\mathcal{N}}
(t,z_{\alpha};q)
\nonumber\\
&=
\underbrace{
\frac{1}{N!}
\frac{(q)_{\infty}^{N}}{(q^{\frac12}t^{-2};q)_{\infty}^{N}}
\oint \prod_{i=1}^{N}\frac{ds_{i}}{2\pi is_{i}}
\prod_{i\neq j}
\frac{\left(\frac{s_{i}}{s_{j}};q \right)_{\infty}}
{\left( q^{\frac12} t^{-2} \frac{s_{i}}{s_{j}};q \right)_{\infty}}
}_{\mathbb{II}^{\textrm{4d $U(N)$}}_{\mathcal{N}}}
\nonumber\\
&\times 
\prod_{k=1}^{N-1}
\left[
\frac{1}{k!} 
\frac{(q^{\frac12}t^2;q)_{\infty}^k}
{(q^{\frac12}t^{-2};q)_{\infty}^k}
\sum_{m_{1}^{(k)}, \cdots, m_{k}^{(k)}} 
\oint 
\prod_{i=1}^{k} 
\frac{ds_{i}^{(k)}}{2\pi is_{i}^{(k)}}
\prod_{i\neq j}
\frac{
\left( \frac{s_{i}^{(k)}}{s_{j}^{(k)}};q \right)_{\infty}
}
{
\left( q^{\frac12} t^{-2} \frac{s_{i}^{(k)}}{s_{j}^{(k)}};q \right)_{\infty}
}
\right]
\nonumber\\
&\times 
\prod_{i=1}^{N}\prod_{j=1}^{N-1}
\frac{
(q^{\frac34+\frac{|m_{j}^{(N-1)}|}{2}} t^{-1}s_{i}^{\pm} s_{j}^{(N-1)\mp};q)_{\infty}
}
{
(q^{\frac14+\frac{|m_{j}^{(N-1)}|}{2}} t s_{i}^{\pm} s_{j}^{(N-1)\mp};q)_{\infty}
}
\prod_{k=1}^{N-2}
\prod_{i=1}^{k}
\prod_{j=1}^{k+1}
\frac{
(q^{\frac34+\frac{| m_{i}^{(k)}- m_{j}^{(k+1)} |}{2}} t^{-1} s_{i}^{(k)\pm} s_{j}^{(k+1)\mp} ;q)_{\infty}
}
{
(q^{\frac14+\frac{| m_{i}^{(k)}- m_{j}^{(k+1)} |}{2}} t s_{i}^{(k)\pm} s_{j}^{(k+1)\mp} ;q)_{\infty}
}
\nonumber\\
&\times 
q^{\frac{N}{4}\sum_{i=1}^{N-1} |m_{i}^{(N-1)}|+\sum_{k=1}^{N-2}\sum_{i=1}^{k}\sum_{j=1}^{k+1}\frac{|m_{i}^{(k)}-m_{j}^{(k+1)}|}{4}
-\sum_{k=1}^{N-1}\sum_{i<j}\frac{|m_{i}^{(k)}-m_{j}^{(k)}|}{2}
}
\nonumber\\
&\times 
t^{-N\sum_{i=1}^{N-1}|m_{i}^{(N-1)}|-\sum_{k=1}^{N-2}\sum_{i=1}^{k}\sum_{j=1}^{k+1}|m_{i}^{(k)}-m_{j}^{(k+1)}|
+2\sum_{k=1}^{N-1}\sum_{i<j}|m_{i}^{(k)}-m_{j}^{(k)}|  }
\nonumber\\
&\times 
\prod_{k=1}^{N-1}
\left(\frac{z_{k}}{z_{k+1}} \right)^{\sum_{i=1}^{k}m_{i}^{(k)}}
\cdot 
\left(\frac{z_{N}}{z_{1}} \right)^{m_{1}^{(1)}+\sum_{i=1}^{N-1}m_{i}^{(N-1)}}. 
\end{align}

The half-index (\ref{uN_to_u1a}) would agree with 
the half-index of Dirichlet boundary $\mathcal{D}'$ for 4d $\mathcal{N}=4$ $U(N)$ gauge theory:
\begin{align}
\label{uN_to_u1b}
\mathbb{II}^{\textrm{4d $U(N)$}}_{\mathcal{D}'}(t,z_{\alpha};q)
&=
\frac{
(q)_{\infty}^{N}
}
{
(q^{\frac12}t^{-2};q)_{\infty}^{N}
}
\prod_{i\neq j}
\frac{
\left(q \frac{z_{i}}{z_{j}};q \right)_{\infty}
}
{
\left( q^{\frac12} t^{-2} \frac{z_{i}}{z_{j}} ;q\right)_{\infty}
}.
\end{align}
In fact, we have checked that 
they coincide with each other for $N=1, 2$ up to certain orders of $q$ by using Mathematica (see Appendix \ref{app_bc}).

%

\subsection{4d $U(N)|$3d $(M)-[2M]$}
\label{sec_4duN_3duM2M}
\begin{figure}
\begin{center}
\includegraphics[width=9cm]{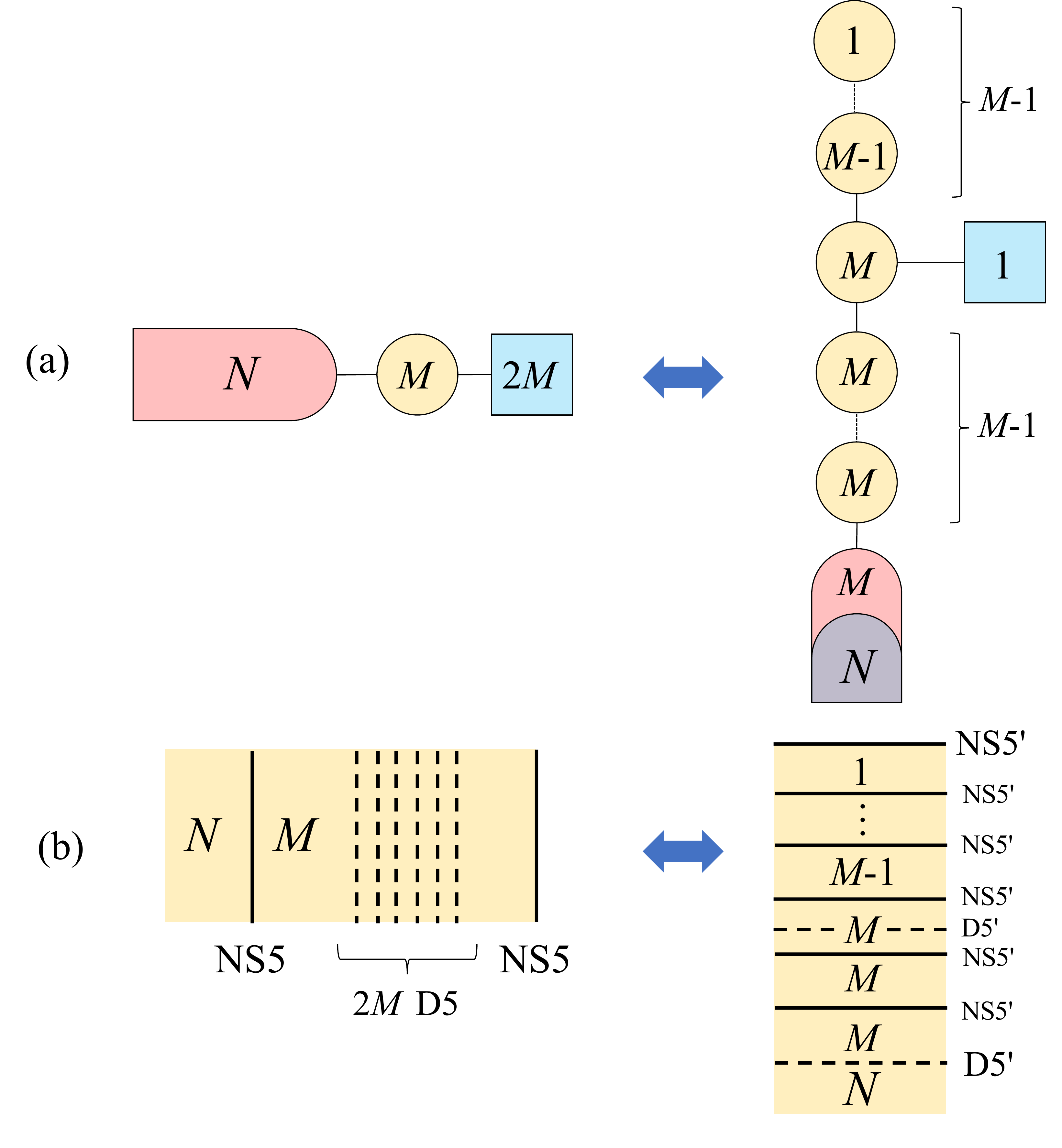}
\caption{
(a) The quiver diagrams of the enriched Neumann b.c. 4d $U(N)|$3d $(M)-[2M]$ and its mirror. 
(b) The corresponding brane constructions. 
}
\label{fig4duN3duM2M}
\end{center}
\end{figure}

Let us consider the enriched Neumann b.c. $\mathcal{N}$ for 
4d $\mathcal{N}=4$ $U(N)$ gauge theory 
coupled to balanced 3d $\mathcal{N}=4$ $U(M)$ gauge theory with $2M$ hypermultiplets. 
We represent this boundary condition by 4d $U(N)|$3d $(M)-[2M]$. 
The corresponding quiver diagram and brane construction are shown in Figure \ref{fig4duN3duM2M}. 
In contrast to the enriched boundary conditions in section \ref{sec_4duN_3duM}, 
the dual boundary conditions can admit both 4d and 3d gauge symmetries. 
The dualities of boundary conditions produce a generalization of mirror symmetry for $(N)-[2N]$ discussed in section \ref{sec_3dNw2N}.

The enriched Neumann b.c. 4d $U(N)|$3d $(M)-[2M]$ 
consists of 
the Neumann b.c. $\mathcal{N}$ for 4d $\mathcal{N}=4$ $U(N)$ gauge theory 
coupled to 3d $\mathcal{N}=4$ $U(M)$ balanced gauge theory with $2M$ fundamental hypermultiplets 
via 3d boundary hypermultiplets transforming as $({\bf N}, \overline{\bf M})$ $\oplus$ $(\overline{\bf N}, {\bf M})$ 
under the $U(N)\times U(M)$ gauge symmetry. 
Here we assume that $N\ge M$. 

The half-index for the enriched Neumann b.c. 4d $U(N)|$3d $(M)-[2M]$ reads
\begin{align}
\label{uN_M2Ma}
&\mathbb{II}^{\textrm{4d $U(N)|(M)-[2M]$}}_{\mathcal{N}}(t,x_{\alpha},z_{\alpha};q)
\nonumber\\
&=
\underbrace{
\frac{1}{N!}\frac{(q)_{\infty}^{N}}{(q^{\frac12}t^{-2};q)_{\infty}^{N}}
\prod_{i=1}^{N}\oint \frac{ds_{i}}{2\pi is_{i}}
\prod_{i\neq j}
\frac{\left(\frac{s_{i}}{s_{j}};q \right)_{\infty}}
{
\left(q^{\frac12} t^{-2}\frac{s_{i}}{s_{j}};q \right)_{\infty}
}
}_{\mathbb{II}_{\mathcal{N}}^{\textrm{4d $U(N)$}}}
\nonumber\\
&\times 
\frac{1}{M!}
\frac{(q^{\frac12}t^2;q)_{\infty}^{M}}
{(q^{\frac12}t^{-2};q)_{\infty}^{M}}
\sum_{m_{1},\cdots,m_{M}\in \mathbb{Z}}
\oint \prod_{i=N+1}^{N+M} \frac{ds_{i}}{2\pi is_{i}}
\prod_{i\neq j} 
\frac{
\left(1-q^{\frac{|m_{i}-m_{j}|}{2}} s_{i}^{\pm} s_{j}^{\mp} \right)
\left( q^{\frac{1+|m_{i}-m_{j}|}{2}} t^2 s_{i}^{\pm} s_{j}^{\mp};q \right)_{\infty}
}
{
\left( q^{\frac{1+|m_{i}-m_{j}|}{2}} t^{-2} s_{i}^{\pm} s_{j}^{\mp};q \right)_{\infty}
}
\nonumber\\
&\times 
\prod_{i=1}^{N} \prod_{j=N+1}^{N+M} 
\frac{
\left( q^{\frac34+\frac{|m_{j-N}|}{2}} t^{-1} s_{i}^{\pm} s_{j}^{\mp};q \right)_{\infty}
}
{
\left( q^{\frac14+\frac{|m_{j-N}|}{2}} t s_{i}^{\pm} s_{j}^{\mp};q \right)_{\infty}
}
\cdot 
\prod_{i=N+1}^{N+M}\prod_{\alpha=1}^{2M}
\frac{
\left(q^{\frac34+\frac{|m_{i-N}|}{2}} t^{-1} s_{i}^{\pm} x_{\alpha}^{\pm};q \right)_{\infty}
}
{
\left(q^{\frac14+\frac{|m_{i-N}|}{2}} t s_{i}^{\pm} x_{\alpha}^{\pm};q \right)_{\infty}
}
\nonumber\\
&\times 
q^{\frac{N+2M}{4}\sum_{i=1}^{M}|m_{i}|-\sum_{i<j}\frac{|m_{i}-m_{j}|}{2} }
\cdot 
t^{-(N+2M)\sum_{i=1}^{M}|m_{i}|+2\sum_{i<j}|m_{i}-m_{j}|}
\cdot 
\left(\frac{z_{1}}{z_{2}} \right)^{\sum_{i=1}^{M}m_{i}}. 
\end{align}
The second and third line 
describe the half-index of Neumann b.c. $\mathcal{N}$ for 4d $U(N)$ gauge theory 
and the full-index of 3d $U(M)$ vector multiplet respectively. 
The fourth line counts the boundary bi-fundamental hypermultiplet 
and the fundamental hypermultiplet in $(M)-[2M]$. 
The last line counts the monopole operator 
with the canonical R-charge
\begin{align}
\Delta(m)&=
\frac{N+2M}{2}\sum_{i=1}^{M}|m_{i}|-\sum_{i<j} |m_{i}-m_{j}|.
\end{align}
This is generalized from the formula (\ref{Nw2N_monodim}) 
by including the additional contributions from the boundary bi-fundamental hypermultiplet 
coupled to 4d $U(N)$ SYM theory.

The S-dual boundary condition can be read off from 
the brane configuration in Figure \ref{fig4duN3duM2M}. 
The 4d gauge symmetry must break down to $U(M)$ 
due to the presence of the D5$'$-brane interface. 
When $N=M$, there is a defect 3d twisted hypermultiplet at the D5$'$-brane interface 
transforming under fundamental representation under the $U(N)$ gauge group. 
For $N>M$, there is no fundamental twisted hypermultiplet at the defect, 
however, there are contributions to the half-index from the broken $U(N)$ gauge theory 
associated with the Nahm pole of rank $(N-M)$ \cite{Gaiotto:2019jvo}.

The surviving 4d $U(M)$ gauge theory should obey the Neumann b.c. $\mathcal{N}'$ due to the NS5$'$-brane. 
It further couples to 3d twisted quiver gauge theory through the 3d bi-fundamental twisted hypermultiplet. 
The 3d twisted quiver gauge theory has 
$U(M)^M\times U(M-1)\times U(M-2)\times U(1)$ gauge symmetry 
and a single $U(1)$ flavor node for the $M$-th $U(M)$ gauge node, 
which we denote by 
$\widetilde{\begin{smallmatrix}
(M)^{M-1}&-&(M)&-&(M-1)&-&(M-2)&-&\cdots&-&(1)\\
&&|&&&&&&&\\
&&[1]&&&&&&&\\
\end{smallmatrix}}$
.

We expect that the half-index (\ref{uN_M2Ma}) agrees with 
\begin{align}
\label{uN_M2Mb}
&\mathbb{II}^{\textrm{4d $U(N)\rightarrow U(M)|
\widetilde{\begin{smallmatrix}
(M)^{M-1}&-&(M)&-&(M-1)&-&(M-2)&-&\cdots&-&(1)\\
&&|&&&&&&&\\
&&[1]&&&&&&&\\
\end{smallmatrix}}
$}}_{\mathcal{N}'}(t,x_{\alpha},z_{\alpha};q)
\nonumber\\
&=
\underbrace{
\frac{1}{M!}\frac{(q)_{\infty}^{M}}
{(q^{\frac12}t^2;q)_{\infty}^{M}}
\oint \prod_{i=1}^{M} \frac{ds_{i}}{2\pi is_{i}} 
\prod_{i\neq j} 
\frac{\left(\frac{s_{i}}{s_{j}};q \right)_{\infty}}
{\left( q^{\frac12}t^2 \frac{s_{i}}{s_{j}};q \right)_{\infty}}
}_{\mathbb{II}_{\mathcal{N}'}^{\textrm{4d $U(M)$}}} 
\cdot 
\underbrace{
\prod_{k=1}^{N-M}
\frac{
(q^{\frac{k+1}{2}} t^{-2(k-1)};q )_{\infty}
}
{
(q^{\frac{k}{2}} t^{-2k};q )_{\infty}
}
}_{\mathbb{II}_{\textrm{Nahm}'}^{\textrm{4d $U(N-M)$}}}
\nonumber\\
&\times 
\left( \frac{1}{M!} \frac{(q^{\frac12} t^{-2};q)_{\infty}^{M} }
{(q^{\frac12}t^2;q)_{\infty}^{M}} \right)^{M} 
\prod_{k=1}^{M} 
\left[
\sum_{m_{1}^{(k)},\cdots,m_{M}^{(k)}\in \mathbb{Z} }
\oint \prod_{i=1}^{M} \frac{ds_{i}^{(k)}}{2\pi is_{i}^{(k)}}
\right.
\nonumber\\
&\times 
\left. 
\prod_{i<j} 
\frac{
(1-q^{\frac{|m_{i}^{(k)}-m_{j}^{(k)}|}{2}} s_{i}^{(k)\pm} s_{j}^{(k)\mp})
(q^{\frac{1+|m_{i}^{(k)}-m_{j}^{(k)}|}{2}} t^{-2} s_{i}^{(k)\pm} s_{j}^{(k)\mp};q )_{\infty}
}
{
(q^{\frac{1+|m_{i}^{(k)}-m_{j}^{(k)}|}{2}} t s_{i}^{(k)\pm} s_{j}^{(k)\mp};q )_{\infty}
}
\right]
\nonumber\\
&\times 
\prod_{k=M+1}^{2M-1}
\left[
\frac{1}{(2M-k)!}
\frac{(q^{\frac12}t^{-2};q)_{\infty}^{2M-k}}
{(q^{\frac12}t^2;q)_{\infty}^{2M-k}}
\sum_{m_{1}^{(k)},\cdots, m_{2M-k}^{(k)}\in \mathbb{Z}} 
\oint \prod_{i=1}^{2M-k} \frac{ds_{i}^{(k)}}{2\pi is_{i}^{(k)}}
\right. 
\nonumber\\
&\times 
\left. 
\prod_{i<j} 
\frac{
(1-q^{\frac{|m_{i}^{(k)}-m_{j}^{(k)}|}{2}} s_{i}^{(k)\pm}  s_{j}^{(k)\mp} )
(q^{\frac{1+|m_{i}^{(k)}-m_{j}^{(k)}|}{2}} t^{-2} s_{i}^{(k)\pm} s_{j}^{(k)\mp};q )_{\infty}
}
{
(q^{\frac{1+|m_{i}^{(k)}-m_{j}^{(k)}|}{2}} t s_{i}^{(k)\pm} s_{j}^{(k)\mp};q )_{\infty}
}
\right]
\nonumber\\
&\times 
\prod_{i=1}^{M} 
\frac{
(q^{\frac34+\frac{N-M}{4}} t^{1-(N-M)} s_{i}^{\pm} z_{1}^{\pm};q )_{\infty}
}
{
(q^{\frac14+\frac{N-N_{c}}{4}} t^{-1-(N-M)} s_{i}^{\pm} z_{1}^{\pm};q )_{\infty}
}
\cdot 
\prod_{i=1}^{M}\prod_{j=1}^{M} 
\frac{
( q^{\frac34+\frac{|m_{j}^{(1)}|}{2}} t s_{i}^{\pm} s_{j}^{(1)\mp};q )_{\infty}
}
{
( q^{\frac14+\frac{|m_{j}^{(1)}|}{2}} t^{-1} s_{i}^{\pm} s_{j}^{(1)\mp};q )_{\infty}
}
\nonumber\\
&\times 
\prod_{k=1}^{2M-2}
\prod_{i} \prod_{j}
\frac{
( q^{\frac34+\frac{|m_{i}^{(k)}-m_{j}^{(k+1)}|}{2}} t s_{i}^{(k)\pm} s_{j}^{(k+1)\mp};q )_{\infty}
}
{
( q^{\frac14+\frac{|m_{i}^{(k)}-m_{j}^{(k+1)}|}{2}} t^{-1} s_{i}^{(k)\pm} s_{j}^{(k+1)\mp};q )_{\infty}
}
\cdot 
\prod_{i=1}^{M} 
\frac{
( q^{\frac34+\frac{|m_{i}^{(M)}|}{2}} t s_{i}^{(M)\pm} z_{2}^{\pm};q )_{\infty}
}
{
( q^{\frac14+\frac{|m_{i}^{(M)}|}{2}} t^{-1} s_{i}^{(M)\pm} z_{2}^{\pm};q )_{\infty}
}
\nonumber\\
&\times 
q^{\frac{M}{4}\sum_{i=1}^{M}|m_{i}^{(1)}| + \sum_{k=1}^{2M-2} \sum_{i} \sum_{j} \frac{|m_{i}^{(k)}-m_{j}^{(k+1)}|}{4} 
+ \frac14\sum_{i=1}^{M}|m_{i}^{(M)}| - \sum_{k=1}^{2M-1}\sum_{i<j}\frac{|m_{i}^{(k)}-m_{j}^{(k)}|}{2} }
\nonumber\\
&\times 
t^{ M\sum_{i=1}^{M}|m_{i}^{(1)}| +\sum_{k=1}^{2M-2}\sum_{i}\sum_{j}|m_{i}^{(k)}-m_{j}^{(k+1)}| 
+\sum_{i=1}^{M}|m_{i}^{(M)}| -2\sum_{k=1}^{2M-1}\sum_{i<j}|m_{i}^{(k)}-m_{j}^{(k)}| }
\nonumber\\
&\times 
\prod_{k=1}^{N_{c}}
\left(\frac{x_{k}}{x_{k+1}} \right)^{\sum_{i=1}^{M} m_{i}^{(k)}} 
\cdot 
\sum_{k=M+1}^{2M-1}
\left(\frac{x_{k}}{x_{k+1}} \right)^{\sum_{i=1}^{2M-k}m_{i}^{(k)}}
\cdot 
\left( \frac{x_{2M}}{x_{1}} \right)^{\sum_{i=1}^{M}m_{i}^{(1)} +m_{1}^{(2M-1)}}.
\end{align}
The second line includes the half-index of Nahm$'$ boundary condition 
for 4d $U(N-M)$ gauge theory corresponding to the defect of D5$'$-brane  
and the half-index of Neumann b.c. $\mathcal{N}'$ for 4d $U(N)$ gauge theory corresponding to the NS5$'$-brane 
on which semi-infinite $N$ D3-branes terminate. 
The contributions from third to sixth line are the full-index of 3d twisted vector multiplets. 
The first factors in the seventh line describe the defect twisted hypermultiplet for $N=M$ 
and the local operators appearing from the broken $U(N)$ gauge theory for $N>M$. 
The remaining contributions are the 3d twisted hypers and monopole operator with conformal dimension
\begin{align}
\label{uN_M2M_monodim}
\Delta(m)&=
\frac{M}{2}\sum_{i=1}^{M}|m_{i}^{(1)}| + \sum_{k=1}^{2M-2} \sum_{i} \sum_{j} \frac{|m_{i}^{(k)}-m_{j}^{(k+1)}|}{2} 
+ \frac12\sum_{i=1}^{M}|m_{i}^{(M)}| 
- \sum_{k=1}^{2M-1}\sum_{i<j} |m_{i}^{(k)}-m_{j}^{(k)}|
\end{align}
where the first terms are the contributions from the bi-fundamental twisted hyper coupled to 4d $U(M)$ and 3d $U((M)$ 
gauge symmetries, 
the second and third are the contributions from the bi-fundamental twisted hyper in the 3d quiver gauge theory 
and the last terms are the contributions from the 3d twisted vector multiplets. 

As shown in Appendix \ref{app_bc}, 
we have checked that 
the indices (\ref{uN_M2Ma}) and (\ref{uN_M2Mb}) 
agree with each others for $(N,M)$ 
$=$ $(1,1)$, $(2,1)$, $(2,2)$ and $(3,2)$ 
up to certain orders of $q$.

\subsection{4d $U(N)|T[SU(M)]$}
\label{sec_un_tsun}
\begin{figure}
\begin{center}
\includegraphics[width=11cm]{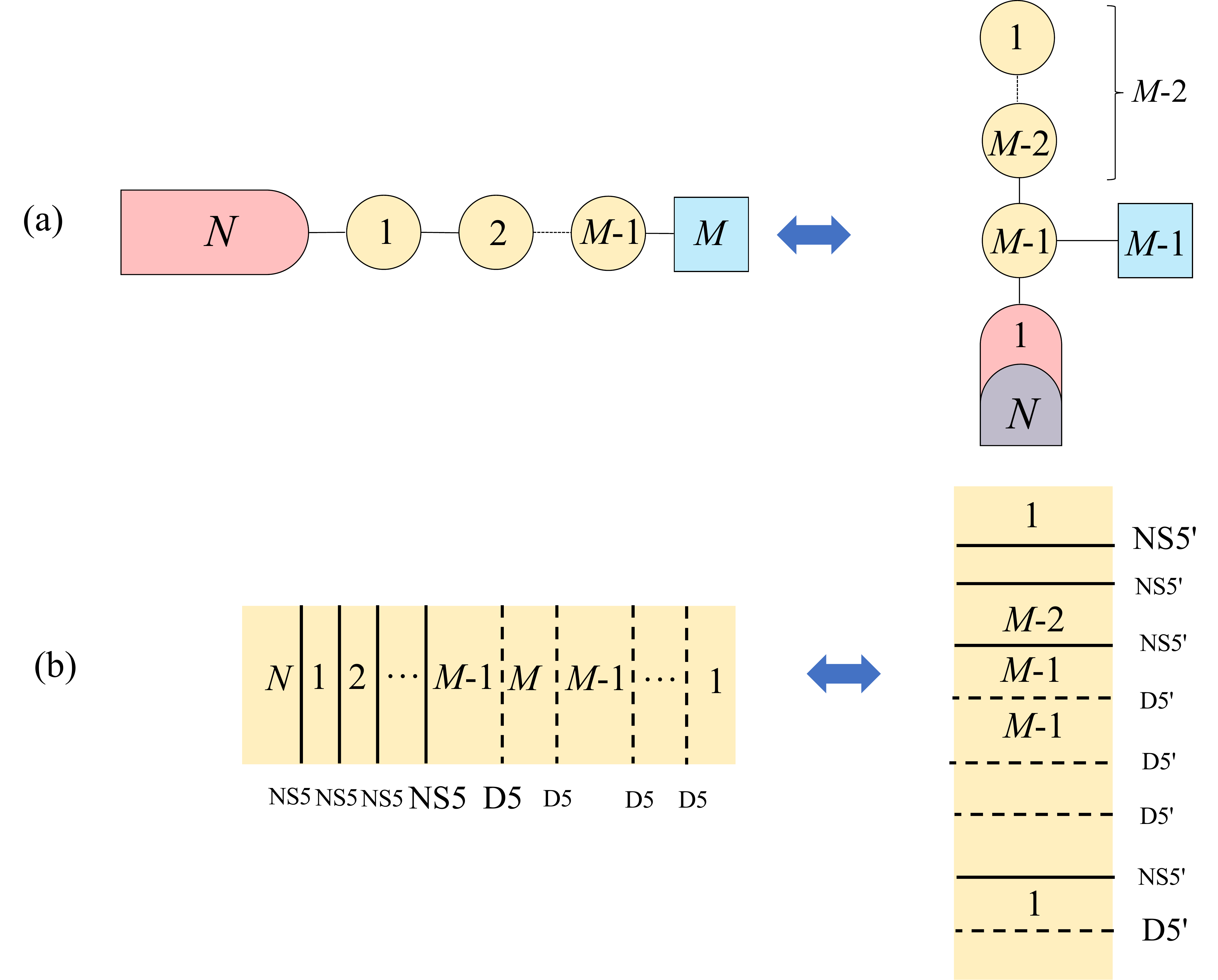}
\caption{
(a) The quiver diagrams of the enriched Neumann b.c. 4d $U(N)|T[SU(M)]$ and its mirror. 
(b) The corresponding brane constructions. 
}
\label{fig4duNtsuM}
\end{center}
\end{figure}

Let us study the enriched Neumann b.c. 4d $U(N)|T[SU(M)]$ for 4d $U(N)$ SYM theory. 
The corresponding quiver diagram and brane configuration are depicted in Figure \ref{fig4duNtsuM}. 
We have already examined the case with $M=2$ in section \ref{sec_4duN_3duM2M}. 
Although $T[SU(N)]$ is self-mirror, the enriched boundary condition 4d $U(N)|T[SU(M)]$ is not self-mirror. 
We further check the dualities for $M=3$ and propose the generalization.

The enriched Neumann boundary condition 
4d $U(N)|T[SU(M)]$ is constructed from the brane setup in Figure \ref{fig4duNtsuM}, 
which is the Neumann b.c. $\mathcal{N}$ for 4d $U(N)$ SYM theory coupled to $T[SU(M)]$ via the boundary 3d hypermultiplet 
transforming as $({\bf N}, -)$ $\oplus$ $(\overline{\bf N}, +)$ 
under the 4d $U(N)$ gauge and 3d $U(1)$ gauge symmetries.

We can write the half-index as 
\begin{align}
\label{uN_tsuMa}
&
\mathbb{II}^{\textrm{4d $U(N)|T[SU(M)]$}}_{\mathcal{N}}(t,x_{\alpha},z_{\alpha};q)
\nonumber\\
&=
\underbrace{
\frac{1}{N!}\frac{(q)_{\infty}^N}{(q^{\frac12}t^{-2};q)_{\infty}^{N}}
\oint \prod_{i=1}^{N}
\frac{ds_{i}}{2\pi is_{i}}
\prod_{i\neq j}
\frac{
\left( \frac{s_{i}}{s_{j}};q \right)_{\infty}
}
{
\left(q^{\frac12} t^{-2} \frac{s_{i}}{s_{j}};q \right)_{\infty}
}
}_{\mathbb{II}_{\mathcal{N}}^{\textrm{4d $U(N)$}}}
\nonumber\\
&\times 
\prod_{k=1}^{M-1}
\left[
\frac{1}{k!}
\frac{(q^{\frac12}t^2;q)_{\infty}^k}{(q^{\frac12}t^{-2};q)_{\infty}^k}
\sum_{m_{1}^{(k)},\cdots, m_{k}^{(k)}\in \mathbb{Z}} 
\oint \prod_{i=1}^{k} 
\frac{ds_{i}^{(k)}}{2\pi is_{i}^{(k)}}
\right. 
\nonumber\\
&\times 
\left. 
\prod_{i<j} 
\frac{
\left( 1-q^{\frac{|m_{i}^{(k)}-m_{j}^{(k)}|}{2}} s_{i}^{(k)\pm} s_{j}^{(k)\mp} \right)
\left( q^{\frac{1+|m_{i}^{(k)}-m_{j}^{(k)}|}{2}} t^2 s_{i}^{(k)\pm} s_{j}^{(k)\mp};q \right)_{\infty}
}
{
\left( q^{\frac{1+|m_{i}^{(k)}-m_{j}^{(k)}|}{2}} t^{-2} s_{i}^{(k)\pm} s_{j}^{(k)\mp};q \right)_{\infty}
}
\right]
\nonumber\\
&\times 
\prod_{i=1}^{N}
\frac{
(q^{\frac34+\frac{|m_{1}^{(1)}|}{2}} t^{-1} s_{i}^{\pm} s_{j}^{\mp};q)_{\infty}
}
{
(q^{\frac14+\frac{|m_{1}^{(1)}|}{2}} t s_{i}^{\pm} s_{j}^{\mp};q)_{\infty}
}
\cdot 
\prod_{k=1}^{M-2} \prod_{i=1}^{k}\prod_{j=1}^{k+1}
\frac{
(q^{\frac34+\frac{|m_{i}^{(k)}-m_{j}^{(k+1)}|}{2}} t^{-1} s_{i}^{(k)\pm} s_{j}^{(k)\mp};q)_{\infty}
}
{
(q^{\frac14+\frac{|m_{i}^{(k)}-m_{j}^{(k+1)}|}{2}} t s_{i}^{(k)\pm} s_{j}^{(k)\mp};q)_{\infty}
}
\nonumber\\
&\times 
\prod_{i=1}^{M-1}\prod_{\alpha=1}^{M} 
\frac{
(q^{\frac34+\frac{|m_{i}^{(M-1)}|}{2}} t^{-1} s_{i}^{(M-1)\pm} x_{\alpha}^{\pm};q)_{\infty}
}
{
(q^{\frac14+\frac{|m_{i}^{(M-1)}|}{2}} t s_{i}^{(M-1)\pm} x_{\alpha}^{\pm};q)_{\infty}
}
\nonumber\\
&\times 
q^{ \frac{N}{4}|m_{1}^{(1)}|
+\frac14\sum_{k=1}^{M-2}\sum_{i=1}^{k}\sum_{j=1}^{k+1}|m_{i}^{(k)}-m_{j}^{(k+1)}|
+\frac{M}{4}\sum_{i=1}^{M-1}|m_{i}^{(M-1)}|
-\sum_{k=1}^{M-1}\sum_{i<j}\frac{|m_{i}^{(k)}-m_{j}^{(k)}|}{2} }
\nonumber\\
&\times 
t^{-N|m_{1}^{(1)}| -\sum_{k=1}^{M-2}\sum_{i=1}^{k}\sum_{j=1}^{k+1}|m_{i}^{(k)}-m_{j}^{(k+1)}|
-M\sum_{i=1}^{M-1}|m_{i}^{(M-1)}|
+2\sum_{k=1}^{M-1}|m_{i}^{(k)}-m_{j}^{(k)}| }
\nonumber\\
&\times 
\prod_{k=1}^{M-1}
\left( \frac{z_{k}}{z_{k+1}} \right)^{\sum_{i=1}^{k}m_{i}^{(k)}}
\cdot 
\left( 
\frac{z_{M}}{z_{1}}
\right)^{m_{1}^{(1)}+\sum_{i=1}^{M-1}m_{i}^{(M-1)}}.
\end{align}
The terms appearing from the second to fourth line 
describe the half-index of Neumann b.c. $\mathcal{N}$ for 
4d $\mathcal{N}=4$ $U(N)$ SYM theory 
and the full-index for 3d $U(1)\times U(2)$ $\times \cdots \times$ $U(M-1)$ vector multiplet. 
The associated magnetic fluxes are labeled by $\frac{M(M-1)}{2}$ integers $\{ m_{i}^{(k)} \}_{i=1,\cdots, k}$ 
where $k=1,\cdots, M-1$. 
The contributions in the fifth and sixth lines describe 3d $\mathcal{N}=4$ hypermultiplets. 
The other terms count the bare monopole with the R-charge 
\begin{align}
\label{uN_tsuMa_monodim}
\Delta(m)&=
\frac{N}{2}|m_{1}^{(1)}|
+\frac12\sum_{k=1}^{M-2}\sum_{i=1}^{k}\sum_{j=1}^{k+1}|m_{i}^{(k)}-m_{j}^{(k+1)}|
+\frac{M}{2}\sum_{i=1}^{M-1}|m_{i}^{(M-1)}|
-\sum_{k=1}^{M-1}\sum_{i<j} |m_{i}^{(k)}-m_{j}^{(k)}|. 
\end{align}
This formula is generalized from (\ref{tsuN_monodim}) by the additional contributions appearing in the first terms. 
They come from the boundary hypermultiplet 
transforming as $({\bf N}, -)$ $\oplus$ $(\overline{\bf N}, +)$ 
under the 4d $U(N)$ gauge and 3d $U(1)$ gauge symmetries.

The dual quiver diagram and the corresponding brane setup are illustrated in Figure \ref{fig4duNtsuM}. 
It is identifies with the boundary condition for 4d $U(N)$ gauge theory 
including the Nahm$'$ pole of rank $(N-1)$ that breaks the 4d gauge symmetry down to $U(1)$. 
While for $N=1$ one finds a defect twisted hypermultiplet arising from D3-D5$'$ string, 
for $N>1$, the half-index receives contributions from broken gauge theory as discussed in \cite{Gaiotto:2019jvo}. 

In addition, the surviving $U(1)$ gauge theory should satisfy the enriched Neumann b.c. $\mathcal{N}'$ 
corresponding to NS5$'$-brane with a coupling to 3d twisted quiver gauge theory 
$\widetilde{\begin{smallmatrix}
(M-1)&-&(M-2)&-&\cdots&-&(1)\\
|&&&&&&\\
[M-1]&&&&&&\\
\end{smallmatrix}}$ through the 3d boundary twisted hypermultiplet. 

We find the half-index 
\begin{align}
\label{uN_tsuMb}
&
\mathbb{II}^{\textrm{4d $U(N)\rightarrow U(1)|
\widetilde{\begin{smallmatrix}
(M-1)&-&(M-2)&-&\cdots&-&(1)\\
|&&&&&&\\
[M-1]&&&&&&\\
\end{smallmatrix}}
$}}_{\mathcal{N}'}(t,x_{\alpha},z_{\alpha};q)
\nonumber\\
&
=
\underbrace{
\frac{(q)_{\infty}}{(q^{\frac12}t^{2};q)_{\infty}}
\oint \frac{ds_{1}}{2\pi is_{1}}
}_{\mathbb{II}_{\mathcal{N}'}^{\textrm{4d $U(1)$}}}
\cdot 
\underbrace{
\prod_{l=1}^{N-1}
\frac{(q^{\frac{l+1}{2}} t^{-2(l-1)};q)_{\infty}}
{(q^{\frac{l}{2}} t^{-2l};q)_{\infty}}
}_{\mathbb{II}^{\textrm{4d $U(N-1)$}}_{\textrm{Nahm}'}}
\nonumber\\
&\times 
\prod_{k=1}^{M-1}
\left[
\frac{1}{k!}
\frac{(q^{\frac12}t^{-2};q)_{\infty}^k}{(q^{\frac12}t^{2};q)_{\infty}^k}
\sum_{m_{1}^{(k)},\cdots, m_{k}^{(k)}\in \mathbb{Z}} 
\oint \prod_{i=1}^{k} 
\frac{ds_{i}^{(k)}}{2\pi is_{i}^{(k)}}
\right.
\nonumber\\
&\times 
\left. 
\prod_{i<j} 
\frac{
\left( 1-q^{\frac{|m_{i}^{(k)}-m_{j}^{(k)}|}{2}} s_{i}^{(k)\pm} s_{j}^{(k)\mp} \right)
\left( q^{\frac{1+|m_{i}^{(k)}-m_{j}^{(k)}|}{2}} t^{-2} s_{i}^{(k)\pm} s_{j}^{(k)\mp};q \right)_{\infty}
}
{
\left( q^{\frac{1+|m_{i}^{(k)}-m_{j}^{(k)}|}{2}} t^{2} s_{i}^{(k)\pm} s_{j}^{(k)\mp};q \right)_{\infty}
}
\right]
\nonumber\\
&\times 
\frac{
(q^{\frac34+\frac{N-1}{4}} t^{1-(N-1)}s_{1}^{\pm} z_{M}^{\pm};q)_{\infty}
}
{
(q^{\frac14+\frac{N-1}{4}} t^{-1-(N-1)}s_{1}^{\pm} z_{M}^{\pm};q)_{\infty}
}
\cdot 
\prod_{i=1}^{M-1}
\frac{
(q^{\frac34+\frac{|m_{i}^{(M-1)}|}{2}} t s_{1}^{\pm} s_{i}^{(M-1)\mp};q)_{\infty}
}
{
(q^{\frac14+\frac{|m_{i}^{(M-1)}|}{2}} t^{-1} s_{1}^{\pm} s_{i}^{(M-1)\mp};q)_{\infty}
}
\nonumber\\
&\times 
\prod_{i=1}^{M-1} \prod_{\alpha=1}^{M-1} 
\frac{
(q^{\frac34+\frac{|m_{i}^{(M-1)}|}{2}} t s_{i}^{(M-1)\pm}z_{\alpha}^{\pm};q )_{\infty}
}
{
(q^{\frac14+\frac{|m_{i}^{(M-1)}|}{2}} t^{-1} s_{i}^{(M-1)\pm}z_{\alpha}^{\pm};q )_{\infty}
}
\cdot 
\prod_{k=1}^{M-2}
\prod_{i=1}^{k}\prod_{j=1}^{k+1}
\frac{
(q^{\frac34+\frac{| m_{i}^{(k)}-m_{j}^{(k+1)} |}{2}} t s_{i}^{(k)\pm} s_{j}^{(k+1)\mp};q )_{\infty}
}
{
(q^{\frac14+\frac{| m_{i}^{(k)}-m_{j}^{(k+1)} |}{2}} t^{-1} s_{i}^{(k)\pm} s_{j}^{(k+1)\mp};q )_{\infty}
}
\nonumber\\
&\times 
q^{\frac{M}{4}\sum_{i=1}^{M}|m_{i}^{(M-1)}|+\sum_{k=1}^{M-2}\sum_{i=1}^{k}\sum_{j=1}^{k+1}
\frac{|m_{i}^{(k)}-m_{j}^{(k+1)}|}{4}
-\sum_{k=1}^{M-1}\sum_{i<j}\frac{|m_{i}^{(k)}-m_{j}^{(k)}|}{2} }
\nonumber\\
&\times 
t^{M\sum_{i=1}^{M-1}|m_{i}^{(M-1)}|
+\sum_{k=1}^{M-2}\sum_{i=1}^{k}\sum_{j=1}^{k+1}|m_{i}^{(k)}-m_{j}^{(k+1)}|
-2\sum_{k=1}^{M-1}\sum_{i<j}|m_{i}^{(k)}-m_{j}^{(k)}|
 }
 \nonumber\\
 &\times 
 \prod_{k=1}^{M-1}
 \left( \frac{x_{k}}{x_{k+1}} \right)^{\sum_{i=1}^{k}m_{i}^{(k)}}
 \cdot 
 \left( \frac{x_{M}}{x_{1}} \right)^{m_{1}^{(1)}+\sum_{i=1}^{M-1}m_{i}^{(M-1)}}.
\end{align}
The contributions in the second line 
are the half-index of Nahm$'$ b.c. for 4d $U(N-1)$ gauge theory 
and the half-index of Neumann b.c. for 4d $U(1)$ gauge theory. 
The next two lines are the full-index of 
3d $U(M-1)\times U(M-2)$ $\times \cdots \times$ $U(1)$ twisted vector multiplet. 
The first terms in the fifth line correspond to the defect hypermultiplet for $N=1$ 
or the extra fields appearing from the broken gauge theory \cite{Gaiotto:2019jvo}. 
The remaining terms count 3d $\mathcal{N}=4$ twisted hypers and monopole operator of dimension 
\begin{align}
\label{uN_tsuMb_monodim}
\Delta(m)&=
\frac{M}{2}\sum_{i=1}^{M}|m_{i}^{(M-1)}|
+\sum_{k=1}^{M-2}\sum_{i=1}^{k}\sum_{j=1}^{k+1}
\frac{|m_{i}^{(k)}-m_{j}^{(k+1)}|}{2}
-\sum_{k=1}^{M-1}\sum_{i<j} |m_{i}^{(k)}-m_{j}^{(k)}|.
\end{align}

We conjecture that 
the half-indices (\ref{uN_tsuMa}) and (\ref{uN_tsuMb}) gives the same result. 
In fact, we have found that 
for $(N,M)$ $=$ $(1,3)$ and $(2,3)$ 
they agree with each other up to certain orders of $q$ 
(see Appendix \ref{app_bc}).

\section{Dualities of interfaces}
\label{sec_3dm4dS2}

In this section we would like to study the dualities of interfaces 
for a pair of 4d $\mathcal{N}=4$ gauge theories including additional 3d $\mathcal{N}=4$ gauge theories.

\subsection{4d $U(N)|$3d $U(N)^{k-1}|$4d $U(N)$}
\label{sec_uNuN_Nk}
\begin{figure}
\begin{center}
\includegraphics[width=12cm]{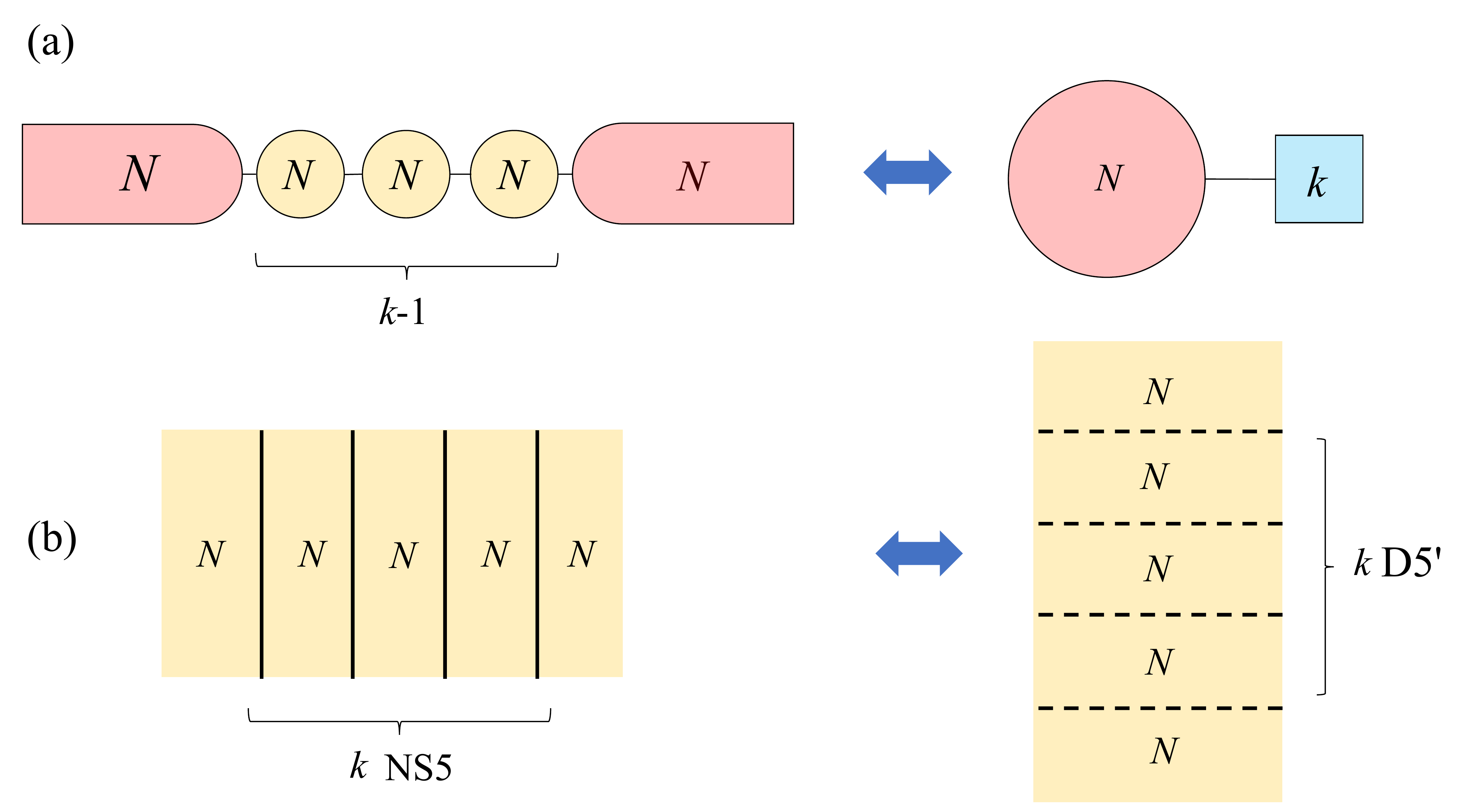}
\caption{
(a) The quiver diagrams of the 4d $U(N)|$3d $U(N)^{k-1}|$4d $U(N)$ interface and its mirror. 
(b) The corresponding brane configurations.  
}
\label{fig4duNuN3duN}
\end{center}
\end{figure}

We study the interface for a pair of two 4d $\mathcal{N}=4$ $U(N)$ gauge theories 
which involves 3d $\mathcal{N}=4$ $U(N)^{k-1}$ quiver gauge theory. 
We denote this by 4d $U(N)|$3d $U(N)^{k-1}|$4d $U(N)$. 
The corresponding quiver diagram and brane configuration are given in Figure \ref{fig4duNuN3duN}. 
It is realized by $N$ D3-branes intersecting with $k$ NS5-branes.  
The dual configuration is realized by $N$ D3-branes intersecting with $k$ D5$'$-branes. 
According to the presence of D5$'$ interface, 
the 4d $U(N)^{k-1}$ gauge symmetry is broken down to a diagonal $U(N)$ 
so that four-dimensional $U(N)$ gauge fields couple to 3d $\mathcal{N}=4$ twisted hypermultiplets. 
The analysis for $k=1$ where the interface has no 3d gauge theory is already given in \cite{Gaiotto:2019jvo}. 
We present more general results and check that the half-indices for the dual interfaces agree with each other.

The interface 4d $U(N)|\textrm{3d $U(N)^{k-1}$}|$4d $U(N)$ has 
a pair of 4d $U(N)$ SYM theories obeying Neumann b.c. $\mathcal{N}$ corresponding to the NS5-branes. 
Both of them are coupled to 3d $U(N)^{k-1}$ vector multiplet at the interface 
through the 3d bi-fundamental hypermultiplets. 
Let us label the magnetic fluxes by $k-1$ sets of $N$-tuple of integers,  
i.e. $N(k-1)$ integers $\{m_{i}^{(l)}\}_{i=1,\cdots,N}$ 
with $l=1,\cdots, k-1$. 

The half-index takes the form
\begin{align}
\label{unun_muna}
&
\mathbb{II}^{\textrm{4d $U(N)|(N)^k|$4d $U(N)$}}_{\mathcal{N}\mathcal{N}}(t,z_{\alpha};q)
\nonumber\\
&=
\underbrace{
\frac{1}{N!}
\frac{(q)_{\infty}^{N}}
{(q^{\frac12}t^{-2};q)_{\infty}^{N}}
\oint \prod_{i=1}^{N} \frac{ds_{i}}{2\pi is_{i}} 
\prod_{i\neq j} 
\frac{
\left(\frac{s_{i}}{s_{j}};q \right)_{\infty}
}
{
\left(q^{\frac12} t^{-2} \frac{s_{i}}{s_{j}};q \right)_{\infty}
}
}_{\mathbb{II}_{\mathcal{N}}^{\textrm{4d $U(1)$}}}
\nonumber\\
&\times 
\left[
\frac{1}{N!} \frac{(q^{\frac12}t^2;q)_{\infty}^{N}}
{(q^{\frac12}t^{-2};q)_{\infty}^{N}}
\right]^{k-1}
\prod_{l=1}^{k-1} 
\sum_{m_{1}^{(l)},\cdots, m_{N}^{(l)}}
\oint \prod_{i=1}^{N} \frac{ds_{i}^{(l)}}{2\pi is_{i}^{(l)}}
\nonumber\\
&\times 
\frac{
\left(1-q^{\frac{|m_{i}^{(l)}-m_{j}^{(l)}|}{2}} s_{i}^{(l)\pm} s_{j}^{(l)\mp} \right)
\left( q^{\frac{1+|m_{i}^{(l)}-m_{j}^{(l)} |}{2}} t^2 s_{i}^{(l)\pm} s_{j}^{(l)\mp};q \right)_{\infty}
}
{
\left( q^{\frac{1+|m_{i}^{(l)}-m_{j}^{(l)} |}{2}} t^{-2} s_{i}^{(l)\pm} s_{j}^{(l)\mp};q \right)_{\infty}
}
\nonumber\\
&\times 
\underbrace{
\frac{1}{N!}
\frac{(q)_{\infty}^{N}}
{(q^{\frac12}t^{-2};q)_{\infty}^{N}}
\oint \prod_{i=N+1}^{2N} \frac{ds_{i}}{2\pi is_{i}} 
\prod_{i\neq j} 
\frac{
\left(\frac{s_{i}}{s_{j}};q \right)_{\infty}
}
{
\left(q^{\frac12} t^{-2} \frac{s_{i}}{s_{j}};q \right)_{\infty}
}
}_{\mathbb{II}_{\mathcal{N}}^{\textrm{4d $U(1)$}}}
\nonumber\\
&\times 
\prod_{i=1}^{N}\prod_{j=1}^{N} 
\frac{
(q^{\frac34+\frac{|m_{i}^{(1)}|}{2}} t^{-1} s_{i}^{\pm} s_{j}^{(1)\mp};q)_{\infty}
}
{
(q^{\frac14+\frac{|m_{i}^{(1)}|}{2}} t s_{i}^{\pm} s_{j}^{(1)\mp};q)_{\infty}
}
\cdot 
\prod_{l=1}^{k-2} \prod_{i=1}^{N}\prod_{j=1}^{N} 
\frac{
(q^{\frac34+\frac{|m_{i}^{(l)}-m_{j}^{(l+1)}|}{2}} t^{-1} s_{i}^{(l)\pm} s_{j}^{(l+1)\mp};q)_{\infty}
}
{
(q^{\frac14+\frac{|m_{i}^{(l)}-m_{j}^{(l+1)}|}{2}} t s_{i}^{(l)\pm} s_{j}^{(l+1)\mp};q)_{\infty}
}
\nonumber\\
&\times 
\prod_{i=1}^{N} \prod_{j=N+1}^{2N} 
\frac{
(q^{\frac34+\frac{|m_{i}^{(k-1)}|}{2}} t^{-1} s_{i}^{(k-1)\pm} s_{j}^{\mp};q)_{\infty}
}
{
(q^{\frac14+\frac{|m_{i}^{(k-1)}|}{2}} t s_{i}^{(k-1)\pm} s_{j}^{\mp};q)_{\infty}
}
\nonumber\\
&\times 
q^{ \sum_{i=1}^{N}\frac{ N|m_{i}^{(1)}|}{4} +\sum_{l=1}^{k-2} \sum_{i=1}^{N}\sum_{j=1}^{N}\frac{|m_{i}^{(l)}-m_{j}^{(l+1)}|}{4}
+\sum_{i=1}^{N}\frac{N|m_{i}^{(k-1)}|}{4}-\sum_{l=1}^{k-1}\sum_{i<j}\frac{|m_{i}^{(l)}-m_{j}^{(l)}|}{2} }
\nonumber\\
&\times 
t^{-\sum_{i=1}^{N}N|m_{i}^{(1)}| -\sum_{l=1}^{k-2}\sum_{i=1}^{N}\sum_{j=1}^{N}|m_{i}^{(l)}-m_{j}^{(l+1)}| 
-\sum_{i=1}^{N}N|m_{i}^{(k-1)}| + 2\sum_{l=1}^{k-1}\sum_{i<j} |m_{i}^{(l)}-m_{j}^{(l)}| }
\nonumber\\
&\times 
\prod_{\alpha=1}^{k-1}
\left(\frac{z_{\alpha}}{z_{\alpha+1}} \right)^{\sum_{i=1}^{N}m_{i}^{(\alpha)}}
\cdot 
\left( 
\frac{z_{M}}{z_{1}}
\right)^{\sum_{i=1}^{N}m_{i}^{(1)}
+\sum_{j=1}^{N}m_{j}^{(k-1)}
}. 
\end{align}
The terms appearing from the second to fifth lines are 
the two half-indices of Neumann b.c. $\mathcal{N}$ for 4d $\mathcal{N}=4$ $U(N)$ SYM theory 
and the full-index of 3d $\mathcal{N}=4$ $U(N)^{k-1}$ vector multiplet. 
The terms in the sixth and seventh lines are the contributions from 
3d $\mathcal{N}=4$ hypermultiplets. 
The remaining terms describe the monopole contributions 
whose canonical R-charge is 
\begin{align}
\Delta(m)&=
\sum_{i=1}^{N}\frac{ N|m_{i}^{(1)}|}{2} +
\sum_{l=1}^{k-2} \sum_{i=1}^{N}\sum_{j=1}^{N}\frac{|m_{i}^{(l)}-m_{j}^{(l+1)}|}{2}
+\sum_{i=1}^{N}\frac{N|m_{i}^{(k-1)}|}{2}-\sum_{l=1}^{k-1}\sum_{i<j} |m_{i}^{(l)}-m_{j}^{(l)}|
\end{align}
where the first and third terms are contributed from bi-fundamental hypers 
coupled to 4d and 3d gauge theories, 
the second terms are the contributions from 3d bi-fundamental hypers between gauge nodes in 
3d quiver gauge theory 
and the last terms are contributed from 3d $U(N)^{k-1}$ vector multiplet. 

Under S-duality 
we find the dual interface which is identified with 
a 4d $\mathcal{N}=4$ $U(N)$ SYM theory with $k$ defects corresponding to $k$ D5$'$-branes, 
which couple the 4d $U(N)$ gauge theory to $k$ 3d $\mathcal{N}=4$ fundamental twisted hypermultiplets. 

The half-index for the dual interface should be computed as
\begin{align}
\label{unun_munb}
&
\mathbb{II}^{\textrm{4d $U(N)+k$ thypers}}_{\mathcal{D}',\cdots,\mathcal{D}'}(t,z_{\alpha};q)
\nonumber\\
&=
\underbrace{
\frac{1}{N!} \frac{(q)_{\infty}^{2N}}{(q^{\frac12}t^2;q)_{\infty}^N (q^{\frac12}t^{-2};q)_{\infty}^N}
\oint \prod_{i=1}^{N} 
\frac{ds_{i}}{2\pi is_{i}} 
\prod_{i\neq j} 
\frac{
\left( \frac{s_{i}}{s_{j}};q \right)_{\infty}
\left( q \frac{s_{i}}{s_{j}};q \right)_{\infty}
}
{
\left(q^{\frac12} t^2 \frac{s_{i}}{s_{j}};q \right)_{\infty}
\left(q^{\frac12} t^{-2} \frac{s_{i}}{s_{j}};q \right)_{\infty}
}
}_{\mathbb{I}^{\textrm{4d $U(N)$}}}
\nonumber\\
&\times 
\prod_{i=1}^{N} \prod_{\alpha=1}^{k}
\underbrace{
\frac{
(q^{\frac34} t s_{i}^{\pm} z_{\alpha}^{\pm};q)_{\infty}
}
{
(q^{\frac14} t^{-1} s_{i}^{\pm} z_{\alpha}^{\pm};q)_{\infty}
}
}_{\mathbb{I}^{\textrm{3d tHM}}(s_{i}z_{\alpha})}.
\end{align}
We expect that 
the half-index (\ref{unun_muna}) is equal to (\ref{unun_munb}). 
In fact, we have confirmed that 
they agree up to certain orders of $q$ for 
$(N,k)$ $=$ $(1,2)$, $(1,3)$, $(2,2)$ and $(2,3)$ 
(see Appendix \ref{app_interface}).

\subsection{4d $U(L)|$3d $U(M)|$4d $U(N)$}
\label{sec_uLuN_uM}
\begin{figure}
\begin{center}
\includegraphics[width=10cm]{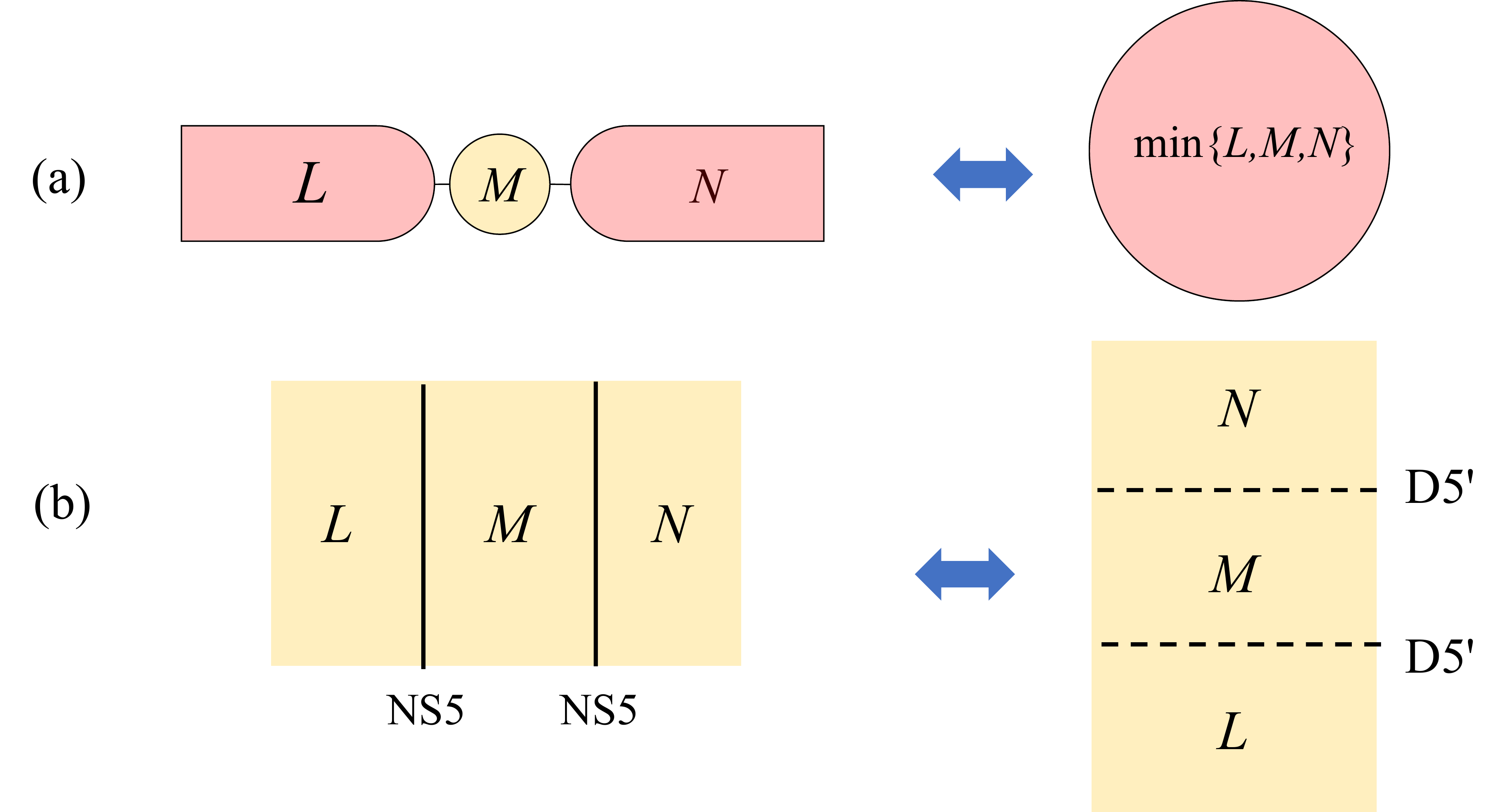}
\caption{
(a) The quiver diagrams of the 4d $U(L)|$3d $U(M)|$ 4d $U(N)$ interface and its mirror. 
(b) The corresponding brane configurations.  
}
\label{fig4duLuN3duM}
\end{center}
\end{figure}

We consider the interface 4d $U(L)|$3d $U(M)|$4d $U(N)$, 
which involves a pair of 4d $\mathcal{N}=4$ $U(L)$ and $U(N)$ gauge theories 
with Neumann b.c. $\mathcal{N}$ and 3d $U(M)$ vector multiplet 
where $L$, $M$ and $N$ are not all equal. 
The corresponding quiver diagram and brane setup are depicted in Figure \ref{fig4duLuN3duM}. 
According to unequal numbers of D3-branes, 
the dual interface can involve singular boundary conditions specified by Nahm poles 
corresponding to the D5$'$-branes. 
We compute the half-indices for a pair of dual interfaces and check the matching. 
We find that the half-indices have additional contributions when 
one of 4d gauge symmetries is smaller than 3d gauge symmetry.

The interface has a pair of 4d $\mathcal{N}=4$ $U(L)$ and $U(N)$ gauge theories 
which obey Neumann b.c. $\mathcal{N}$ at the 3d boundary. 
Each of them couples to 3d $\mathcal{N}=4$ $U(M)$ vector multiplet 
through the 3d boundary bi-fundamental hypermultiplets. 
We label magnetic fluxes for 3d $U(M)$ gauge theory 
by $M$ integers $m_{1},\cdots, m_{M}$. 
The interface includes bare monopole whose 
canonical R-charge is given by 
\begin{align}
\label{uLuN_uMa_monodim}
\Delta(m)&=
\sum_{i=1}^{M}\frac{(L+N)|m_{i}|}{2}-\sum_{i<j} |m_{i}-m_{j}|. 
\end{align}
To ensure that all monopole operators are above the unitarity bound, 
we assume that $L+N\ge 2M-1$. 
In fact, for $L+N< 2M-1$ the half-indices may not be convergent. 
Without loss of generality we further assume that $L\le N$. 

The half-index takes the form 
\begin{align}
\label{uLuN_uMa}
&
\mathbb{II}^{\textrm{4d $U(L)|(M)|$4d $U(N)$}}_{\mathcal{N}\mathcal{N}}(t,z_{\alpha};q)
\nonumber\\
&=
\underbrace{
\frac{1}{L!}
\frac{(q)_{\infty}^L}{(q^{\frac12}t^{-2};q)_{\infty}^L}
\oint \prod_{i=1}^{L} \frac{ds_{i}}{2\pi is_{i}}
\prod_{i\neq j}
\frac{
\left( \frac{s_{i}}{s_{j}};q \right)_{\infty}
}
{
\left( q^{\frac12} t^{-2} \frac{s_{i}}{s_{j}};q \right)_{\infty}
}
}_{\mathbb{II}^{\textrm{4d $U(L)$}}_{\mathcal{N}}}
\nonumber\\
&\times 
\frac{1}{M!} 
\frac{(q^{\frac12}t^{2};q)_{\infty}^M}{(q^{\frac12}t^{-2};q)_{\infty}^M}
\sum_{m_{1},\cdots, m_{M}\in \mathbb{Z}} 
\oint \prod_{i=L+1}^{L+M}
\frac{ds_{i}}{2\pi is_{i}}
\prod_{i<j}
\frac{
\left(1-q^{\frac{|m_{i-L}-m_{j-L}|}{2}} s_{i}^{\pm} s_{j}^{\mp} \right)
\left( q^{\frac{1+|m_{i-L}-m_{j-L}|}{2}} t^2 s_{i}^{\pm} s_{j}^{\mp};q \right)_{\infty}
}
{
\left( q^{\frac{1+|m_{i-L}-m_{j-L}|}{2}} t^2 s_{i}^{\pm} s_{j}^{\mp};q \right)_{\infty}
}
\nonumber\\
&\times 
\underbrace{
\frac{1}{N!}
\frac{(q)_{\infty}^N}{(q^{\frac12}t^{-2};q)_{\infty}^N}
\oint \prod_{i=L+M+1}^{L+M+N} \frac{ds_{i}}{2\pi is_{i}}
\prod_{i\neq j}
\frac{
\left( \frac{s_{i}}{s_{j}};q \right)_{\infty}
}
{
\left( q^{\frac12} t^{-2} \frac{s_{i}}{s_{j}};q \right)_{\infty}
}
}_{\mathbb{II}^{\textrm{4d $U(N)$}}_{\mathcal{N}}}
\nonumber\\
&\times 
\prod_{i=1}^{L}\prod_{j=L+1}^{L+M}
\frac{
(q^{\frac34+\frac{|m_{j-L}|}{2}} t^{-1} s_{i}^{\pm} s_{j}^{\mp};q )_{\infty} 
}
{
(q^{\frac14+\frac{|m_{j-L}|}{2}} t s_{i}^{\pm} s_{j}^{\mp};q )_{\infty} 
}
\cdot 
\prod_{i=L+1}^{L+M} 
\prod_{j=L+M+1}^{L+M+N}
\frac{
(q^{\frac34+\frac{|m_{i-L}|}{2}} t^{-1} s_{i}^{\pm} s_{j}^{\mp};q )_{\infty} 
}
{
(q^{\frac14+\frac{|m_{i-L}|}{2}} t s_{i}^{\pm} s_{j}^{\mp};q )_{\infty} 
}
\nonumber\\
&\times 
q^{\sum_{i=1}^{M}\frac{(L+N)|m_{i}|}{4}-\sum_{i<j}\frac{|m_{i}-m_{j}|}{2}}
\cdot 
t^{-\sum_{i=1}^{M}(L+N)|m_{i}|+2\sum_{i<j}|m_{i}-m_{j}|}
\cdot 
\left( 
\frac{z_{1}}{z_{2}}
\right)^{\sum_{i=1}^{M}m_{i}}. 
\end{align}
The terms from the second to fourth line 
describe the half-indices of Neumann b.c. $\mathcal{N}$ for 4d $\mathcal{N}=4$ $U(L)$ and $U(N)$ gauge theories 
and full-index for 3d $\mathcal{N}=4$ $U(M)$ vector multiplet. 
The first and second terms in the fifth line count 3d $\mathcal{N}=4$ bi-fundamental hypermultiplets 
which couple to 4d $U(L)$ and 3d $U(M)$ gauge theories and 
those which couple to 4d $U(N)$ and 3d $U(M)$ gauge theories respectively. 
The remaining terms count bare monopole operator of dimension (\ref{uLuN_uMa_monodim}).

Now consider the dual interface. 
The corresponding quiver and brane configuration are illustrated in Figure \ref{fig4duLuN3duM}. 
When $M<L$, the 4d $U(L)\times U(M)\times U(N)$ gauge symmetry is broken down to $U(M)$ 
and the whole 4d $U(M)$ gauge theory remains. 
For the two defects corresponding to two D5$'$-branes, 
we would have two Nahm$'$ boundary conditions of rank $L-M$ and rank $N-M$. 
In addition, there are extra contributions to the half-index 
which come from the broken $U(L)$ and $U(N)$ gauge theories.

The half-index (\ref{uLuN_uMa}) for the dual interface with $M<L$ will be computed as
\begin{align}
\label{uLuN_uMb1}
&
\mathbb{II}^{\textrm{4d $U(L)|[M]|$4d $U(N)$}}_{\mathcal{D}'}
(t,z_{\alpha};q)
\nonumber\\
&=
\underbrace{
\frac{1}{M!}
\frac{(q)_{\infty}^{2M}}{(q^{\frac12}t^{2};q)_{\infty}^M (q^{\frac12}t^{-2};q)_{\infty}^M}
\oint \prod_{i=1}^{M}
\frac{ds_{i}}{2\pi is_{i}}
\prod_{i\neq j}
\frac{
\left(
\frac{s_{i}}{s_{j}};q \right)_{\infty}
\left(q \frac{s_{i}}{s_{j}};q \right)_{\infty}
}
{
\left(q^{\frac12} t^{2} \frac{s_{i}}{s_{j}};q \right)_{\infty}
\left(q^{\frac12} t^{-2} \frac{s_{i}}{s_{j}};q \right)_{\infty}
}
}_{\mathbb{I}^{\textrm{4d $U(M)$}}}
\nonumber\\
&\times 
\underbrace{
\prod_{k=1}^{L-M}
\frac{
\left(q^{\frac{k+1}{2}} t^{-2(k-1)};q \right)_{\infty}
}
{
\left(q^{\frac{k}{2}} t^{-2k};q \right)_{\infty}
}
}_{\mathbb{II}^{\textrm{4d $U(L-M)$}}_{\textrm{Nahm}'}}
\cdot 
\underbrace{
\prod_{l=1}^{N-M}
\frac{
\left(q^{\frac{l+1}{2}} t^{-2(l-1)};q \right)_{\infty}
}
{
\left(q^{\frac{l}{2}} t^{-2l};q \right)_{\infty}
}
}_{\mathbb{II}^{\textrm{4d $U(N-M)$}}_{\textrm{Nahm}'}}
\nonumber\\
&\times 
\prod_{i=1}^{M}
\frac{
\left(q^{\frac34+\frac{L-M}{4}} t^{1-(L-M)} s_{i}^{\pm} z_{1}^{\pm};q \right)_{\infty}
}
{
\left(q^{\frac14+\frac{L-M}{4}} t^{-1-(L-M)} s_{i}^{\pm} z_{1}^{\pm};q \right)_{\infty}
}
\cdot 
\prod_{i=1}^{M}
\frac{
\left(q^{\frac34+\frac{N-M}{4}} t^{1-(N-M)} s_{i}^{\pm} z_{1}^{\pm};q \right)_{\infty}
}
{
\left(q^{\frac14+\frac{N-M}{4}} t^{-1-(N-M)} s_{i}^{\pm} z_{1}^{\pm};q \right)_{\infty}
}.
\end{align}
The terms in the second and third line are the full-index for 4d $U(M)$ gauge theory 
and half-indices for Nahm$'$ boundary conditions of rank $(L-M)$ and rank $(N-M)$. 
The terms in the last line describe the extra contributions from broken gauge theories.

We conjecture that 
the half-indices (\ref{uLuN_uMa}) and (\ref{uLuN_uMb1}) give the same answer for $M<L$. 
In fact, we have found that 
they match up to certain order of $q$ 
for $(L,M,N)$ $=$ $(2,1,1)$ and $(2,1,3)$  
(see Appendix \ref{app_interface}).

For $L\le M$, 
the gauge symmetry $U(L)\times U(M)\times U(N)$ is broken to $U(L)$ 
and the $U(L)$ gauge symmetry is kept in the whole 4d theory. 
For $L<M$ the defect of D5$'$-brane which has $L$ and $M$ D3-branes in their its sides 
may give rise to the Nahm$'$ b.c. of rank $L-M$ and the associated extra degrees of freedom at the interface 
which couple to the surviving 4d $U(L)$ gauge theory. 
For $L=M$, it couples 3d fundamental twisted hypermultiplet to the 4d $U(L)$ gauge theory. 
On the other defect of D5$'$-brane which has $M$ and $N$ D3-branes in their its sides 
would lead to the Nahm$'$ b.c. of rank $N-M$ and the associated extra degrees of freedom at the interface. 
These extra operators would couple to the surviving $U(L)$ gauge theory 
however, further degrees of freedom which do not couple to the surviving $U(L)$ gauge theory 
would appear in contrast to the case with $M<L$. 

Thus the half-index will be evaluated as
\begin{align}
\label{uLuN_uMb2}
&
\mathbb{II}^{\textrm{4d $U(L)|[M]|$4d $U(N)$}}_{\mathcal{D}'}
(t,z_{\alpha};q)
\nonumber\\
&=
\underbrace{
\frac{1}{L!}
\frac{(q)_{\infty}^{2L}}{(q^{\frac12}t^{2};q)_{\infty}^L (q^{\frac12}t^{-2};q)_{\infty}^L}
\oint \prod_{i=1}^{L}
\frac{ds_{i}}{2\pi is_{i}}
\prod_{i\neq j}
\frac{
\left(
\frac{s_{i}}{s_{j}};q \right)_{\infty}
\left(q \frac{s_{i}}{s_{j}};q \right)_{\infty}
}
{
\left(q^{\frac12} t^{2} \frac{s_{i}}{s_{j}};q \right)_{\infty}
\left(q^{\frac12} t^{-2} \frac{s_{i}}{s_{j}};q \right)_{\infty}
}
}_{\mathbb{I}^{\textrm{4d $U(L)$}}}
\nonumber\\
&\times 
\underbrace{
\prod_{k=1}^{M-L}
\frac{
\left(q^{\frac{k+1}{2}} t^{-2(k-1)};q \right)_{\infty}
}
{
\left(q^{\frac{k}{2}} t^{-2k};q \right)_{\infty}
}
}_{\mathbb{II}^{\textrm{4d $U(M-L)$}}_{\textrm{Nahm}'}}
\cdot 
\underbrace{
\prod_{l=1}^{N-M}
\frac{
\left(q^{\frac{l+1}{2}} t^{-2(l-1)};q \right)_{\infty}
}
{
\left(q^{\frac{l}{2}} t^{-2l};q \right)_{\infty}
}
}_{\mathbb{II}^{\textrm{4d $U(N-M)$}}_{\textrm{Nahm}'}}
\nonumber\\
&\times 
\prod_{i=1}^{L}
\frac{
\left(q^{\frac34+\frac{M-L}{4}} t^{1-(M-L)} s_{i}^{\pm} z_{1}^{\pm};q \right)_{\infty}
}
{
\left(q^{\frac14+\frac{M-L}{4}} t^{-1-(M-L)} s_{i}^{\pm} z_{1}^{\pm};q \right)_{\infty}
}
\cdot 
\prod_{i=1}^{L}
\frac{
\left(q^{\frac34+\frac{N-M}{4}} t^{1-(N-M)} s_{i}^{\pm} z_{1}^{\pm};q \right)_{\infty}
}
{
\left(q^{\frac14+\frac{N-M}{4}} t^{-1-(N-M)} s_{i}^{\pm} z_{1}^{\pm};q \right)_{\infty}
}
\nonumber\\
&\times 
\frac{
\left(q^{\frac34+\frac{N-M}{4}} t^{1-(N-M)} z_{1}^{\pm} z_{2}^{\mp};q \right)_{\infty}
}
{
\left(q^{\frac14+\frac{N-M}{4}} t^{-1-(N-M)} z_{1}^{\pm} z_{2}^{\mp};q \right)_{\infty}
}. 
\end{align}
The terms in the second and third line are the full-index for 4d $U(L)$ gauge theory 
and half-indices for Nahm$'$ boundary conditions of rank $(M-L)$ and rank $(N-M)$. 
The terms in the fourth line are the extra contributions from broken gauge theories 
which couple to the surviving gauge theory. 
The terms in the last line is those which are neutral under the surviving gauge symmetry. 

For $L\le M$, the half-index (\ref{uLuN_uMa}) will be equal to the half-indices (\ref{uLuN_uMb2}). 
As shown in Appendix \ref{app_interface}, 
we have checked that 
they match up to certain orders of $q$ 
for $(L,M,N)$ $=$ $(1,1,2)$, $(1,2,2)$, $(2,2,3)$, $(1,2,3)$, 
$(1,2,4)$, $(1,2,5)$ and $(2,3,3)$.

\subsection{4d $U(1)|\textrm{3d SQED}_{N_{f}}|$4d $U(1)$}
\label{sec_u1sqednfu1_0}
\begin{figure}
\begin{center}
\includegraphics[width=12cm]{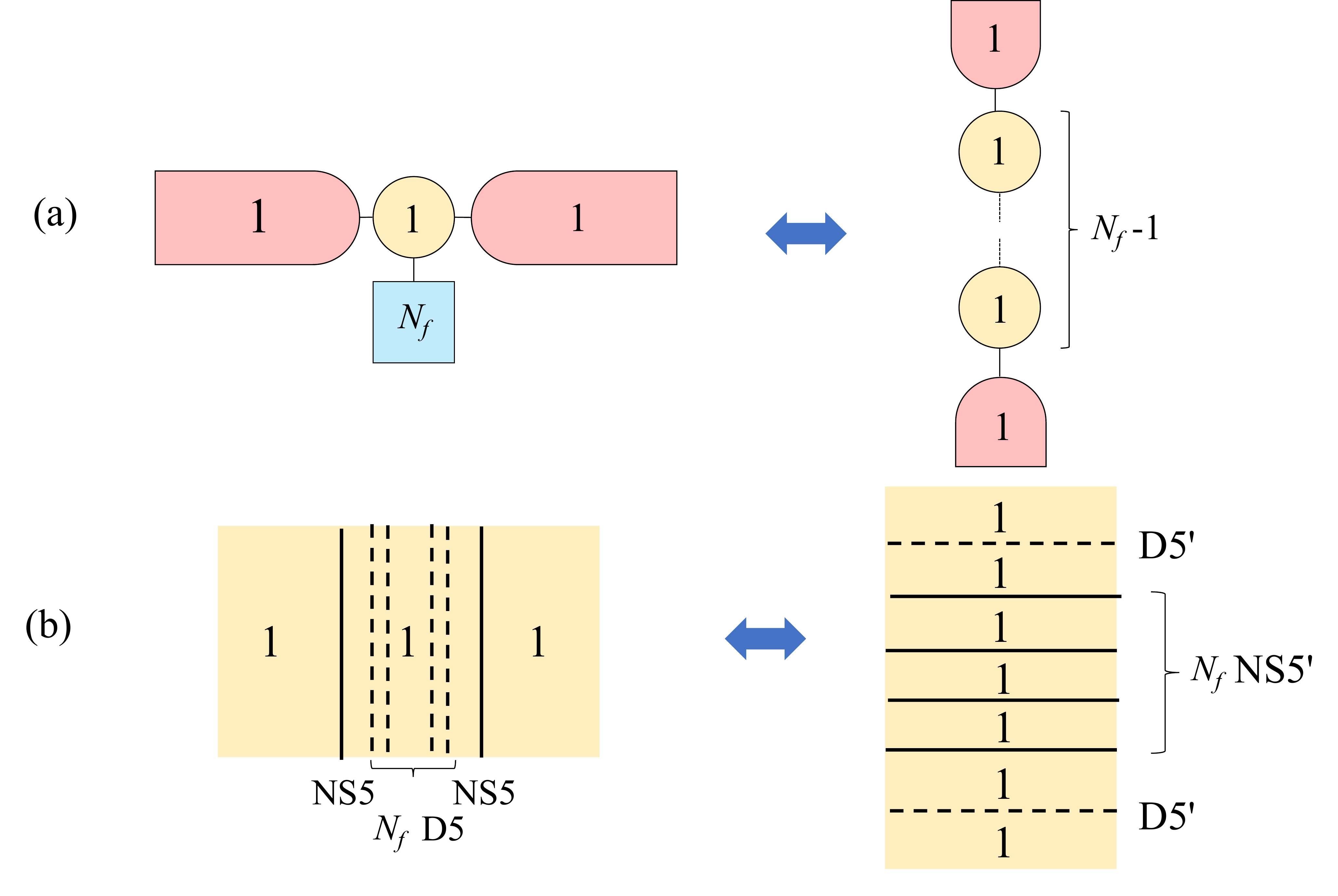}
\caption{
(a) The quiver diagrams of the interface 4d $U(1)|\textrm{3d SQED}_{N_{f}}|$4d $U(1)$ and its mirror. 
(b) The corresponding brane configurations. 
}
\label{fig4du1u13dsqed}
\end{center}
\end{figure}

Let us turn to the interfaces which 
include 3d $\mathcal{N}=4$ gauge theories with flavors. 
We consider the interface which has a pair of 4d $\mathcal{N}=4$ $U(1)$ gauge theories 
with Neumann b.c. $\mathcal{N}$ coupled to 3d $\mathcal{N}=4$ SQED$_{N_{f}}$ 
through the 3d bi-fundamental hypermultiplets. 
We denote this interface by 
4d $U(1)|\textrm{3d SQED}_{N_{f}}|$4d $U(1)$. 
The corresponding quiver diagram and brane configuration are illustrated in Figure \ref{fig4du1u13dsqed}. 
It is realized by a single D3-brane 
intersecting with two NS5-branes and $N_{f}$ D5-branes. 
On the other hand, the S-dual configuration  
contains a single D3-brane interacting with $N_{f}$ NS5$'$-branes 
and two D5$'$-branes. 
For the corresponding dual interface involves  
3d Abelian quiver gauge theories as for the Abelian mirror symmetry, 
however, the topological symmetry in the original interface 
is mapped to the flavor symmetry for the twisted hypers living at defects in the bulk 4d theory 
which are decoupled from the 3d quiver gauge theories. 
We compute the half-indices to check the dualities of the interfaces.

For the interface 4d $U(1)|\textrm{3d SQED}_{N_{f}}|$4d $U(1)$, 
we have a pair of 4d $\mathcal{N}=4$ $U(1)$ gauge theories living in half-spaces  
obeying Neumann b.c. $\mathcal{N}$. 
The interface has 3d $\mathcal{N}=4$ SQED$_{N_f}$ coupled to the 4d theories in either sides.

The half-index of the 4d $U(1)|\textrm{3d SQED}_{N_{f}}|$4d $U(1)$ interface is
\begin{align}
\label{4du1sqednfu1a}
&\mathbb{II}^{\textrm{4d $U(1)|$SQED$_{N_{f}}|$4d $U(1)$}}_{\mathcal{N}\mathcal{N}}(t,x_{\alpha},z_{\alpha};q)
\nonumber\\
&=
\underbrace{
\frac{(q)_{\infty}}{(q^{\frac12}t^{-2};q)_{\infty}} 
\oint \frac{ds_{1}}{2\pi is_{1}}
}_{\mathbb{II}_{\mathcal{N}}^{\textrm{4d $U(1)$}}}
\cdot 
\frac{(q^{\frac12}t^2;q)_{\infty}}
{(q^{\frac12}t^{-2};q)_{\infty}} 
\sum_{m\in \mathbb{Z}} \oint \frac{ds_{2}}{2\pi is_{2}} 
\underbrace{
\frac{(q)_{\infty}}{(q^{\frac12}t^{-2};q)_{\infty}} 
\oint \frac{ds_{3}}{2\pi is_{3}}
}_{\mathbb{II}_{\mathcal{N}}^{\textrm{4d $U(1)$}}}
\nonumber\\
&\times 
\frac{(q^{\frac34+\frac{|m|}{2}} t^{-1}s_{1}^{\pm}s_{2}^{\mp};q)_{\infty}}
{(q^{\frac14+\frac{|m|}{2}} ts_{1}^{\pm} s_{2}^{\mp};q)_{\infty}}
\cdot 
\prod_{\alpha=1}^{N_{f}}
\frac{(q^{\frac34+\frac{|m|}{2}} t^{-1}s_{2}^{\pm} x_{\alpha}^{\pm};q)_{\infty}}
{(q^{\frac14+\frac{|m|}{2}} ts_{2}^{\pm}x_{\alpha}^{\pm};q)_{\infty}}
\cdot 
\frac{(q^{\frac34+\frac{|m|}{2}} t^{-1}s_{2}^{\pm}s_{3}^{\mp};q)_{\infty}}
{(q^{\frac14+\frac{|m|}{2}} ts_{2}^{\pm}s_{3}^{\mp};q)_{\infty}}
\nonumber\\
&\times 
q^{\frac{(N_{f}+2)|m|}{4}} t^{-(N_{f}+2)|m|} z_{1}^m z_{2}^{-m} 
\end{align}
where $x_{\alpha}$ are the fugacities for the $SU(N_{f})$ flavor symmetry of 3d SQED$_{N_{f}}$ 
with $\prod_{\alpha=1}^{N_{f}}x_{\alpha}=1$ 
and $z_{\alpha}$ are the fugacities for the topological symmetries of the 3d SQED$_{N_{f}}$. 
The terms in the second line 
are the two half-indices of Neumann b.c. $\mathcal{N}$ for 4d $\mathcal{N}=4$ $U(1)$ gauge theory 
and the full-index for 3d $\mathcal{N}=4$ $U(1)$ vector multiplet. 
The terms in the next line describe the 3d $\mathcal{N}=4$ hypermultiplets. 
The terms in the last line count monopole operator of dimension $\Delta(m)=\frac{(N_{f}+2)|m|}{2}$. 

The dual interface involves 
a pair of 4d $\mathcal{N}=4$ $U(1)$ gauge theories living in a half-space 
obeying Neumann b.c. $\mathcal{N}'$. 
Unlike the original interface, 
each of 4d theories has a defect that couples a 3d $\mathcal{N}=4$ fundamental twisted hypermultiplet 
to the associated 4d $U(1)$ gauge theories. 
In addition, the interface has a 3d $\mathcal{N}=4$ $U(1)^{N_{f}-1}$ twisted vector multiplet 
coupled to both 4d gauge theories through the 3d $\mathcal{N}=4$ bi-fundamental twisted hypermultiplets. 

The half-index for the dual interface should take the form 
\begin{align}
\label{4du1sqednfu1b}
&\mathbb{II}_{\mathcal{N}'\mathcal{N}'}
^{\textrm{4d $U(1)+\textrm{thyper}|
\widetilde{(1)^{N_{f}-1}}|$4d $U(1)+\textrm{thyper}$}}(t,x_{\alpha},z_{\alpha};q)
\nonumber\\
&=
\underbrace{
\frac{(q)_{\infty}}{(q^{\frac12}t^2;q)_{\infty}} 
\oint \frac{ds_{1}}{2\pi is_{1}}
}_{\mathbb{II}_{\mathcal{N}'}^{\textrm{4d $U(1)$}}} 
\cdot 
\frac{(q^{\frac12}t^{-2};q)_{\infty}^{N_{f}-1}}
{(q^{\frac12}t^2;q)_{\infty}^{N_{f}-1}} 
\sum_{m_{1},\cdots,m_{N_{f}-1}\in \mathbb{Z}}
\oint \prod_{i=1}^{N_{f}-1} \frac{ds_{i}}{2\pi is_{i}}
\underbrace{
\frac{(q)_{\infty}}{(q^{\frac12}t^2;q)_{\infty}} 
\oint \frac{ds_{N_{f}}}{2\pi is_{N_{f}}}
}_{\mathbb{II}_{\mathcal{N}'}^{\textrm{4d $U(1)$}}} 
\nonumber\\
&\times 
\underbrace{
\frac{(q^{\frac34}ts_{1}^{\pm}z_{1}^{\pm};q)_{\infty}}
{(q^{\frac14}t^{-1}s_{1}^{\pm}z_{1}^{\pm};q)_{\infty}}
}_{\mathbb{I}^{\textrm{3d tHM}}(s_{1}z_{1})}
\cdot 
\frac{
(q^{\frac34+\frac{|m_{1}|}{2}} t s_{1}^{\pm} s_{2}^{\mp};q )_{\infty}
}
{
(q^{\frac14+\frac{|m_{1}|}{2}} t^{-1} s_{1}^{\pm} s_{2}^{\mp};q )_{\infty}
}
\cdot 
\prod_{i=2}^{N_{f}-2}
\frac{
(q^{\frac34+\frac{|m_{i}-m_{i+1}|}{2}} t s_{i}^{\pm} s_{i+1}^{\mp} ;q )_{\infty}
}
{
(q^{\frac14+\frac{|m_{i}-m_{i+1}|}{2}} t^{-1} s_{i}^{\pm} s_{i+1}^{\mp} ;q )_{\infty}
}
\nonumber\\
&\times 
\frac{
(q^{\frac34+\frac{|m_{N_{f}-1}|}{2} } t s_{N_{f}-1}^{\pm} s_{N_{f}}^{\mp};q)_{\infty}
}
{
(q^{\frac14+\frac{|m_{N_{f}-1}|}{2} } t^{-1} s_{N_{f}-1}^{\pm} s_{N_{f}}^{\mp};q)_{\infty}
}
\cdot 
\underbrace{
\frac{(q^{\frac34}ts_{N_{f}}^{\pm}z_{2}^{\pm};q)_{\infty}}
{(q^{\frac14}t^{-1}s_{N_{f}}^{\pm}z_{2}^{\pm};q)_{\infty}}
}_{\mathbb{I}^{\textrm{3d tHM}}(s_{1}z_{1})}
\nonumber\\
&\times 
q^{\frac{|m_{1}|+|m_{N_{f}-1}|+\sum_{i=1}^{N_{f}-2}|m_{i}-m_{i+1}|}{4}} 
\cdot 
t^{|m_{1}|+|m_{N_{f}-1}|+\sum_{i=1}^{N_{f}-2}|m_{i}-m_{i+1}|}
\cdot 
\prod_{i=1}^{N_{f}-1}
\left(\frac{x_{i}}{x_{i+1}}\right)^{m_{i}} 
\cdot 
\left(\frac{x_{N_{f}}}{x_{1}}\right)^{m_{1}+m_{N_{f}-1}}
\end{align}
where $z_{\alpha}$ is the fugacities for the flavor symmetry of defect 3d twisted hypermultiplets 
and $x_{\alpha}$ is the fugacities for the topological symmetry of the 3d $U(1)^{N_{f}-1}$ twisted vector multiplet. 
The contributions in the second line are 
the square of half-index of Neumann b.c. $\mathcal{N}'$ for 4d $U(1)$ gauge theory 
and the full-index for 3d $U(1)^{N_{f}-1}$ twisted vector multiplet. 
The terms in the third and fourth lines correspond to the contributions of 3d $\mathcal{N}=4$ twisted hypermultiplets. 
The terms in the last line count monopole operator of dimension $\Delta(m)$ $=$ $\frac{1}{2}
(|m_{1}|+|m_{N_{f}-1}|+\sum_{i=1}^{N_{f}-2}|m_{i}-m_{i+1}|)$.

It is expected that the half-indices (\ref{4du1sqednfu1a}) and (\ref{4du1sqednfu1b}) agree with each other. 
We have confirmed the matching for $N_{f}=1,2,3$ up to certain orders of $q$ 
(see Appendix \ref{app_interface}).

\subsection{4d $U(N)|\textrm{3d SQED}_{N_{f}}|$4d $U(M)$}
\label{sec_uNsqednfuM_0}
\begin{figure}
\begin{center}
\includegraphics[width=12cm]{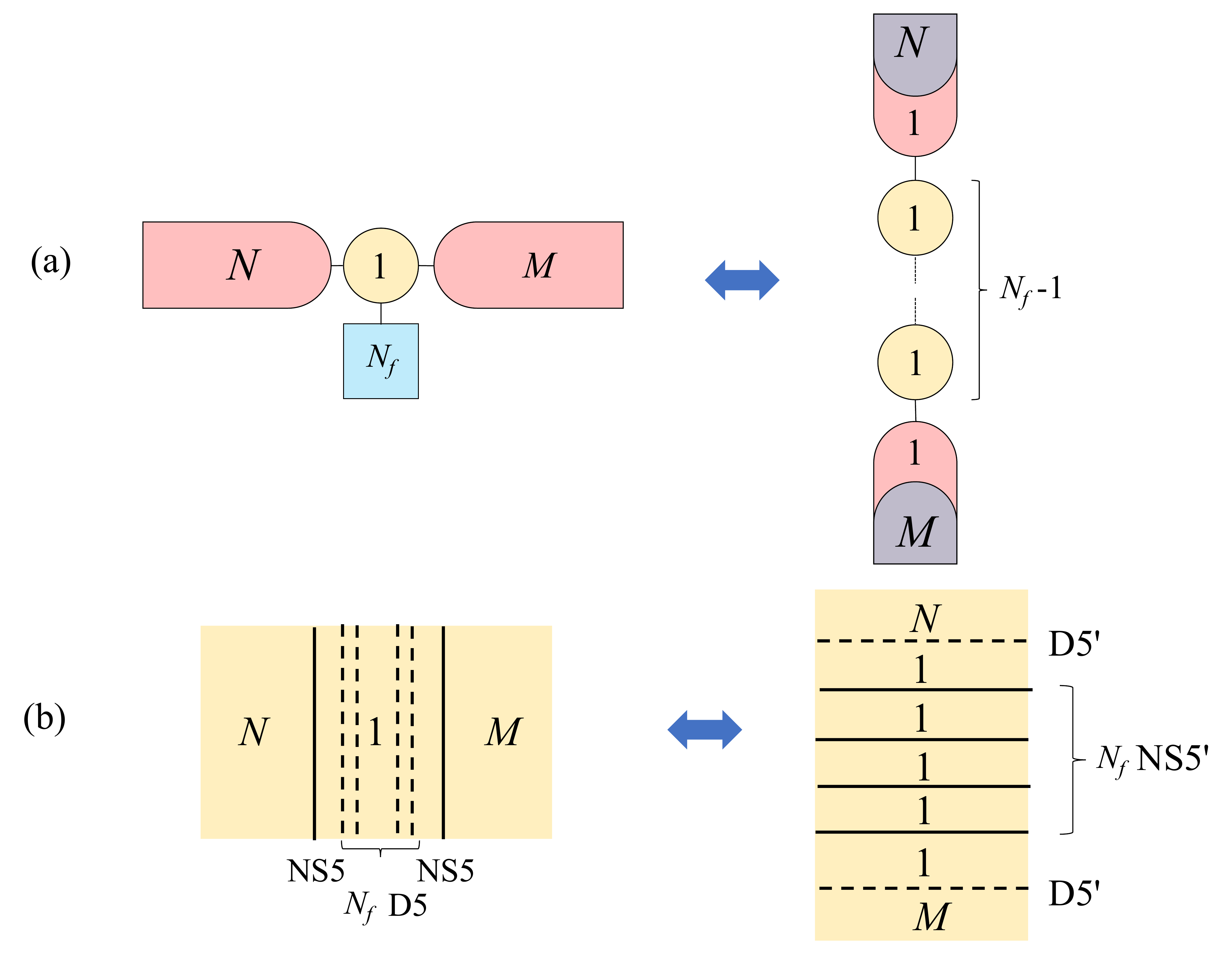}
\caption{
(a) The quiver diagrams of the interface 4d $U(N)|\textrm{3d SQED}_{N_{f}}|$4d $U(M)$ and its mirror. 
(b) The corresponding brane configurations. 
}
\label{fig4duNuM3dsqed}
\end{center}
\end{figure}

We next turn to the interface 4d $U(N)|\textrm{3d SQED}_{N_{f}}|$4d $U(M)$ 
where $N,M$ $>1$.  
The corresponding quiver diagram and brane construction are drawn in Figure \ref{fig4duNuM3dsqed}. 
Unlike the interface discussed in section \ref{sec_u1sqednfu1_0}, 
the dual interface may include singular boundary conditions for 4d $U(N)$ and $U(M)$ gauge theories 
which are characterized by the Nahm poles of rank $(N-1)$ and rank $(M-1)$. 
We calculate the half-indices for a pair of dual interfaces and find the matching.

The interface involves a pair of 4d $\mathcal{N}=4$ $U(N)$ and $U(M)$ SYM theories satisfying Neumann b.c. $\mathcal{N}$ 
corresponding to the NS5-branes. 
Both of 4d theories couple to 3d $\mathcal{N}=4$ SQED$_{N_{f}}$ through 
3d bi-fundamental hypermultiplets. 
The magnetic fluxes for 3d SQED$_{N_{f}}$ can be labeled by an integer $m$.

The half-index of the 4d $U(N)|\textrm{3d SQED}_{N_{f}}|$4d $U(M)$ interface is
\begin{align}
\label{4duNsqedNfuMa}
&\mathbb{II}_{\mathcal{N}\mathcal{N}}^{\textrm{4d $U(N)|\textrm{SQED}_{N_{f}}|$4d $U(M)$}}(t,x_{\alpha},z_{\alpha};q)
\nonumber\\
&=
\underbrace{
\frac{1}{N!}\frac{(q)_{\infty}^{N}}{(q^{\frac12}t^{-2};q)_{\infty}^{N}}
\oint \prod_{i=1}^{N} \frac{ds_{i}}{2\pi is_{i}}
\prod_{i\neq j}\frac{\left(\frac{s_{i}}{s_{j}};q\right)_{\infty}}
{\left(q^{\frac12}t^{-2}\frac{s_{i}}{s_{j}};q\right)_{\infty}}
}_{\mathbb{II}_{\mathcal{N}}^{\textrm{4d $U(N)$}}}
\nonumber\\
&\times 
\frac{(q^{\frac12}t^2;q)_{\infty}}
{(q^{\frac12}t^{-2};q)_{\infty}}\sum_{m\in \mathbb{Z}}
\oint \frac{ds_{N+1}}{2\pi is_{N+1}} 
\nonumber\\
&\times 
\underbrace{
\frac{1}{M!}
\frac{(q)_{\infty}^M}{(q^{\frac12}t^{-2};q)_{\infty}^M}
\oint \prod_{i=N+2}^{N+M+1} \frac{ds_{i}}{2\pi is_{i}} 
\prod_{i\neq j}\frac{\left(\frac{s_{i}}{s_{j}};q\right)_{\infty}}
{\left(q^{\frac12}t^{-2}\frac{s_{i}}{s_{j}};q\right)_{\infty}}
}_{\mathbb{II}_{\mathcal{N}}^{\textrm{4d $U(M)$}}}
\nonumber\\
&\times 
\prod_{i=1}^{N}
\frac{(q^{\frac34+\frac{|m|}{2}} t^{-1}s_{i}^{\pm} s_{N+1}^{\mp};q)_{\infty}}
{(q^{\frac14+\frac{|m|}{2}} ts_{i}^{\pm}s_{N+1}^{\mp};q)_{\infty}}
\cdot 
\prod_{\alpha=1}^{N_{f}}
\frac{(q^{\frac34+\frac{|m|}{2}} t^{-1}s_{N+1}^{\pm}x_{\alpha}^{\pm};q)_{\infty}}
{(q^{\frac14+\frac{|m|}{2}}ts_{N+1}^{\pm} x_{\alpha}^{\pm};q)_{\infty}} 
\cdot 
\prod_{i=N+2}^{N+M+1}
\frac{(q^{\frac34+\frac{|m|}{2}} t^{-1}s_{N+1}^{\pm}s_{i}^{\mp};q)_{\infty}}
{(q^{\frac14+\frac{|m|}{2}} t s_{N+1}^{\pm} s_{i}^{\mp};q)_{\infty}}
\nonumber\\
&\times 
q^{\frac{(N+M+N_{f})|m|}{4}} t^{-(N+M+N_{f})|m|} z_{1}^m z_{2}^{-m}
\end{align}
where the fugacities $x_{\alpha}$ with $\prod_{\alpha=1}^{N_{f}}x_{\alpha}=1$ are associated with 
the $SU(N_{f})$ flavor symmetry of SQED$_{N_{f}}$ 
while the fugacities $z_{\alpha}$ are associated with 
the topological symmetry of SQED$_{N_{f}}$. 
The terms from the second to fourth line describe gauge multiplets in the interface, 
including the half-indices of Neumann b.c. $\mathcal{N}$ for 4d $\mathcal{N}=4$ $U(N)$ and $U(M)$ gauge theories 
and the full-index for 3d $\mathcal{N}=4$ $U(1)$ vector multiplet. 
The terms in the fifth line correspond to the 3d $\mathcal{N}=4$ hypermultiplets. 
The terms in the last line describe monopole operator 
with the R-charge $\Delta(m)$ $=$ $\frac{(N+M+N_{f})|m|}{2}$. 

Under the action of S-duality 
one finds the dual interface which also has a pair of 4d $\mathcal{N}=4$ gauge theories. 
The D5$'$-brane intersecting with $N$ D3-brane breaks $U(N)\times U(1)$ gauge symmetry down to $U(1)$ 
while the other D5$'$-brane intersecting with $M$ D3-branes breaks $U(M)\times U(1)$ gauge symmetry down to $U(1)$. 
The reductions of 4d gauge symmetries are described by the 
boundary conditions specified by the two Nahm poles of rank $(N-1)$ and rank $(M-1)$. 
Correspondingly, extra contributions would appear at the defects from the broken gauge theories \cite{Gaiotto:2019jvo}. 

According to the presence of NS5$'$-branes, 
the surviving 4d $\mathcal{N}=4$ $U(1)$ gauge theories further satisfy Neumann b.c. $\mathcal{N}'$. 
They are couple to 3d $\mathcal{N}=4$ $\widetilde{U(1)^{N_{f}-1}}$ quiver gauge theory 
via 3d $\mathcal{N}=4$ bi-fundamental twisted hypermultiplets. 

The half-index should take the form
\begin{align}
\label{4duNsqedNfuMb}
&\mathbb{II}^{\textrm{4d $U(N)\rightarrow U(1)|\widetilde{(1)^{N_{f}-1}}|$4d $U(M)\rightarrow U(1)$}}
_{\mathcal{N}'\mathcal{N}'}(t,x_{\alpha},z_{\alpha};q)
\nonumber\\
&=
\underbrace{
\frac{(q)_{\infty}}{(q^{\frac12}t^{2};q)_{\infty}} 
\oint \frac{ds_{1}}{2\pi is_{1}}
}_{\mathbb{II}_{\mathcal{N}'}^{\textrm{4d $U(1)$}}}
\cdot 
\underbrace{
\prod_{k=1}^{N-1}
\frac{(q^{\frac{k+1}{2}} t^{-2(k-1)};q)_{\infty}}
{(q^{\frac{k}{2}} t^{-2k};q)_{\infty}}
}_{\mathbb{II}_{\textrm{Nahm}'}^{\textrm{4d $U(N-1)$}}}
\cdot 
\frac{(q^{\frac12}t^{-2};q)_{\infty}^{N_{f}-1}}{(q^{\frac12}t^2;q)_{\infty}^{N_{f}-1}}
\sum_{m_{1},\cdots,m_{N_{f}-1}\in \mathbb{Z}}
\oint \prod_{i=2}^{N_{f}}\frac{ds_{i}}{2\pi is_{i}}
\nonumber\\
&\times 
\underbrace{
\prod_{l=1}^{M-1}
\frac{(q^{\frac{l+1}{2}} t^{-2(l-1)};q)_{\infty}}{(q^{\frac{l}{2}} t^{-2l};q)_{\infty}}
}_{\mathbb{II}_{\textrm{Nahm}'}^{\textrm{4d $U(M-1)$}}}
\cdot 
\underbrace{
\frac{(q)_{\infty}}{(q^{\frac12}t^2;q)_{\infty}}\oint \frac{ds_{N_{f}+1}}{2\pi is_{N_{f}+1}}
}_{\mathbb{II}_{\mathcal{N}'}^{\textrm{4d $U(1)$}}}
\nonumber\\
&\times 
\frac{
\left( q^{\frac34+\frac{N-1}{4}} t^{1-(N-1)} s_{1}^{\pm} z_{1}^{\pm};q \right)_{\infty}
}
{
\left(q^{\frac14+\frac{N-1}{4}} t^{-1-(N-1)} s_{1}^{\pm} z_{1}^{\pm};q \right)_{\infty}
}
\cdot 
\frac{
\left( q^{\frac34+\frac{|m_{1}|}{2}} t s_{1}^{\pm} s_{2}^{\mp};q \right)_{\infty}
}
{
\left( q^{\frac14+\frac{|m_{1}|}{2}} t^{-1} s_{1}^{\pm} s_{2}^{\mp};q \right)_{\infty}
}
\cdot 
\prod_{i=2}^{N_{f}-1}
\frac{
\left( q^{\frac34+\frac{|m_{i}-m_{i+1}|}{2}} t s_{i}^{\pm} s_{i+1}^{\mp}; q \right)_{\infty}
}
{
\left( q^{\frac14+\frac{|m_{i}-m_{i+1}|}{2}} t^{-1} s_{i}^{\pm} s_{i+1}^{\mp};q \right)_{\infty}
}
\nonumber\\
&\times 
\frac{
\left( q^{\frac34+\frac{|m_{N_{f}-1}|}{2}} t s_{N_{f}}^{\pm} s_{N_{f}+1}^{\mp};q \right)_{\infty}
}
{
\left(q^{\frac14+\frac{|m_{N_{f}-1}|}{2}} t^{-1} s_{N_{f}}^{\pm} s_{N_{f}-1}^{\mp};q \right)_{\infty}
}
\cdot 
\frac{
\left(q^{\frac34+\frac{M-1}{4}} t^{1-(M-1)} s_{N_{f}+1}^{\pm} z_{2}^{\pm}; q \right)_{\infty}
}
{
\left(q^{\frac14+\frac{M-1}{4}} t^{-1-(M-1)} s_{N_{f}+1}^{\pm} z_{2}^{\pm};q \right)_{\infty}
}
\nonumber\\
&\times 
q^{\frac{|m_{1}|+|m_{N_{f}-1}|+\sum_{i=1}^{N_{f}-2}|m_{i}-m_{i+1}|}{4}}
t^{|m_{1}|+|m_{N_{f}-1}|+\sum_{i=1}^{N_{f}-2}|m_{i}-m_{i+1}|} 
\prod_{\alpha=1}^{N_{f}-1}
\left(\frac{x_{\alpha}}{x_{\alpha+1}}\right)^{m_{\alpha}}
\cdot 
\left(\frac{x_{N_{f}}}{x_{1}}\right)^{m_{1}+m_{N_{f}-1}}
\end{align}
where $z_{1}, z_{2}$ are associated with the extra local operators at the two defects 
and $x_{\alpha}$ are associated with the topological symmetry 
for 3d $\widetilde{U(1)^{N_{f}-1}}$ quiver gauge theory. 
The terms from the second to third line 
includes the square of half-indices of Neumann b.c. $\mathcal{N}'$ for 4d $\mathcal{N}=4$ $U(1)$ gauge theory, 
the two half-indices of Nahm$'$ b.c. of rank $(N -1)$ and rank $(M-1)$, 
and the full-index for 3d $\mathcal{N}=4$ $U(1)$ vector multiplet. 
The terms from the fourth to fifth line describe 3d $\mathcal{N}=4$ twisted hypermultiplets. 
The contributions in the last line count bare monopole of dimension 
$\Delta(m)$ $=$ $\frac12( |m_{1}|+|m_{N_{f}-1}|+\sum_{i=1}^{N_{f}-2}|m_{i}-m_{i+1}| )$. 
 
We expect that the half-indices (\ref{4duNsqedNfuMa}) and (\ref{4duNsqedNfuMb}) agree with each other. 
We have shown the matching of indices for $(N,M,N_{f})$ $=$  
$(2,3,1)$, $(2,3,2)$ and $(2,3,3)$ 
up to certain orders of $q$ in Appendix \ref{app_interface}.

\subsection{4d $U(1)+N$ hypers$|$3d SQED$_{N_{f}}|$4d $U(1)+M$ hypers}
\label{sec_3dinterface_hmplus}
\begin{figure}
\begin{center}
\includegraphics[width=12cm]{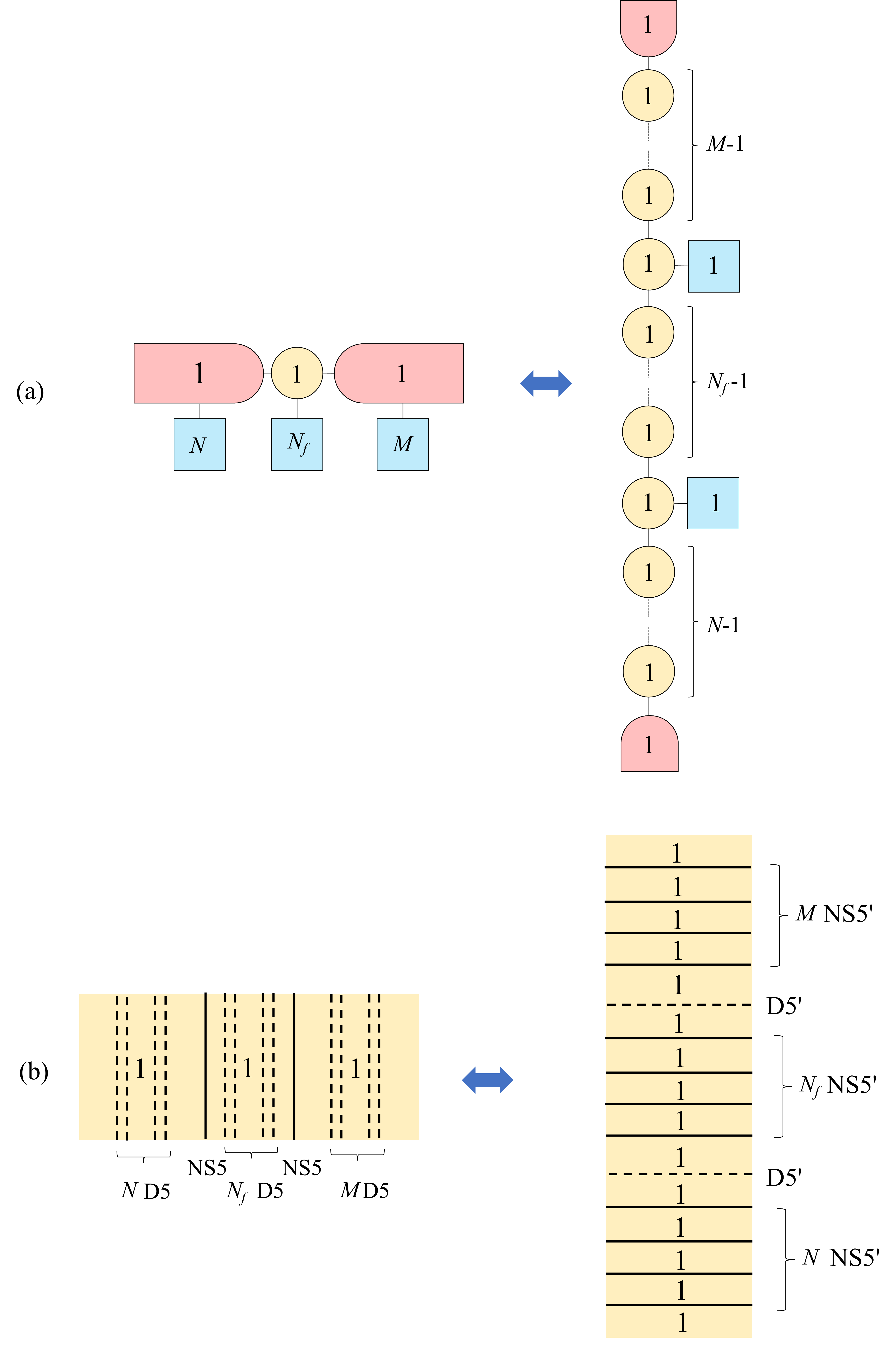}
\caption{
(a) The quiver diagrams of the interface 4d $U(1)+N$ hypers$|$3d SQED$_{N_{f}}|$4d $U(1)+M$ hypers and its mirror. 
(b) The corresponding brane configurations.  
}
\label{fig4du1u1hmNM3dsqed}
\end{center}
\end{figure}

We investigate the interface with a pair of 4d $\mathcal{N}=4$ $U(1)$ gauge theories 
where one of them has $N$ D5-brane defects 
and the other has $M$ D5-brane defects and 
the both 4d theories obey the Neumann b.c. $\mathcal{N}$ with a coupling to 3d $\mathcal{N}=4$ SQED$_{N_{f}}$. 
\footnote{
Although we have restricted to the Abelian interface in this section, 
the generalization to the non-Abelian interface is straightforward by using the results so far. 
}
We denote this interface by 
4d $U(1)+N$ hypers$|$3d SQED$_{N_{f}}|$4d $U(1)+M$ hypers. 
The corresponding quiver diagram and brane construction are shown in Figure \ref{fig4du1u1hmNM3dsqed}.
The $N$ and $M$ D5-branes introduce $N$ and $M$ fundamental hypermultiplets living at the defects.

In this case, there are a 4d $\mathcal{N}=4$ $U(1)$ gauge theory with $N$ defects 
which couple the 4d theory to the $N$ fundamental 3d hypermultiplets 
and another 4d  $\mathcal{N}=4$ $U(1)$ gauge theory with $M$ defects 
which couple the 4d theory to the $M$ fundamental 3d hypermultiplets. 
A pair of 4d gauge theories 
obeys Neumann b.c. $\mathcal{N}$ and interacts with 3d $\mathcal{N}=4$ SQED$_{N_{f}}$ 
through a coupling given by 3d $\mathcal{N}=4$ bi-fundamental hypermultiplets. 
The interface has 4d $U(1)\times U(1)$ gauge symmetry and 3d $U(1)$ gauge symmetry. 
We can label the magnetic fluxes for 3d $U(1)$ gauge theory by an integer $m$.

Then the half-index for the interface 4d $U(1)+N$ hypers$|$3d SQED$_{N_{f}}|$4d $U(1)+M$ hypers 
takes the form 
\begin{align}
\label{u1u1hmNM_u1a}
&
\mathbb{II}^{\textrm{4d $U(1)+N$ hypers$|$SQED$_{N_{f}}|$4d $U(1)+M$ hypers}}
_{\mathcal{N}\mathcal{N}}(t,x_{\alpha},z_{\alpha};q)
\nonumber\\
&=
\underbrace{
\frac{(q)_{\infty}}{(q^{\frac12}t^{-2};q)_{\infty}} \oint \frac{ds_{1}}{2\pi is_{1}}
}_{\mathbb{II}^{\textrm{4d $U(1)$}}_{\mathcal{N}}}
\cdot 
\frac{(q^{\frac12}t^2;q)_{\infty}}
{(q^{\frac12}t^{-2};q)_{\infty}}
\sum_{m\in \mathbb{Z}} \oint \frac{ds_{2}}{2\pi is_{2}}
\cdot 
\underbrace{
\frac{(q)_{\infty}}{(q^{\frac12}t^{-2};q)_{\infty}} 
\oint \frac{ds_{3}}{2\pi is_{3}}
}_{\mathbb{II}_{\mathcal{N}}^{\textrm{4d $U(1)$}}}
\nonumber\\
&\times 
\prod_{\alpha=1}^{N}
\underbrace{
\frac{(q^{\frac34}t^{-1}s_{1}^{\pm}x_{\alpha}^{\pm};q)_{\infty}}
{(q^{\frac14}t^{-1}s_{1}^{\pm}x_{\alpha}^{\pm};q)_{\infty}}
}_{\mathbb{I}^{\textrm{3d HM}}(s_{1}x_{\alpha})}
\cdot 
\frac{
(q^{\frac34+\frac{|m|}{2}} t^{-1}s_{1}^{\pm} s_{2}^{\mp};q)_{\infty}
}
{
(q^{\frac14+\frac{|m|}{2}} t s_{1}^{\pm} s_{2}^{\mp};q)_{\infty}
}
\cdot 
\prod_{\alpha=N+1}^{N+N_{f}}
\frac{(q^{\frac34+\frac{|m|}{2}} t^{-1} s_{2}^{\pm} x_{\alpha}^{\pm};q)_{\infty}}
{(q^{\frac14+\frac{|m|}{2}} t s_{2}^{\pm} x_{\alpha}^{\pm};q)_{\infty}}
\nonumber\\
&\times 
\frac{
(q^{\frac34+\frac{|m|}{2}} t^{-1}s_{2}^{\pm} s_{3}^{\mp};q)_{\infty}
}
{
(q^{\frac14+\frac{|m|}{2}} t s_{2}^{\pm} s_{3}^{\mp};q)_{\infty}
}
\cdot 
\prod_{\alpha=N+N_{f}+1}^{N+N_{f}+M}
\underbrace{
\frac{(q^{\frac34}t^{-1}s_{3}^{\pm}x_{\alpha}^{\pm};q)_{\infty}}
{(q^{\frac14}t s_{3}^{\pm} x_{\alpha}^{\pm};q)_{\infty}}
}_{\mathbb{I}^{\textrm{3d HM}}(s_{3}x_{\alpha})}
\nonumber\\
&\times 
q^{\frac{(N+N_{f}+M)|m|}{4}} 
\cdot 
t^{-(N+N_{f}+M)|m|}
\cdot 
\left( \frac{z_{1}}{z_{2}} \right)^{m}
\end{align}
where we have used $\{ x_{\alpha}\}_{\alpha=1,\cdots, N}$ 
for the fugacities of the $SU(N)$ flavor symmetries 
of the defect hypermultiplets, 
$\{x_{\alpha}\}_{\alpha=N+1,\cdots, N+N_{f}}$ 
for the fugacities of the $SU(N_{f})$ flavor symmetry 
of 3d SQED$_{N_{f}}$, 
$\{x_{\alpha}\}_{\alpha=N+N_{f}+1,\cdots, N+N_{f}+M}$ 
for the fugacities of the $SU(M)$ flavor symmetries 
of the defect hypermultiplets. 
The fugacities $z_{\alpha}$ are associated to the topological symmetry of 3d SQED$_{N_{f}}$. 
The contributions in the second line 
are the two half-indices of Neumann b.c. $\mathcal{N}$ for 4d $U(1)$ gauge theory 
and the full-index for 3d $U(1)$ vector multiplet. 
The terms appearing in the next two lines are the contributions from the 3d hypermultiplets. 
The terms in the last line count bare monopole of dimension $\Delta(m)=\frac{(N+N_{f}+M)|m|}{2}$. 

From the S-dual configuration of the brane setup in Figure \ref{fig4du1u1hmNM3dsqed}, 
we can read off the dual interface. 
It has a pair of 4d $\mathcal{N}=4$ $U(1)$ gauge theories 
which satisfy Neumann b.c. $\mathcal{N}'$ and 
couple to 3d quiver gauge theory 
$\widetilde{\begin{smallmatrix}
(1)^{N-1}&-&(1)&-&(1)^{N_{f}-1}&-&(1)&-&(1)^{M-1}\\
&&|&&&&|&& \\
&&[1]&&&&[1]&& \\
\end{smallmatrix}}$ 
through the 3d bu-fundamental twisted hypermultiplets. 
We can label the magnetic fluxes for the 3d quiver gauge theory 
by $N+N_{f}+M-1$ integers 
$m_{1},\cdots, m_{N+N_{f}+M-1}$.

The half-index for the dual configuration is given by 
\begin{align}
\label{u1u1hmNM_u1b}
&
\mathbb{II}^{\textrm{4d $U(1)|
\widetilde{\begin{smallmatrix}
(1)^{N-1}&-&(1)&-&(1)^{N_{f}-1}&-&(1)&-&(1)^{M-1}\\
&&|&&&&|&& \\
&&[1]&&&&[1]&& \\
\end{smallmatrix}}
|$4d $U(1)$}}_{\mathcal{N}'\mathcal{N}'}
(t,x_{\alpha},z_{\alpha};q)
\nonumber\\
&=
\underbrace{
\frac{(q)_{\infty}}{(q^{\frac12}t^{2};q)_{\infty}}
\oint \frac{ds_{1}}{2\pi is_{1}}
}_{\mathbb{II}_{\mathcal{N}'}^{\textrm{4d $U(1)$}}}
\nonumber\\
&\times 
\frac{(q^{\frac12}t^{-2};q)_{\infty}^{N+N_{f}+M-1}}
{(q^{\frac12}t^{2};q)_{\infty}^{N+N_{f}+M-1}}
\sum_{m_{1},\cdots,m_{N+N_{f}+M-1}\in \mathbb{Z}}
\oint \prod_{i=2}^{N+N_{f}+M}
\frac{ds_{i}}{2\pi is_{i}}
\nonumber\\
&\times 
\underbrace{
\frac{(q)_{\infty}}{(q^{\frac12}t^2;q)_{\infty}}
\oint \frac{ds_{N+N_{f}+M+1}}{2\pi is_{N+N_{f}+M+1}}
}_{\mathbb{II}_{\mathcal{N}'}^{\textrm{4d $U(1)$}}}
\nonumber\\
&\times 
\frac{
(q^{\frac34+\frac{|m_{1}|}{2}} t s_{1}^{\pm} s_{2}^{\mp};q )_{\infty}
}
{
(q^{\frac14+\frac{|m_{1}|}{2}} t^{-1} s_{1}^{\pm} s_{2}^{\mp};q )_{\infty}
}
\cdot 
\prod_{i=2}^{N+N_{f}+M-1}
\frac{
(q^{\frac34 +\frac{|m_{i-1}-m_{i}|}{2} } t s_{i}^{\pm} s_{i+1}^{\mp};q)_{\infty}
}
{
(q^{\frac14 +\frac{|m_{i-1}-m_{i}|}{2} } t^{-1} s_{i}^{\pm} s_{i+1}^{\mp};q)_{\infty}
}
\cdot 
\frac{
(q^{\frac34 +\frac{|m_{N}|}{2} } t s_{N+1}^{\pm} z_{1}^{\pm};q)_{\infty}
}
{
(q^{\frac14 +\frac{|m_{N}|}{2} } t^{-1} s_{N+1}^{\pm} z_{1}^{\pm};q)_{\infty}
}
\nonumber\\
&\times 
\frac{
(q^{\frac34 +\frac{|m_{N+N_{f}}|}{2} } t s_{N+N_{f}+1}^{\pm} z_{2}^{\pm};q)_{\infty}
}
{
(q^{\frac14 +\frac{|m_{N+N_{f}}|}{2} } t^{-1} s_{N+N_{f}+1}^{\pm} z_{2}^{\pm};q)_{\infty}
}
\cdot 
\frac{
(q^{\frac34 +\frac{|m_{N+N_{f}+M-1}|}{2} } t s_{N+N_{f}+M}^{\pm} s_{N+N_{f}+M+1}^{\mp};q)_{\infty}
}
{
(q^{\frac14 +\frac{|m_{N+N_{f}+M-1}|}{2} } t^{-1} s_{N+N_{f}+M}^{\pm} s_{N+N_{f}+M+1}^{\mp};q)_{\infty}
}
\nonumber\\
&\times 
q^{ \frac{|m_{1}|}{4}+\frac{|m_{N}|}{4}+\frac{|m_{N+N_{f}}|}{4}+\frac{|m_{N+N_{f}+M-1}|}{4}+\sum_{i=1}^{N+N_{f}+M-2}\frac{|m_{i}-m_{i+1}|}{4} }
\nonumber\\
&\times 
t^{|m_{1}|+|m_{N}|+|m_{N+N_{f}}|+|m_{N+N_{f}+M-1}|+\sum_{i=1}^{N+N_{f}+M-2}|m_{i}-m_{i+1}|}
\nonumber\\
&\times 
\prod_{\alpha=1}^{N+N_{f}+M} x_{\alpha}^{m_{N+N_{f}+M-\alpha+1}-m_{N+N_{f}+M-\alpha}}
\end{align}
where $m_{0}$ $=$ $m_{N+N_{f}+M}$ $\equiv$ $0$. 
The fugacities $x_{\alpha}$ and $z_{\alpha}$ are now associated to 
the topological and flavor symmetries for 
3d quiver gauge theory $\widetilde{\begin{smallmatrix}
(1)^{N-1}&-&(1)&-&(1)^{N_{f}-1}&-&(1)&-&(1)^{M-1}\\
&&|&&&&|&& \\
&&[1]&&&&[1]&& \\
\end{smallmatrix}}$. 
The terms from the second to fourth line are 
the two half-indices of Neumann b.c. $\mathcal{N}$ for 4d $U(1)$ gauge theory 
and the full-index for 3d $U(1)^{N+N_{f}+M-1}$ twisted vector multiplet. 
The terms appearing in the next two lines describe the contributions from 3d twisted hypermultiplets. 
The remaining terms count bare monopole of dimension 
\begin{align}
\label{u1u1hmNM_u1b_monodim}
\Delta(m)&=
\frac{|m_{1}|}{2}+\frac{|m_{N}|}{2}+\frac{|m_{N+N_{f}}|}{2}+\frac{|m_{N+N_{f}+M-1}|}{2}+\sum_{i=1}^{N+N_{f}+M-2}\frac{|m_{i}-m_{i+1}|}{2}. 
\end{align}

We expect that the half-index (\ref{u1u1hmNM_u1a}) is equal to (\ref{u1u1hmNM_u1b}). 
In fact, we have checked that 
they match up to certain orders of $q$ for 
$(N,M,N_{f})$ $=$ $(2,2,0)$, $(2,2,1)$, $(3,3,0)$ 
(see Appendix \ref{app_interface}).

\subsection*{Acknowledgements}
The author would like to thank Davide Gaiotto, Miroslav Rapcak, Junya Yagi and Masahito Yamazaki for useful discussions and comments. 
This work is supported in part by Perimeter Institute for Theoretical Physics and 
JSPS Overseas Research fellowships. Research at
Perimeter Institute is supported by the Government of Canada through the Department of
Innovation, Science and Economic Development and by the Province of Ontario through the
Ministry of Research, Innovation and Science.

\appendix
\section{Series expansions of indices}
\label{app_exp}
We only shows the several terms in the expansions of indices by using Mathematica. 

\subsection{3d full-indices}

\subsubsection{Wilson lines in SQED$_{1}$}
\label{app_wil}
We have checked that the 3d full-index (\ref{sqed1fullw1}) 
for the SQED$_{1}$ with a Wilson line operator $\mathcal{W}_{n}$ of charge $n$ 
and the 3d full-index (\ref{thmv1}) 
for the twisted hypermultiplet with a vortex line $\mathcal{V}_{n}$ 
agree up to $\mathcal{O}(q^{10})$ for $n=1,\cdots,5$. 
\begin{align}
\begin{array}{c|c|c}
\mathbb{I}^{\textrm{3d SQED$_{1}$}}_{\mathcal{W}_{n}}
&
\mathbb{I}^{\textrm{tHM}}_{\mathcal{V}_{n}}
&
\textrm{series expansions}
\\ \hline
n=1
&
n=1
&
{\scriptscriptstyle 
tq^{\frac14} 
+(x+x^{-1})q
-t^2(x+x^{-1})q^{\frac32}
+t^{-1}(x^{-2}+1+x^2)q^{\frac74}
+(x^{-1}+x)q^2+\cdots
}
\\ \hline 
n=2
&
n=2
&
{\scriptscriptstyle 
t^2 q^{\frac12}
+t(x^{-1}+x)q^{\frac74}
-t^{-3}(x^{-1}+x)q^{\frac94}
+t(x^{-1}+x)q^{\frac{11}{4}}
+(x^{-2}+1+x^2)q^3+\cdots
}
\\ \hline 
n=3
&
n=3
&
{\scriptscriptstyle 
t^3 q^{\frac34}
+t^{2}(x^{-1}+x)q^{\frac52}
-t^{4}(x^{-1}+x)q^3
+t^2 (x^{-1}+x)q^{\frac72}
-t^{4}(x^{-1}+x)q^{4}
+\cdots
}
\\ \hline 
n=4
&
n=4
&
{\scriptscriptstyle 
t^4q
+t^{3}(x^{-1}+x)q^{\frac{4}{13}}
-t^{5}(x^{-1}+x)q^{\frac{15}{4}}
+t^{3}(x^{-1}+x)q^{\frac{17}{4}}
-t^{5}(x^{-1}+x)q^{\frac{19}{4}}
+\cdots
}
\\ \hline 
n=5
&
n=5
&
{\scriptscriptstyle 
t^5 q^{\frac54}
+t^{4}(x^{-1}+x)q^{4}
-t^{6}(x^{-1}+x)q^{\frac{9}{2}}
+t^{4}(x^{-1}+x)q^5
-t^{6}(x^{-1}+x)q^{\frac{11}{2}}
+\cdots
}
\end{array}
\end{align}

\subsubsection{Abelian mirror symmetry}
\label{app_abe}
We have confirmed that  3d $\mathcal{N}=4$ full-indices for the following Abelian gauge theories agree up to $\mathcal{O}(q^{5})$.  
\begin{align}
\begin{array}{c|c|c}
\textrm{theories}
&
\textrm{mirror theories}
&
\textrm{series expansions}
\\ \hline
\scriptstyle{T[SU(2)]}
&
\scriptstyle{\widetilde{T[SU(2)]}}
&
{\scriptscriptstyle 
1+t^{-2}(1+\frac{x_{1}}{x_{2}}+\frac{x_{2}}{x_{1}}+t^4(1+\frac{z_{1}}{z_{2}}+\frac{z_{2}}{z_{1}}))q^{\frac12}
+\cdots
}
\\ \hline 
{\scriptstyle\textrm{SQED$_{3}$}}
&
{\scriptstyle \widetilde{[1]-(1)^2-[1]}}
&
{\scriptscriptstyle 
(1+(t^{-2}+t^2 x_{1}^{-2}x_{2}^{-1}x_{3}^{-1}(x_{1}+x_{2})(x_{1}+x_{3})(x_{2}+x_{3})))q^{\frac12}
+t^{-3}z_{1}^{-1}z_{2}^{-1}(z_{1}^2+z_{2}^2)q^{\frac34}
+\cdots
}
\\ \hline 
{\scriptstyle\textrm{SQED$_{4}$}}
&
{\scriptstyle \widetilde{[1]-(1)^3-[1]}}
&
{\scriptscriptstyle 
1+
(t^{-2}+x_{1}^{-1}x_{2}^{-1}x_{3}^{-1}x_{4}^{-1}t^2 
(x_{2}x_{3}x_{4}
(x_{2}+x_{3}+x_{4})+x_{1}^2 (x_{3}x_{4}+x_{2}
(x_{3}+x_{4}))}
\\ 
&
&
{\scriptscriptstyle+
x_{1}
(x_{2}^2(x_{3}+x_{4})+x_{3}x_{4}(x_{3}+x_{4})+x_{2} 
(x_{3}^2+3 x_{3}x_{4}+x_{4}^2))))q^{\frac12}
+\cdots}
\end{array}
\end{align}

\subsubsection{Non-Abelian mirror symmetry}
\label{app_nonabe}
We have confirmed that 
the 3d $\mathcal{N}=4$ full-indices for the following mirror pairs of 
non-Abelian gauge theories agree up to $\mathcal{O}(q^{3})$.  
\begin{align}
\begin{array}{c|c|c}
\textrm{theories}
&
\textrm{mirror theories}
&
\textrm{series expansions}
\\ \hline
\scriptstyle{(2)-[4]}
&
\scriptstyle{\widetilde{\begin{smallmatrix}
\\
(1)-&(2)&-(1)\\
&|&\\
&[2]&\\
\end{smallmatrix}
}}
&
{\scriptscriptstyle 
1+t^{-2}x_{1}^{-2}x_{2}^{-1}x_{3}^{-1}x_{4}^{-1}z_{1}^{-1}z_{2}^{-1}
(t^4 x_{1}^2 (x_{3}x_{4}+x_{2}(x_{3}+x_{4}))z_{1}z_{2}
+x_{1}(t^4 x_{2}^{2}(x_{3}+x_{4})z_{1}z_{2}
}
\\ 
&
&
{\scriptscriptstyle
+t^{4}x_{3}x_{4}(x_{3}+x_{4})z_{1}z_{2}
+x_{2}(t^4 x_{3}^{2}z_{1}z_{2}+t^{4}x_{4}^2 z_{1}z_{2}+x_{3}x_{4} (z_{1}^2+z_{1}z_{2}+3t^4z_{1}z_{2}+z_{2}^2
))
)
)q^{\frac12}}
\\ \hline
{\scriptstyle T[SU(3)]}
&
{\scriptstyle \widetilde{T[SU(3)]}}
&
{\scriptscriptstyle 
1+(
t^2 x_{1}^{-1}x_{2}^{-1}x_{3}^{-1} (x_{1}+x_{2}) (x_{1}+x_{3}) (x_{2}+x_{3})
+t^{-2}z_{1}^{-1}z_{2}^{-1}z_{3}^{-1}(z_{1}+z_{2})(z_{1}+z_{3})(z_{2}+z_{3})
)q^{\frac12}
+\cdots
}
\\ \hline 
{\scriptstyle 
\begin{smallmatrix}
\\
(1)&-&(2)&-&(1)\\
|&&|&&|\\
[1]&&[2]&&[1]\\
\end{smallmatrix}
}
&
{\scriptstyle 
\widetilde{
\begin{smallmatrix}
\\
(1)&-&(2)&-&(1)\\
|&&|&&|\\
[1]&&[2]&&[1]\\
\end{smallmatrix}}
}
&
{\scriptscriptstyle 
1+
t^{-2}
\left(
3+t^4\left( 3+x_{2}x_{3}^{-1}+x_{3}x_{2}^{-1} \right)
+z_{2}z_{3}^{-1}+z_{3}z_{2}^{-1}
\right)
q^{\frac12}
}
\\ 
&
&
{\scriptscriptstyle
+\left(
t^{3}
\frac{
\left(x_{2}+x_{3}\right)
\left(x_{1}+x_{4}\right)
\left(x_{2}x_{3}+x_{1}x_{4}\right)}
{
x_{1}x_{2}x_{3}x_{4}
}
+
t^{-3}
\frac{
\left(z_{2}+z_{3}\right)
\left(z_{1}+z_{4}\right)
\left(z_{2}z_{3}+z_{1}z_{4}\right)
}
{
z_{1}z_{2}z_{3}z_{4}
}
\right)q^{\frac34}+\cdots
}
\\ \hline
\scriptstyle{
\begin{smallmatrix}
\\
(1)&-&(2)&-&(1)\\
&&|&&|\\
&&[2]&&[1]\\
\end{smallmatrix}
}
&
\scriptstyle{\widetilde{\begin{smallmatrix}
\\
(2)-&(1)\\
|&|\\
[3]&[1]\\
\end{smallmatrix}
}}
&
{\scriptscriptstyle 
1+t^{-2}
\left(
3+t^4 \frac{\left(x_{1}+x_{2}\right)^2}{x_{1}+x_{2}}
+z_{1}\left(z_{2}^{-1}+z_{3}^{-1}\right)
+\frac{z_{2}}{z_{3}}
+\frac{z_{3}}{z_{2}}
+\frac{z_{2}+z_{3}}{z_{1}}
\right)q^{\frac12}
}
\\ 
&
&
{\scriptscriptstyle 
+\left(
t^{3}
\left(
\frac{x_{1}}{x_{3}}
+\frac{x_{2}}{x_{3}}
+\frac{x_{3}}{x_{1}}
+\frac{x_{3}}{x_{2}}
\right)
+t^{-3}
\left(
\frac{z_{1}}{z_{4}}
+\frac{z_{2}}{z_{4}}
+\frac{z_{3}}{z_{4}}
+\frac{z_{4}}{z_{1}}
+\frac{z_{4}}{z_{2}}
+\frac{z_{4}}{z_{3}}
\right)
\right)q^{\frac34}+\cdots
}
\\ \hline
\scriptstyle{(1)-(2)-[4]}
&
\scriptstyle{\widetilde{\begin{smallmatrix}
\\
(2)-&(2)&-(1)\\
|&|&\\
[2]&[1]&\\
\end{smallmatrix}
}}
&
{\scriptscriptstyle 
1+t^{-2}
\frac{
t^{4}x_{2}x_{3}x_{4}(x_{2}+x_{3}+x_{4})z_{1}z_{2}
+t^4 x_{1}^2(x_{3}x_{4}+x_{2}(x_{3}+x_{4}))z_{1}z_{2}
+x_{1}(t^4 x_{2}^2(x_{3}+x_{4})z_{1}z_{2}
}
{t^2 x_{1}x_{2}x_{3}x_{4}z_{1}z_{2}}
}
\\ 
&
&
{\scriptscriptstyle
\frac{
+t^4 x_{3}x_{4} (x_{3}+x_{4}) z_{1}z_{2}
+x_{2}(t^4 x_{3}^2z_{1}z_{2}+t^4 x_{4}^2 z_{1}z_{2}+x_{3}x_{4} (z_{1}^2+(2;3t^4)z_{1}z_{2}+z_{2}^2))
)
}
{}
q^{\frac12}+\cdots
}
\end{array}
\end{align}

\subsubsection{Seiberg-like duality}
\label{app_seiberg}
We have checked that 
the pairs of full-indices for Seiberg-like dual 3d $\mathcal{N}=4$ gauge theories agree up to $\mathcal{O}(q^{3})$.  
\begin{align}
\begin{array}{c|c|c}
\textrm{ugly}
&
\textrm{good}
&
\textrm{series expansions}
\\ \hline
\scriptstyle{(2)-[3]}
&
\scriptstyle{\textrm{SQED$_3 +$ tHM}}
&
{\scriptscriptstyle 
1+t^{-1}\frac{(z_{1}^{2}+z_{2}^2)}{z_{1}z_{2}}q^{\frac14}
+\left(
\frac{t^2 (x_{1}+x_{2}) (x_{1};x_[3])(x_{2}+x_{3})}
{x_{1}x_{2}x_{3}}
+t^{-2}\frac{
(z_{1}^2+z_{2}^2)^2
}
{
z_{1}^2 z_{2}^2
}
\right)q^{\frac12}+\cdots
}
\\ \hline 
{\scriptstyle (3)-[5]}
&
{\scriptstyle (2)-[5]+\textrm{tHM}}
&
{\scriptscriptstyle 
1+t^{-1}\frac{(z_{1}^2+z_{2}^2)}{z_{1}z_{2}}q^{\frac14}
}
\\
&&
{\scriptscriptstyle 
+
\left(
t^2 
\frac{
x_{2}x_{3}x_{4}x_{5}
(x_{2}+x_{3}+x_{4}+x_{5})
+x_{1}^2 (x_{2}x_{4}x_{5}+x_{3}x_{4}x_{5}+x_{2}x_{3}(x_{4}+x_{5}))
 +x_{1}(x_{3}x_{4}x_{5}(x_{3}+x_{4}+x_{5})
}
{x_{1}x_{2}x_{3}x_{4}x_{5}}
\right.
}
\\
&&
{\scriptscriptstyle
 \left. 
\frac{
 +x_{2}^2(x_{4}x_{5}+x_{3}(x_{4}+x_{5}))
 +x_{2}(x_{3}^2(x_{4}+x_{5})+x_{4}x_{5}(x_{4}+x_{5}) 
 +x_{3}(x_{4}^2+4 x_{4}x_{5}+x_{5}^2))
 }{}
 +t^{-2}\frac{(z_{1}^2+z_{2}^2)^2}{z_{1}^2z_{2}^2}
\right)q^{\frac12}+\cdots
 }
\end{array}
\end{align}

\subsection{4d half-indices}

\subsubsection{Hafl-BPS boundary conditions}
\label{app_bc}
We have checked that the following pairs of half-indices of dual half-BPS boundary conditions in 
4d $\mathcal{N}=4$ gauge theories agree up to certain orders of $q$. 

\begin{align}
\begin{array}{c|c|c}
\textrm{half-BPS b.c.}
&
\textrm{series expansions}
&
\textrm{up to orders} 
\\ \hline
\scriptstyle{\textrm{4d $U(1)|$3d $U(1)$}}
&
{\scriptscriptstyle 
1+\frac{(z_{1}^2+z_{2}^2)}{t z_{1}z_{2}}q^{\frac14}
+\frac{(z_{1}+z_{2})^2}{t^2 z_{1}^2 z_{2}^2}q^{\frac12}
+\frac{
(z_{1}^2+z_{2}^2)(z_{1}^4-(-1+t^4)z_{1}^2z_{2}^2+z_{2}^4)
}{t^3 z_{1}^3 z_{2}^3}q^{\frac34}
+\cdots
}
& 
\mathcal{O}(q^{10})
\\
\scriptstyle{\textrm{4d $U(2)|$3d $U(1)$}}
&
{\scriptscriptstyle 
1+\frac{(z_{1}+z_{2})^2}{t^2 z_{1}z_{2}}q^{\frac12}
+\frac{
z_{1}^4-(-2+t^4)z_{1}^3 z_{2}-2(-2+t^4)z_{1}^{2}z_{2}^2-(-2+t^4)z_{1}z_{2}^3+z_{2}^4
}
{t^4 z_{1}^2z_{2}^2}
q
+\cdots
}
& 
\mathcal{O}(q^{5})
\\
\scriptstyle{\textrm{4d $U(3)|$3d $U(1)$}}
&
{\scriptscriptstyle 
1+\frac{2}{t^2}q^{\frac12}
+\frac{z_{1}^2+z_{2}^2}{t^3 z_{1}z_{2}}q^{\frac34}
+\left(-2+\frac{4}{t^4}\right)q
-\frac{(-2+t^4)(z_{1}^2+z_{2}^2)}
{t^5 z_{1}z_{2}}q^{\frac54}
+\cdots
}
& 
\mathcal{O}(q^{5})
\\
\scriptstyle{\textrm{4d $U(3)|$3d $U(2)$}}
&
{\scriptscriptstyle 
1+\frac{z_{1}^2+z_{2}^2}{t z_{1}z_{2}}q^{\frac14}
+\frac{z_{1}^4+3z_{1}^2z_{2}^2+z_{2}^4}{t^2 z_{1}^2 z_{2}^2}q^{\frac12}
+\frac{(z_{1}^2+z_{2}^2)(z_{1}^4-(-3+t^4)z_{1}^2z_{2}^2+z_{2}^4)}
{t^3 z_{1}^3 z_{2}^3}q^{\frac34}
+\cdots
}
& 
\mathcal{O}(q^{3})
\\
\scriptstyle{\textrm{4d $U(4)|$3d $U(2)$}}
&
{\scriptscriptstyle 
1+\frac{(z_{1}+z_{2})^2}{t^2 z_{1}z_{2}}q^{\frac12}
+\frac{
z_{1}^4-(-3+t^4)z_{1}^3z_{2}-2(-3+t^4)z_{1}^2 z_{2}^2
-(-3+t^4)z_{1}z_{2}^3+z_{2}^4
}
{t^4 z_{1}^2z_{2}^2}q
+\cdots
}
& 
\mathcal{O}(q^{2})
\\
\scriptstyle{\textrm{4d $U(5)|$3d $U(2)$}}
&
{\scriptscriptstyle 
1+\frac{2}{t^2}q^{\frac12}
+\frac{(z_{1}^2+z_{2}^2)}{t^3 z_{1}z_{2}}q^{\frac34}
+\left( -2+\frac{5}{t^4} \right)q
+\frac{(-3+t^4)(z_{1}+z_{2}^2)}{t^5 z_{1}z_{2}}q^{\frac54}
+\cdots
}
& 
\mathcal{O}(q^{\frac32})
\\ \hline
\scriptstyle{\textrm{4d $U(3)|$3d $U(2)\times U(1)$}}
&
{\scriptscriptstyle 
1+\frac{
\left( 
z_{1}^2 (z_{2}+z_{3})+z_{2}z_{3}(z_{2}+z_{3})+z_{1}(z_{2}^2+3z_{2}z_{3}+z_{3}^2)
\right)
}
{t^2 z_{1}z_{2}z_{3}}q^{\frac12}
+\cdots
}
& 
\mathcal{O}(q^{2})
\\ \hline
\scriptstyle{\textrm{4d $U(1)|$3d $(1)-[2]$}}
&
{\scriptscriptstyle 
1+
\frac{
2x_{1}x_{2}+t^4(x_{1}+x_{2})^2
}
{t^2 x_{1}x_{2}}q^{\frac12}
+
\frac{
z_{1}^2+z_{2}^2
}
{t^3 z_{1}z_{2}}q^{\frac34}
+\frac{
3x_{1}^2x_{2}^2-3t^4 x_{1}^2 x_{2}^2+t^8 (x_{1}^2+x_{1}x_{2}+x_{2}^2)^2
}
{t^4 x_{1}^2 x_{2}^2}q
+\cdots
}
& 
\mathcal{O}(q^{3})
\\
\scriptstyle{\textrm{4d $U(2)|$3d $(1)-[2]$}}
&
{\scriptscriptstyle 
1+
\frac{2x_{1}x_{2}+t^4 (x_{1}+x_{2})^2}{t^2 x_{1}x_{2}}q^{\frac12}
+\left(
-2+\frac{t^4 (x_{1}^2+x_{1}x_{2}+x_{2}^2)^2}{x_{1}^2 x_{2}^2}
+\frac{4+\frac{z_{1}}{z_{2}}+\frac{z_{2}}{z_{1}}}
{t^4}
\right)q
+\cdots
}
& 
\mathcal{O}(q^{3})
\\
\scriptstyle{\textrm{4d $U(2)|$3d $(2)-[4]$}}
&
{\scriptscriptstyle 
1+\left(
\frac{2}{t^2}
+\frac{t^2
(x_{2}x_{3}x_{4}(x_{2}+x_{3}+x_{4})
+x_{1}^2(x_{3}x_{4}+x_{2}(x_{3}+x_{4}))
}
{}
\right.
}
& 
\mathcal{O}(q^{\frac32})
\\
&
{\scriptscriptstyle 
\left. 
\frac{
+x_{1}(x_{2}^2 (x_{3}+x_{4})
+x_{3}x_{4}(x_{3}+x_{4})
+x_{2}(x_{3}^2+4x_{3}x_{4}+x_{4}^2)
)
}
{x_{1}x_{2}x_{3}x_{4}}
\right)
q^{\frac12}
+\cdots
}
& \\
\scriptstyle{\textrm{4d $U(3)|$3d $(2)-[4]$}}
&
{\scriptscriptstyle 
1+\left(
\frac{2}{t^2}
+\frac{t^2
(x_{2}x_{3}x_{4}(x_{2}+x_{3}+x_{4})
+x_{1}^2(x_{3}x_{4}+x_{2}(x_{3}+x_{4}))
}
{}
\right.
}
& 
\mathcal{O}(q^{\frac32})
\\
&
{\scriptscriptstyle 
\left. 
\frac{
+x_{1}(x_{2}^2 (x_{3}+x_{4})
+x_{3}x_{4}(x_{3}+x_{4})
+x_{2}(x_{3}^2+4x_{3}x_{4}+x_{4}^2)
)
}
{x_{1}x_{2}x_{3}x_{4}}
\right)
q^{\frac12}
+\cdots
}
& \\ \hline
\scriptstyle{\textrm{4d $U(1)|T[SU(3)]$}}
&
{\scriptscriptstyle 
1+
\frac{
3+t^4
\left( 
3+x_{1}
\left(
\frac{1}{x_{2}}
+\frac{1}{x_{3}}
\right)
+\frac{x_{2}}{x_{3}}
+\frac{x_{3}}{x_{2}}
+\frac{x_{2}+x_{3}}{x_{1}}
\right)
+\frac{z_{1}}{z_{2}}
+\frac{z_{2}}{z_{1}}
}
{t^2}
q^{\frac12}
+\cdots
}
& 
\mathcal{O}(q^{2})
\\
\scriptstyle{\textrm{4d $U(2)|T[SU(3)]$}}
&
{\scriptscriptstyle 
1+
\frac{
3+t^4\left(
3+x_{1}\left(\frac{1}{x_{2}}+\frac{1}{x_{3}} \right)
+\frac{x_{2}}{x_{3}}
+\frac{x_{3}}{x_{2}}
+\frac{x_{2}+x_{3}}{x_{1}}
\right)
+\frac{z_{1}}{z_{2}}
+\frac{z_{2}}{z_{1}}
}
{t^2}
q^{\frac12}
+\cdots
}
& 
\mathcal{O}(q^{2})
\\
\end{array}
\end{align}

\subsubsection{Hafl-BPS interfaces}
\label{app_interface}
We have confirmed that 
the following pairs of half-indices of 3d dual interfaces agree up to certain orders of $q$. 

\begin{align}
\begin{array}{c|c|c}
\textrm{half-BPS interfaces}
&
\textrm{series expansions}
&
\textrm{up to orders} 
\\ \hline
\scriptstyle{\textrm{4d $U(1)|$3d $U(1)|$4d $U(1)$}}
&
{\scriptscriptstyle 
1+
\frac{z_{1}^2+(3+t^4)z_{1}z_{2}+z_{2}^2}
{t^2 z_{1}z_{2}}
q^{\frac12}
+\frac{
z_{1}^4-(-3+t^4)z_{1}^3 z_{2}
+(6-3t^4+tt^8)z_{1}^2z_{2}^2
-(-3+t^4)z_{1}z_{2}^3+z_{2}^4
}
{t^4 z_{1}^2 z_{2}^2}q
+\cdots
}
& 
\mathcal{O}(q^{5})
\\
\scriptstyle{\textrm{4d $U(1)|$3d $U(1)^2|$4d $U(1)$}}
&
{\scriptscriptstyle 
1+
\frac{
z_{1}^2 (z_{2}+z_{3})
+z_{2}z_{3}(z_{2}+z_{3})
+z_{1}(z_{2}^2+(4+t^4)z_{2}z_{3}+z_{3}^2)
}
{t^2 z_{1}z_{2}z_{3}}
q^{\frac12}
+\cdots
}
& 
\mathcal{O}(q^{2})
\\
\scriptstyle{\textrm{4d $U(2)|$3d $U(2)|$4d $U(2)$}}
&
{\scriptscriptstyle 
1+
\frac{z_{1}^2+(3+t^4)z_{1}z_{2}+z_{2}^2}{t^2 z_{1}z_{2}}q^{\frac12}
+\frac{z_{1}^4+4z_{1}^3z_{2}+2(5+t^8)z_{1}^2 z_{2}^2+4z_{1}z_{2}^3+z_{2}^4}{t^4 z_{1}^2 z_{2}^2}q
+\cdots
}
& 
\mathcal{O}(q^{3})
\\ 
\scriptstyle{\textrm{4d $U(2)|$3d $U(2)^2|$4d $U(2)$}}
&
{\scriptscriptstyle 
1+
\frac{
z_{1}^2 (z_{2}+z_{3})+z_{2}z_{3}(z_{2}+z_{3})
+z_{1}(z_{2}^2+(4+t^4)z_{2}z_{3}+z_{3}^2)
}
{t^2 z_{1}z_{2}z_{3}}q^{\frac12}
+\cdots
}
& 
\mathcal{O}(q^{1})
\\
\hline 
\scriptstyle{\textrm{4d $U(1)|$3d $U(1)|$4d $U(2)$}}
&
{\scriptscriptstyle 
1+\frac{3+t^4}{t^2}q^{\frac12}
+\frac{z_{1}^2+z_{2}^2}{t^3 z_{1}z_{2}}q^{\frac34}
+\left(
-2+\frac{7}{t^4}+t^4
\right)q
-
\frac{
(-3+t^4)(z_{1}^2+z_{2}^2)
}
{t^5 z_{1}z_{2}}q^{\frac54}
+\cdots
}
& 
\mathcal{O}(q^{3})
\\
\scriptstyle{\textrm{4d $U(2)|$3d $U(1)|$4d $U(2)$}}
&
{\scriptscriptstyle 
1+\frac{3+t^4}{t^2}q^{\frac12}
+\left(
-1+t^4+\frac{8+\frac{z_{1}}{z_{2}}+\frac{z_{2}}{z_{1}}}{t^4}
\right)q
+\frac{
-(-3+t^4)z_{1}^2+(16-5t^4+t^{12})z_{1}z_{2}
-(-3+t^4)z_{2}^2
}
{t^6 z_{1}z_{2}}
q^{\frac32}
+\cdots
}
& 
\mathcal{O}(q^{3})
\\
\scriptstyle{\textrm{4d $U(2)|$3d $U(1)|$4d $U(3)$}}
&
{\scriptscriptstyle 
1+\frac{3+t^4}{t^2}q^{\frac12}
+\left(-1+\frac{8}{t^4}+t^4 \right)q
+\frac{z_{1}^2+z_{2}^2}{t^5 z_{1}z_{2}}q^{\frac54}
+\frac{17-4t^4+t^{12}}{t^6}q^{\frac32}
+\cdots
}
& 
\mathcal{O}(q^{3})
\\
\scriptstyle{\textrm{4d $U(1)|$3d $U(2)|$4d $U(2)$}}
&
{\scriptscriptstyle 
1+\frac{z_{1}^2+z_{2}^2}{t z_{1}z_{2}}q^{\frac14}
+\frac{z_{1}^4+(4+t^4)z_{1}^2z_{2}^2+z_{2}^4}
{t^2 z_{1}^2 z_{2}^2}q^{\frac12}
+\frac{z_{1}^6+5 z_{1}^4z_{2}^2+5z_{1}^2z_{2}^4+z_{2}^6}
{t^3 z_{1}^3z_{2}^3}q^{\frac34}
+\cdots
}
& 
\mathcal{O}(q^{3})
\\
\scriptstyle{\textrm{4d $U(2)|$3d $U(2)|$4d $U(3)$}}
&
{\scriptscriptstyle 
1+\left(\frac{3}{t^2}+t^2\right)q^{\frac12}
+\left(\frac{z_{1}}{t^3 z_{2}}
+\frac{z_{2}}{t^3 z_{1}}
\right)q^{\frac34}
+\left(
\frac{9}{t^4}+2t^4
\right)q
+
\frac{
4(z_{1}^2+z_{2}^2)
}
{t^5 z_{1} z_{2}}q^{\frac54}
+\cdots
}
& 
\mathcal{O}(q^{2})
\\
\scriptstyle{\textrm{4d $U(1)|$3d $U(2)|$4d $U(3)$}}
&
{\scriptscriptstyle 
1
+\frac{
z_{1}^2+(3+t^4)z_{1}z_{2}+z_{2}^2
}
{t^2 z_{1}z_{2}}q^{\frac12}
+
\frac{z_{1}^4+4 z_{1}^3 z_{2}+
(9-t^4+t^8)z_{1}^2 z_{2}^2
+4 z_{1}z_{2}^3+z_{2}^4
}{t^4 z_{1}^2 z_{2}^2}q
+\cdots
}
& 
\mathcal{O}(q^{2})
\\
\scriptstyle{\textrm{4d $U(1)|$3d $U(2)|$4d $U(4)$}}
&
{\scriptscriptstyle 
1
+
\frac{3+t^4}{t^2}q^{\frac12}
+\frac{z_{1}^2+z_{2}^2}{t^3 z_{1}z_{2}}q^{\frac34}
+\left(-1+\frac{8}{t^4}+t^4 \right)q
+\frac{4(z_{1}^2+z_{2}^2)}{t^5 z_{1}z_{2}}q^{\frac54}
+\cdots
}
& 
\mathcal{O}(q^{2})
\\
\scriptstyle{\textrm{4d $U(1)|$3d $U(2)|$4d $U(5)$}}
&
{\scriptscriptstyle 
1
+
\frac{3+t^4}{t^2}q^{\frac12}
+\left(
-1+t^4+\frac{8+\frac{z_{1}}{z_{2}}+\frac{z_{2}}{z_{1}}}{t^4}
\right)q
+\frac{
4z_{1}^2+(17-4t^4+t^{12})z_{1}z_{2}+4z_{2}^2
}
{t^6 z_{1}z_{2}}q^{\frac32}
+\cdots
}
& 
\mathcal{O}(q^{2})
\\
\scriptstyle{\textrm{4d $U(2)|$3d $U(3)|$4d $U(3)$}}
&
{\scriptscriptstyle 
1
+
\frac{z_{1}^2+z_{2}^2}{t z_{1}z_{2}}q^{\frac14}
+\frac{z_{1}^4+(4+t^4)z_{1}^2 z_{2}^2+z_{2}^4}{t^2 z_{1}^2 z_{2}^2}
q^{\frac12}
+
\frac{
z_{1}^6 +5z_{1}^4z_{2}^2
+5z_{1}^2 z_{2}^4+z_{2}^6
}
{t^3 z_{1}^3 z_{2}^3}q^{\frac34}
+\cdots
}
& 
\mathcal{O}(q^{\frac32})
\\ \hline
\scriptstyle{\textrm{4d $U(1)|$3d SQED$_{1}|$4d $U(1)$}}
&
{\scriptscriptstyle 
1
+\frac{3+2t^4}{t^2}q^{\frac12}
+\frac{z_{1}^2+z_{2}^2}{t^3 z_{1}z_{2}}q^{\frac34}
+\left(
-2+\frac{6}{t^4}+3t^4
\right)q
-\frac{(-3+t^4)(z_{1}^2+z_{2}^2)}
{t^5 z_{1}z_{2}}q^{\frac54}
+\cdots
}
& 
\mathcal{O}(q^{3})
\\
\scriptstyle{\textrm{4d $U(1)|$3d SQED$_{2}|$4d $U(1)$}}
&
{\scriptscriptstyle 
1
+\left(
\frac{3}{t^2}
+t^2 \left(
3+\frac{x_{1}}{x_{2}}
+\frac{x_{2}}{x_{1}}
\right)
\right)q^{\frac12}
+
\left(
-1+\frac{x_{1}}{x_{2}}
+\frac{x_{2}}{x_{1}}
+t^4\left(
6+\frac{x_{1}^2}{x_{2}^2}
+\frac{3 x_{1}}{x_{2}}
+\frac{3 x_{2}}{x_{1}}
+\frac{x_{2}^2}{x_{1}^2}
\right)
+\frac
{6+\frac{z_{1}}{z_{2}}+\frac{z_{2}}{z_{1}}
}{t^4}
\right)q
+\cdots
}
& 
\mathcal{O}(q^{3})
\\
\scriptstyle{\textrm{4d $U(1)|$3d SQED$_{3}|$4d $U(1)$}}
&
{\scriptscriptstyle 
1
+\left(
\frac{3}{t^2}
+t^2
\left(
4+x_{1}\left(\frac{1}{x_{2}}+\frac{1}{x_{3}}\right)
+\frac{x_{2}}{x_{3}}
+\frac{x_{3}}{x_{2}}
+\frac{x_{2}+x_{3}}{x_{1}}
\right)
\right)q^{\frac12}
+\cdots
}
& 
\mathcal{O}(q^{3})
\\
\hline 
\scriptstyle{\textrm{4d $U(2)|$3d SQED$_{1}|$4d $U(3)$}}
&
{\scriptscriptstyle 
1
+
\frac{3+2t^4}{t^2}q^{\frac12}
+\frac{8+3t^8}{t^4}q
+\frac{z_{1}^2+(17-t^4+4t^{12})z_{1}z_{2}+z_{2}^2}
{t^6 z_{1}z_{2}}q^{\frac32}
+\cdots
}
& 
\mathcal{O}(q^{3})
\\
\scriptstyle{\textrm{4d $U(2)|$3d SQED$_{2}|$4d $U(3)$}}
&
{\scriptscriptstyle 
1
+\left(
\frac{3}{t^2}
+t^2\left(
3+\frac{x_{1}}{x_{2}}+\frac{x_{2}}{x_{1}}
\right)
\right)q^{\frac12}
+
\left(
1+\frac{8}{t^4}
+\frac{x_{1}}{x_{2}}
+\frac{x_{2}}{x_{1}}
+t^4 
\left(
6+\frac{x_{1}^2}{x_{2}^2}
+\frac{3 x_{1}}{x_{2}}
+\frac{3 x_{2}}{x_{1}}
+\frac{x_{2}^2}{x_{1}^2}
\right)
\right)q
+\cdots
}
& 
\mathcal{O}(q^{3})
\\
\scriptstyle{\textrm{4d $U(2)|$3d SQED$_{3}|$4d $U(3)$}}
&
{\scriptscriptstyle 
1
+\left(
\frac{3}{t^2}
+t^2 
\left(
4+x_{1}
\left(
\frac{1}{x_{2}}
+\frac{1}{x_{3}}
\right)
+\frac{x_{2}}{x_{3}}
+\frac{x_{3}}{x_{2}}
+\frac{x_{2}+x_{3}}{x_{1}}
\right)
\right)q^{\frac12}
+\cdots
}
& 
\mathcal{O}(q^{3})
\\
\hline 
\scriptstyle{\textrm{4d $U(1)_{2HM}|$3d $U(1)|$4d $U(1)_{2HM}$}}
&
{\scriptscriptstyle 
1
+
\frac{3+t^4\left(5+\frac{x_{1}}{x_{2}}
+\frac{x_{2}}{x_{1}}+\frac{x_{3}}{x_{4}}+\frac{x_{4}}{x_{3}} \right)
+\frac{z_{1}}{z_{2}}+
\frac{z_{2}}{z_{1}}
}{t^2}q^{\frac12}
+\cdots
}
& 
\mathcal{O}(q^{2})
\\ 
\scriptstyle{\textrm{4d $U(1)_{2HM}|$3d SQED$_{1}|$4d $U(1)_{2HM}$}}
&
{\scriptscriptstyle 
1
+
\left(
\frac{3}{t^2}+t^2
\left(
6+\frac{x_{1}}{x_{2}}
+\frac{x_{2}}{x_{1}}
+\frac{x_{3}}{x_{4}}
+\frac{x_{4}}{x_{3}}
\right)
\right)q^{\frac12}
+\cdots
}
& 
\mathcal{O}(q^{2})
\\ 
\scriptstyle{\textrm{4d $U(1)_{3HM}|$3d $U(1)|$4d $U(1)_{3HM}$}}
&
{\scriptscriptstyle 
1
+
\frac{
3+t^4
\left(
7+x_{1}
\left(
\frac{1}{x_{2}}+\frac{1}{x_{3}}
\right)
+\frac{x_{2}}{x_{3}}
+\frac{x_{3}}{x_{2}}
+\frac{x_{2}+x_{3}}{x_{1}}
+\frac{x_{4}}{x_{5}}
+\frac{x_{5}}{x_{4}}
+\frac{x_{4}}{x_{6}}
+\frac{x_{5}}{x_{6}}
+\frac{x_{6}}{x_{4}}
+\frac{x_{6}}{x_{5}}
\right)
+\frac{z_{1}}{z_{2}}
+\frac{z_{2}}{z_{1}}
}
{t^2}
q^{\frac12}
+\cdots
}
& 
\mathcal{O}(q^{2})
\\ 
\end{array}
\end{align}

\bibliographystyle{utphys}
\bibliography{ref}

\end{document}